\documentclass{article}
\usepackage{amssymb}

\usepackage{psfig}
\usepackage{graphicx}
\usepackage{amsmath}

\input{tcilatex}

\begin{document}

\title{Soliton self-modulation of the turbulence amplitude and plasma rotation}
\author{Florin Spineanu and Madalina Vlad \\
\textit{Association Euratom-C.E.A. sur la Fusion, C.E.A.-Cadarache,} \\
\textit{F-13108 Saint-Paul-lez-Durance, France} \\
and \\
\textit{Association Euratom-NASTI Romania,} \\
\textit{National Institute for Laser, Plasma and Radiation Physics,} \\
\textit{P.O.Box MG-36, Magurele, Bucharest, Romania} \\
E-mails: \textit{florin@drfc.cad.cea.fr}, \textit{madi@drfc.cad.cea.fr}}
\maketitle

\begin{abstract}
The space-uniform amplitude envelope of the Ion Temperature Gradient driven
turbulence is unstable to small perturbations and evolves to nonuniform,
soliton-like modulated profiles. The induced poloidal asymmetry of \ the
transport fluxes can generate spontaneous poloidal spin-up of the tokamak
plasma.
\end{abstract}

\tableofcontents

\section{Introduction}

The ion temperature gradient turbulence, considered a major source of
anomalous transport in the tokamak plasma, is characterized by the
coexistence of irregular patterns (randomly fluctuating field) and
intermittent robust cuasi-coherent structures. In closely related fluid
models (for exemple, in the physics of atmosphere) modulational
instabilities are known to produce solitary structures on the envelope of
the fluctuating field. In the case of the tokamak, this can be particularly
important since a poloidally nonuniform amplitude of the turbulence
generates nonuniform transport rates. For a sufficiently high nonuniformity
the torque arising via the mechanism initially mentioned by Stringer \cite
{Stringer} may overcame the rotation damping due to the poloidal magnetic
pumping. In this work we show that this can indeed be the case by proving
that the uniform poloidal envelope of the ITG turbulence is an unstable
state. It appears that the ITG turbulence has intrinsic resources to
generate poloidal rotation via a combination of mechanisms which is not
directly related to the Reynolds stress, inverse cascade, direct-ion loss,
or other classical sources of rotation. We provide an essentially
mathematical explanation of the instability of the solution consisting of
poloidally uniform envelope of the turbulence. Based on the
geometrico-algebraic method of solving integrable nonlinear differential
equations on periodic domains we invoke an existing result, that any
perturbation removes the degeneracies due to coincident eigenvalues in the
main spectrum of the Lax operator, thus changing the topology of the
hyperelliptic Riemann surface that provides the solution. The perturbed
solution separates exponentially (in function space) from the initial one
(uniform envelope) and this yields an exponential growth of the poloidal
nonuniformity. The magnitude of the torque may be comparable to the poloidal
magnetic damping.

There are many works in plasma physics related to the soliton dynamics \cite
{Diamond}, \cite{Itoh}, \cite{Spatschek1}, \cite{Spatschek2}, \cite{Horton1}%
, etc. In a recent paper the focusing solution has been used to study the
formation of coherent motion or intermittent patterns (\emph{streamers}) 
\cite{Diamond}. Our work is basically a mathematical approach, using well
developed technics related to the Inverse Scattering Transform method for
periodic domains. However, more physical analysis may become possible
combining the knowledge of the spectral properties of the ion turbulence
with the nonlinear stability properties.

This work consists of several parts that may appear as developing
separately: the derivation of the ion equation (barotropic equation), the
multiple space-time scales analysis, the solution of the Nonlinear
Schrodinger Equation and the stability of the solution, the torque arising
from the Stringer mechanism and the possibility of the rotation. For being
self-contained the paper also includes a review of the multiple space-time
analysis (from the atmospheric physics applications) and of the
geometric-algebraic method of integration of the Nonlinear Schrodinger
Equation. Even if these parts could be found in basic works (see the
references) they are included both for clarity and for their extreme
importance for further applications related to our problem or independent of
this.

\section{The slab model of the ion mode instability}

We consider cylindrical geometry with circular magnetic surfaces. Locally
the model can be reduced to a slab geometry with $\left( x,y\right) $
cartezian coordinates replacing respectively the radius and poloidal angle
coordinates $(r,\theta )$. At equilibrium the plasma parameters are constant
on the magnetic surfaces. The effects of the toroidicity and of the particle
drifts are not included instead the nonlinearity related to the ion
polarization drift is fully retained. The plasma model consists of the
continuity equations and the equations of motion for electrons and ions: 
\begin{equation}
\frac{\partial n_{\alpha }}{\partial t}+\mathbf{\nabla \cdot }\left(
n_{\alpha }\mathbf{v}_{\alpha }\right) =0  \label{contei}
\end{equation}
\begin{equation}
m_{\alpha }n_{\alpha }\left( \frac{\partial \mathbf{v}_{\alpha }}{\partial t}%
+\left( \mathbf{v}_{\alpha }\cdot \mathbf{\nabla }\right) \mathbf{v}_{\alpha
}\right) =-\mathbf{\nabla }p_{\alpha }-\mathbf{\nabla \cdot \pi }_{\alpha
}+e_{\alpha }n_{\alpha }\left( \mathbf{E+v}_{\alpha }\mathbf{\times B}%
\right) +\mathbf{R}_{\alpha }  \label{momeq}
\end{equation}
where $\alpha =e,i$. The friction forces $\mathbf{R}_{e}=-\mathbf{R}%
_{i}=-n\left| e\right| \mathbf{J}_{\parallel }/\sigma _{\parallel }$ , which
are important for the parallel electron momentum balance, vanish for
infinite plasma conductivity, which we will assume. The collisional
viscosity $\mathbf{\pi }_{e,i}$ will be neglected as well. However we will
need to include it later when we will consider the balance of the forces
contributing to the poloidal rotation. The electron and ion temperatures are
considered constant along the magnetic lines $\nabla _{\parallel }T_{e,i}=0$%
. The equilibrium quantities are perturbed by the wave potential $\phi $: $%
n=n_{0}+\widetilde{n}$. A sheared poloidal plasma rotation is included, and
we later will make explicit the corresponding part in the potential, $\phi
_{0}$.

The momentum conservation equations are used to determine the perpendicular
velocities of the electrons and ions. The parallel momentum conservation
equation for electrons, in the absence of \emph{dissipation} or \emph{drifts}
gives the adiabatic distribution of the density fluctuation. The velocities
are introduced in the continuity equations to find the dynamical equations
for the density and electric potential.

From the equations of motion for the \emph{ions} the velocities are obtained
in the form: 
\begin{equation}
\mathbf{v}_{i}=\mathbf{v}_{\perp i}=\mathbf{v}_{dia,i}+\mathbf{v}_{E}+%
\mathbf{v}_{pol,i}  \notag
\end{equation}
where the ion diamagnetic velocity is 
\begin{equation*}
\mathbf{v}_{dia,i}=\frac{1}{n_{i}}\frac{1}{m_{i}\Omega _{i}}\widehat{\mathbf{%
n}}\times \mathbf{\nabla }p_{i}
\end{equation*}
The versor of the magnetic field is $\widehat{\mathbf{n}}$; the versors
along the transversal coordinate axis $\left( x,y\right) $ will be noted $%
\left( \widehat{\mathbf{e}}_{x},\widehat{\mathbf{e}}_{y}\right) $. The 
\textbf{ion-polarization velocity} is: 
\begin{eqnarray}
\mathbf{v}_{pol,i} &=&-\frac{1}{\left| e\right| n_{i}B}n_{i}m_{i}\left( 
\frac{\partial \mathbf{v}_{E}}{\partial t}+\left( \mathbf{v}_{E}\cdot 
\mathbf{\nabla }\right) \mathbf{v}_{E}\right) \times \widehat{\mathbf{n}}
\label{ionpol} \\
&=&-\frac{1}{B\Omega _{i}}\left( \frac{\partial }{\partial t}+\left( \mathbf{%
v}_{E}\cdot \mathbf{\nabla }_{\perp }\right) \right) \mathbf{\nabla }_{\perp
}\phi  \notag
\end{eqnarray}
Using the notation $\frac{d}{dt}\equiv \frac{\partial }{\partial t}+\left( 
\mathbf{v}_{E}\cdot \mathbf{\nabla }_{\perp }\right) $ the perpendicular ion
velocity can be written 
\begin{eqnarray}
\mathbf{v}_{\perp ,i} &=&\mathbf{v}_{dia,i}+\mathbf{v}_{pol,i}+\frac{-%
\mathbf{\nabla }\phi \times \widehat{\mathbf{n}}}{B}  \label{vperpi} \\
&=&\frac{T_{i}}{\left| e\right| B}\frac{1}{n_{i}}\frac{dn_{i}}{dr}\widehat{%
\mathbf{e}}_{y}-\frac{1}{B\Omega _{i}}\frac{d}{dt}\left( \mathbf{\nabla }%
_{\perp }\phi \right) +\frac{-\mathbf{\nabla }\phi \times \widehat{\mathbf{n}%
}}{B}  \notag
\end{eqnarray}

The equation for the velocity of the \emph{electrons} is 
\begin{equation*}
\mathbf{v}_{e}=\mathbf{v}_{\perp ,e}+\mathbf{v}_{\parallel ,e}
\end{equation*}
\begin{eqnarray}
\mathbf{v}_{\perp e} &=&\mathbf{v}_{E}+\mathbf{v}_{dia,e}  \label{vperpe} \\
&=&\frac{-\mathbf{\nabla }\phi \times \widehat{\mathbf{n}}}{B}+\frac{1}{n}%
\frac{1}{m_{e}\Omega _{e}}\left( \widehat{\mathbf{n}}\times \mathbf{\nabla }%
p\right)  \notag \\
&=&\frac{-\mathbf{\nabla }\phi \times \widehat{\mathbf{n}}}{B}+\frac{T_{e}}{%
-\left| e\right| B}\frac{1}{n_{0}}\frac{dn_{0}}{dr}\widehat{\mathbf{e}}_{y}
\end{eqnarray}

We assume neutrality $n_{e}=n_{i}=n$ and introduce the expressions of the
velocities in the continuity equations for ions and for electrons.

The ion continuity equation is 
\begin{equation*}
\frac{\partial n}{\partial t}+\mathbf{\nabla \cdot }\left( n\mathbf{v}%
_{\perp ,i}\right) =0
\end{equation*}

The electron continuity equation is 
\begin{equation*}
\frac{\partial n}{\partial t}+\mathbf{\nabla }_{\perp }\mathbf{\cdot }\left(
n\mathbf{v}_{\perp ,e}\right) +\nabla _{\parallel }\left( nv_{\parallel
e}\right) =0
\end{equation*}
The equations are substracted , to obtain 
\begin{eqnarray*}
&&-\nabla _{\parallel }\left( nv_{\parallel e}\right) \\
&&-\frac{1}{B\Omega _{i}}\mathbf{\nabla }_{\perp }\cdot \left( n\frac{d}{dt}%
\mathbf{\nabla }_{\perp }\phi \right) \\
&&+\mathbf{\nabla }_{\perp }\cdot \left( n\mathbf{v}_{dia,i}\right) -\mathbf{%
\nabla }_{\perp }\cdot \left( n\mathbf{v}_{dia,e}\right) \\
&=&0
\end{eqnarray*}
From the last term in the left we get: 
\begin{eqnarray*}
\frac{dn_{0}}{dx}\widehat{\mathbf{e}}_{x}\cdot \mathbf{v}_{dia,i}+\left( 
\mathbf{\nabla }_{\perp }\widetilde{n}\right) \cdot \mathbf{v}_{dia,i} &=& \\
&=&-\frac{T_{i}}{T_{e}}\left( \frac{T_{e}}{\left| e\right| B}\frac{1}{n_{0}}%
\frac{dn_{0}}{dx}\right) \frac{\partial \widetilde{n}}{\partial y}
\end{eqnarray*}
Including the similar term for the electrons, we obtain 
\begin{eqnarray*}
&&-\nabla _{\parallel }\left( nv_{\parallel e}\right) \\
&&-\frac{1}{B\Omega _{i}}\mathbf{\nabla }_{\perp }\cdot \left( n\frac{d}{dt}%
\mathbf{\nabla }_{\perp }\phi \right) \\
&&-\frac{T_{i}}{T_{e}}\left( \frac{T_{e}}{\left| e\right| B}\frac{1}{n_{0}}%
\frac{dn_{0}}{dx}\right) \frac{\partial \widetilde{n}}{\partial y}-\left( 
\frac{T_{e}}{\left| e\right| B}\frac{1}{n_{0}}\frac{dn_{0}}{dx}\right) \frac{%
\partial \widetilde{n}}{\partial y} \\
&=&0
\end{eqnarray*}
or: 
\begin{equation}
-\left( 1+\frac{T_{i}}{T_{e}}\right) \left( \frac{T_{e}}{\left| e\right| B}%
\frac{1}{n_{0}}\frac{dn_{0}}{dx}\right) \frac{\partial \widetilde{n}}{%
\partial y}+\frac{1}{B\Omega _{i}}\mathbf{\nabla }_{\perp }\cdot \left( n%
\frac{d}{dt}\mathbf{\nabla }_{\perp }\phi \right) +\nabla _{\parallel
}\left( nv_{\parallel e}\right) =0  \label{eq2}
\end{equation}
From the continuity equation for electrons 
\begin{equation}
\frac{\partial n}{\partial t}+\mathbf{\nabla }_{\perp }\cdot \left( n\mathbf{%
v}_{\perp }\right) +\mathbf{\nabla }_{\parallel }\cdot \left( n\mathbf{v}%
_{\parallel }\right) =0  \label{eq1}
\end{equation}
we obtain 
\begin{eqnarray}
&&\frac{\partial \widetilde{n}}{\partial t}+\frac{1}{B}\frac{\partial \phi }{%
\partial y}\frac{dn_{0}}{dx}++\mathbf{v}_{dia,e}\cdot \mathbf{\nabla }%
_{\perp }\widetilde{n}+V_{0}\frac{\partial \widetilde{n}}{\partial y}+
\label{eqcon5} \\
&&+n_{0}\left( \mathbf{\nabla }_{\parallel }\cdot \mathbf{v}_{\parallel
}\right) +v_{\parallel }\nabla _{\parallel }\widetilde{n}  \notag \\
&=&0  \notag
\end{eqnarray}
where the seed poloidal velocity is $\mathbf{V}_{0}\left( x\right) =-\frac{%
d\phi _{0}\left( x\right) }{Bdx}\widehat{\mathbf{e}}_{y}$. The parallel
momenttum balance gives the parallel electron velocity 
\begin{equation}
\mathbf{v}_{\parallel e}=-\frac{\sigma _{\parallel }}{e^{2}}\mathbf{\nabla }%
_{\parallel }\left( \left| e\right| \phi +T_{e}\ln \left( n/n_{0}\right)
\right)  \label{eq3}
\end{equation}
In the absence of friction $\left( \sigma \rightarrow \infty \right) $ and
of particle drifts the electron response is adiabatic 
\begin{equation*}
\frac{\widetilde{n}}{n_{0}}=-\frac{\left| e\right| \phi }{T_{e}}
\end{equation*}
and the potential is determined from Eq.(\ref{eq2}). To develop separately
the ion-polarization drift term, we introduce the notation: 
\begin{eqnarray*}
W &\equiv &\frac{1}{B\Omega _{i}}\mathbf{\nabla }_{\perp }\mathbf{\cdot }%
\left[ \left( n_{0}+\widetilde{n}\right) \left( \frac{\partial }{\partial t}+%
\mathbf{v}_{E}\cdot \mathbf{\nabla }_{\perp }\right) \mathbf{\nabla }_{\perp
}\phi \right] \\
&=&\frac{1}{B\Omega _{i}}\mathbf{\nabla }_{\perp }\mathbf{\cdot }\left[
\left( n_{0}+\widetilde{n}\right) \mathbf{I}\right]
\end{eqnarray*}
where we make explicit the electric potential $\phi _{0}$ associated to the
initial plasma poloidal rotation, $V_{0}\widehat{\mathbf{e}}_{y}$. 
\begin{equation*}
\mathbf{I}\equiv \left( \frac{\partial }{\partial t}+\left( V_{0}\widehat{%
\mathbf{e}}_{y}+\widetilde{\mathbf{v}}\right) \cdot \mathbf{\nabla }_{\perp
}\right) \mathbf{\nabla }_{\perp }\left( \phi _{0}+\widetilde{\phi }\right)
\end{equation*}
We have 
\begin{eqnarray*}
\mathbf{I} &=&\frac{\partial }{\partial t}\left( \mathbf{\nabla }_{\perp
}\phi _{0}\right) + \\
&&+\frac{\partial }{\partial t}\left( \mathbf{\nabla }_{\perp }\widetilde{%
\phi }\right) + \\
&&+\left[ V_{0}\frac{\partial }{\partial y}+\left( \widetilde{\mathbf{v}}%
\cdot \mathbf{\nabla }_{\perp }\right) \right] \left( \mathbf{\nabla }%
_{\perp }\phi _{0}+\mathbf{\nabla }_{\perp }\widetilde{\phi }\right)
\end{eqnarray*}
or 
\begin{eqnarray*}
\mathbf{I} &=&\frac{\partial }{\partial t}\left( \widehat{\mathbf{e}}%
_{x}BV_{0}\right) + \\
&&+\frac{\partial }{\partial t}\left( \mathbf{\nabla }_{\perp }\widetilde{%
\phi }\right) + \\
&&+\left[ V_{0}\frac{\partial }{\partial y}+\left( \widetilde{\mathbf{v}}%
\cdot \mathbf{\nabla }_{\perp }\right) \right] \left( \mathbf{\nabla }%
_{\perp }\phi _{0}+\mathbf{\nabla }_{\perp }\widetilde{\phi }\right) \\
&=&\frac{\partial }{\partial t}\left( \widehat{\mathbf{e}}_{x}BV_{0}\right)
+V_{0}B\frac{\partial V_{0}}{\partial y}\widehat{\mathbf{e}}_{x}+ \\
&&+\frac{\partial }{\partial t}\left( \mathbf{\nabla }_{\perp }\widetilde{%
\phi }\right) +\left( \widetilde{\mathbf{v}}\cdot \mathbf{\nabla }_{\perp
}\right) BV_{0}\widehat{\mathbf{e}}_{x}+ \\
&&+V_{0}\frac{\partial }{\partial y}\left( \mathbf{\nabla }_{\perp }%
\widetilde{\phi }\right) +\left( \widetilde{\mathbf{v}}\cdot \mathbf{\nabla }%
_{\perp }\right) \left( \mathbf{\nabla }_{\perp }\widetilde{\phi }\right)
\end{eqnarray*}
with the relations 
\begin{eqnarray*}
\mathbf{\nabla }_{\perp }\widetilde{\phi } &=&B\left( \widetilde{\mathbf{v}}%
\times \widehat{\mathbf{n}}\right) \\
&=&\left( B\widetilde{v}_{y}\right) \widehat{\mathbf{e}}_{x}+\left( -B%
\widetilde{v}_{x}\right) \widehat{\mathbf{e}}_{y}
\end{eqnarray*}
\begin{equation*}
\widetilde{v}_{y}=\frac{1}{B}\frac{\partial \widetilde{\phi }}{\partial x}\;%
\text{and}\;\widetilde{v}_{x}=-\frac{1}{B}\frac{\partial \widetilde{\phi }}{%
\partial y}
\end{equation*}
After very simple calculations we obtain: 
\begin{eqnarray}
\frac{1}{B}I_{x} &=&\frac{\partial V_{0}}{\partial t}+\widetilde{v}_{y}\frac{%
\partial V_{0}}{\partial y}+V_{0}\frac{\partial V_{0}}{\partial y}
\label{ir} \\
&&+\frac{\partial \widetilde{v}_{y}}{\partial t}+\widetilde{v}_{x}\frac{%
\partial V_{0}}{\partial x}+V_{0}\frac{\partial \widetilde{v}_{y}}{\partial y%
}  \notag \\
&&+\widetilde{v}_{x}\frac{\partial \widetilde{v}_{y}}{\partial x}+\widetilde{%
v}_{y}\frac{\partial \widetilde{v}_{y}}{\partial y}  \notag
\end{eqnarray}
and 
\begin{eqnarray}
\frac{1}{B}I_{y} &=&-\frac{\partial \widetilde{v}_{x}}{\partial t}-V_{0}%
\frac{\partial \widetilde{v}_{x}}{\partial y}  \label{iy} \\
&&-\widetilde{v}_{x}\frac{\partial \widetilde{v}_{x}}{\partial x}-\widetilde{%
v}_{y}\frac{\partial \widetilde{v}_{x}}{\partial y}  \notag
\end{eqnarray}
It will be useful to calculate the derivaties of these quantities 
\begin{eqnarray}
\frac{1}{B}\frac{\partial I_{x}}{\partial x} &=&\frac{\partial }{\partial x}%
\frac{\partial V_{0}}{\partial t}+\frac{\partial \widetilde{v}_{y}}{\partial
x}\frac{\partial V_{0}}{\partial y}+\widetilde{v}_{y}\frac{\partial ^{2}V_{0}%
}{\partial y\partial x}+\frac{\partial V_{0}}{\partial x}\frac{\partial V_{0}%
}{\partial y}+V_{0}\frac{\partial ^{2}V_{0}}{\partial x\partial y}
\label{didr} \\
&&+\frac{\partial \widetilde{v}_{y}}{\partial t}\frac{\partial V_{0}}{%
\partial x}+\widetilde{v}_{x}\frac{\partial ^{2}V_{0}}{\partial x^{2}}+\frac{%
\partial }{\partial t}\frac{\partial \widetilde{v}_{y}}{\partial x}+  \notag
\\
&&+\frac{\partial V_{0}}{\partial x}\frac{\partial \widetilde{v}_{y}}{%
\partial y}+V_{0}\frac{\partial ^{2}\widetilde{v}_{y}}{\partial x\partial y}+
\notag \\
&&+\frac{\partial \widetilde{v}_{x}}{\partial x}\frac{\partial \widetilde{v}%
_{x}}{\partial y}+\widetilde{v}_{x}\frac{\partial ^{2}\widetilde{v}_{y}}{%
\partial x^{2}}+  \notag \\
&&+\frac{\partial \widetilde{v}_{y}}{\partial x}\frac{\partial \widetilde{v}%
_{y}}{\partial y}+\widetilde{v}_{y}\frac{\partial ^{2}\widetilde{v}_{y}}{%
\partial x\partial y}  \notag
\end{eqnarray}
and 
\begin{eqnarray}
\frac{1}{B}\frac{\partial I_{y}}{\partial y} &=&-\frac{\partial V_{0}}{%
\partial y}\frac{\partial \widetilde{v}_{x}}{\partial y}  \label{didy} \\
&&-\frac{\partial }{\partial t}\frac{\partial \widetilde{v}_{x}}{\partial y}%
-V_{0}\frac{\partial ^{2}\widetilde{v}_{y}}{\partial y^{2}}-  \notag \\
&&-\frac{\partial \widetilde{v}_{x}}{\partial y}\frac{\partial \widetilde{v}%
_{x}}{\partial x}-\widetilde{v}_{x}\frac{\partial ^{2}\widetilde{v}_{x}}{%
\partial x\partial y}-  \notag \\
&&-\frac{\partial \widetilde{v}_{y}}{\partial y}\frac{\partial \widetilde{v}%
_{x}}{\partial y}-\widetilde{v}_{y}\frac{\partial ^{2}\widetilde{v}_{x}}{%
\partial y^{2}}  \notag
\end{eqnarray}

The quantity denoted by $W$ takes the form 
\begin{eqnarray*}
W &=&\frac{1}{B\Omega _{i}}\left( \frac{dn_{0}}{dx}\right) I_{x}+\frac{1}{%
B\Omega _{i}}n_{0}\left( \frac{\partial I_{x}}{\partial x}\right) + \\
&&+\frac{1}{B\Omega _{i}}\left( \frac{\partial \widetilde{n}}{\partial x}%
\right) I_{x}+\frac{1}{B\Omega _{i}}\widetilde{n}\left( \frac{\partial I_{x}%
}{\partial x}\right) + \\
&&+\frac{1}{B\Omega _{i}}n_{0}\left( \frac{\partial I_{y}}{\partial y}%
\right) +\frac{1}{B\Omega _{i}}\left( \frac{\partial \widetilde{n}}{\partial
y}\right) I_{y}+\frac{1}{B\Omega _{i}}\widetilde{n}\left( \frac{\partial
I_{y}}{\partial y}\right)
\end{eqnarray*}

\subsection{\protect\bigskip The mode evolution in a fixed plasma rotation
profile}

We will assume that the mode evolves initially without perturbing the
equilibrium profiles, in particular the seed poloidal rotation. This allows
us to simplify the expressions above, taking: 
\begin{eqnarray*}
\frac{\partial V_{0}}{\partial y} &=&0 \\
\frac{\partial V_{0}}{\partial t} &=&0
\end{eqnarray*}
Then the first lines in the formulas Eqs.(\ref{ir}), (\ref{didr}), (\ref
{didy}) are zero. Let us consider in the expression of $W$ the part $W_{0}$
which \textbf{does not contain the fluctuating density} $\widetilde{n}$.
Writting 
\begin{equation*}
W=W_{0}+\widetilde{W}
\end{equation*}
we have 
\begin{eqnarray*}
W_{0} &\equiv &\frac{1}{\Omega _{i}}\left( \frac{dn_{0}}{dx}\right) \left[ 
\widetilde{v}_{x}\frac{\partial V_{0}}{\partial x}+V_{0}\frac{\partial 
\widetilde{v}_{y}}{\partial y}+\widetilde{v}_{x}\frac{\partial \widetilde{v}%
_{y}}{\partial x}+\widetilde{v}_{y}\frac{\partial \widetilde{v}_{y}}{%
\partial y}+\frac{\partial \widetilde{v}_{y}}{\partial t}\right] + \\
&&+\frac{1}{\Omega _{i}}n_{0}\mathbf{\nabla }_{\perp }\cdot \left( I_{x}%
\widehat{\mathbf{e}}_{x}+I_{y}\widehat{\mathbf{e}}_{y}\right)
\end{eqnarray*}
and 
\begin{equation*}
\widetilde{W}=\frac{1}{B\Omega _{i}}\left( \frac{\partial \widetilde{n}}{%
\partial x}\right) I_{x}+\frac{1}{B\Omega _{i}}\widetilde{n}\left( \frac{%
\partial I_{x}}{\partial x}\right) +\frac{1}{B\Omega _{i}}\left( \frac{%
\partial \widetilde{n}}{\partial y}\right) I_{y}+\frac{1}{B\Omega _{i}}%
\widetilde{n}\left( \frac{\partial I_{y}}{\partial y}\right)
\end{equation*}
Replacing the perturbed velocity by the perturbed potential, writting all
terms and summing, we get: 
\begin{eqnarray*}
W_{0} &=&\left( \frac{\partial }{\partial t}+V_{0}\frac{\partial }{\partial y%
}\right) \frac{1}{B}\frac{\partial \widetilde{\phi }}{\partial x}+\left( -%
\frac{1}{B}\frac{\partial \widetilde{\phi }}{\partial y}\right) \frac{dV_{0}%
}{dx}+\left( \frac{-\mathbf{\nabla }_{\perp }\phi \times \widehat{\mathbf{n}}%
}{B}\cdot \mathbf{\nabla }_{\perp }\right) \frac{\partial \widetilde{\phi }}{%
\partial x}+ \\
&&+\frac{1}{\Omega _{i}}n_{0}\left\{ \frac{1}{B}V_{0}\frac{\partial }{%
\partial y}\nabla _{\perp }^{2}\widetilde{\phi }+\right. \\
&&\hspace{1cm}+\frac{1}{B}\left( \frac{-\mathbf{\nabla }_{\perp }\phi \times 
\widehat{\mathbf{n}}}{B}\cdot \mathbf{\nabla }_{\perp }\right) \nabla
_{\perp }^{2}\widetilde{\phi }- \\
&&\hspace{1cm}-\frac{1}{B}\frac{\partial \widetilde{\phi }}{\partial y}\frac{%
d^{2}V_{0}}{dx^{2}}+ \\
&&\hspace{1cm}\left. +\frac{1}{B}\frac{\partial }{\partial t}\left( \nabla
_{\perp }^{2}\widetilde{\phi }\right) \right\}
\end{eqnarray*}
Collecting all what we have at this moment the ion continuity equation (\ref
{eq2}) becomes: 
\begin{eqnarray*}
&&\nabla _{\parallel }\left( nv_{\parallel e}\right) + \\
&&-\left( 1+\frac{T_{i}}{T_{e}}\right) \left( \frac{T_{e}}{\left| e\right| B}%
\frac{1}{n_{0}}\frac{dn_{0}}{dx}\right) \frac{\partial \widetilde{n}}{%
\partial y}+ \\
&&+\frac{1}{B\Omega _{i}}n_{0}\left\{ \left( \frac{\partial }{\partial t}%
+V_{0}\frac{\partial }{\partial y}\right) \nabla _{\perp }^{2}\widetilde{%
\phi }-V_{0}^{^{\prime \prime }}\frac{\partial \widetilde{\phi }}{\partial y}%
+\left( \frac{-\mathbf{\nabla }_{\perp }\phi \times \widehat{\mathbf{n}}}{B}%
\cdot \mathbf{\nabla }_{\perp }\right) \nabla _{\perp }^{2}\widetilde{\phi }%
\right\} \\
&&\hspace{2.5cm}+\frac{1}{B\Omega _{i}}\left( \frac{dn_{0}}{dx}\right) \left[
\left( \frac{\partial }{\partial t}+V_{0}\frac{\partial }{\partial y}\right) 
\widetilde{v}_{y}+\widetilde{v}_{x}\frac{\partial \widetilde{v}_{y}}{%
\partial x}+\widetilde{v}_{y}\frac{\partial \widetilde{v}_{y}}{\partial y}+%
\widetilde{v}_{x}\frac{\partial V_{0}}{\partial x}\right] \\
&&\hspace{2.5cm}+\widetilde{W}+\left\{ \text{terms of the first line in the
expressions of }I_{x}\text{ and derivatives}\right\} \\
&=&0
\end{eqnarray*}

In the above equation (which is exact) we shall make the following
approximations:

\begin{itemize}
\item  neglect the term containing the parallel electron velocity, assuming
infinite electric conductivity;

\item  neglect the term which contains $dn_{0}/dx$ since it is in the ratio $%
k:1/L_{n}$ with the other terms, and we consider 
\begin{equation*}
kL_{n}\ll 1
\end{equation*}

\item  neglect $\widetilde{W}$ ; these terms are in the ratio $\widetilde{n}%
/n_{0}$ with the terms which are retained;

\item  neglect of the terms in the first lines of the expressions for $I_{x}$
and $dI_{x}/dx$, $dI_{y}/dy$. (These are the terms in the curly brakets, the
last line). As explained above, we assume that the mode evolves in a
background of fixed rotation profile, $V_{0}\left( x\right) $.
\end{itemize}

The resulting equation is 
\begin{eqnarray*}
&&-B\Omega _{i}\left( 1+\frac{T_{i}}{T_{e}}\right) \left( \frac{T_{e}}{%
\left| e\right| B}\frac{1}{n_{0}}\frac{dn_{0}}{dx}\right) \frac{\partial }{%
\partial y}\left( \frac{\widetilde{n}}{n_{0}}\right) \hspace{1cm} \\
&&+\left( \frac{\partial }{\partial t}+V_{0}\frac{\partial }{\partial y}%
\right) \nabla _{\perp }^{2}\widetilde{\phi }-V_{0}^{^{\prime \prime }}\frac{%
\partial \widetilde{\phi }}{\partial y}+\left( \frac{-\mathbf{\nabla }%
_{\perp }\phi \times \widehat{\mathbf{n}}}{B}\cdot \mathbf{\nabla }_{\perp
}\right) \nabla _{\perp }^{2}\widetilde{\phi } \\
&=&0
\end{eqnarray*}
and, replace the adiabatic form of the density perturbation 
\begin{equation*}
\frac{\widetilde{n}}{n_{0}}=-\frac{\left| e\right| \widetilde{\phi }}{T_{e}}
\end{equation*}
We define 
\begin{equation*}
\beta \equiv B\Omega _{i}\left( 1+\frac{T_{i}}{T_{e}}\right) \left( \frac{%
T_{e}}{\left| e\right| B}\frac{1}{n_{0}}\frac{dn_{0}}{dx}\right) \frac{%
\left| e\right| }{T_{e}}=\Omega _{i}\left( 1+\frac{T_{i}}{T_{e}}\right) 
\frac{1}{L_{n}}
\end{equation*}
and obtain 
\begin{equation*}
\beta \frac{\partial \widetilde{\phi }}{\partial y}+\left( \frac{\partial }{%
\partial t}+V_{0}\frac{\partial }{\partial y}\right) \nabla _{\perp }^{2}%
\widetilde{\phi }-V_{0}^{^{\prime \prime }}\frac{\partial \widetilde{\phi }}{%
\partial y}+\left( \frac{-\mathbf{\nabla }_{\perp }\phi \times \widehat{%
\mathbf{n}}}{B}\cdot \mathbf{\nabla }_{\perp }\right) \nabla _{\perp }^{2}%
\widetilde{\phi }=0
\end{equation*}
which is the barotropic equation.

\subsection{Nondimensional form of the equation}

We consider that the ion mode extends in the spatial ($x$) direction over a
length $L$. A typical value for the sheared poloidal rotation is noted $%
U_{0} $. We make the replacements 
\begin{equation*}
y\rightarrow Ly
\end{equation*}
\begin{equation*}
t\rightarrow t\frac{L}{U_{0}}
\end{equation*}
\begin{equation*}
\widetilde{\phi }\rightarrow \phi \frac{T_{e}}{\left| e\right| }
\end{equation*}
\begin{equation*}
V_{0}\rightarrow UU_{0}
\end{equation*}
such that from now on $y,t,\phi $ and $U$ are nondimensional quantities. We
also change the radial coordinate into a dimensionless variable 
\begin{equation*}
x\rightarrow Lx
\end{equation*}
and rewrite the equation 
\begin{eqnarray*}
&&\left( \Omega _{i}\left( 1+\frac{T_{i}}{T_{e}}\right) \frac{1}{L_{n}}\frac{%
L^{2}}{U_{0}}\right) \frac{\partial \phi }{\partial y}+ \\
&&+\left( \frac{\partial }{\partial t}+U\frac{\partial }{\partial y}\right)
\nabla _{\perp }^{2}\phi \\
&&-\frac{d^{2}U}{dx^{2}}\frac{\partial \phi }{\partial y}+ \\
&&+\left( \frac{1}{U_{0}}\frac{T_{e}}{\left| e\right| }\frac{1}{B}\frac{1}{L}%
\right) \left[ \left( -\mathbf{\nabla }_{\perp }\phi \times \widehat{\mathbf{%
n}}\right) \cdot \mathbf{\nabla }_{\perp }\right] \nabla _{\perp }^{2}\phi \\
&=&0
\end{eqnarray*}
The coefficients are 
\begin{equation*}
\beta ^{\prime }\equiv \Omega _{i}\left( 1+\frac{T_{i}}{T_{e}}\right) \frac{1%
}{L_{n}}\frac{L^{2}}{U_{0}}\;\;\text{non-dimensional}
\end{equation*}
\begin{equation*}
\varepsilon \equiv \frac{1}{U_{0}}\frac{T_{e}}{\left| e\right| }\frac{1}{B}%
\frac{1}{L}\;\;\text{non-dimensional}
\end{equation*}

For an order of magnitude, $\varepsilon $ is the ratio of the diamagnetic
electron velocity to the rotation velocity $U_{0}$ multiplied by the ratio
of the density gradient length to the length of the spatial domain. This
quantity, $\varepsilon $ is in general smaller than unity.

The quantity $\beta ^{\prime }$ is the ratio of the ion cyclotron frequency
to the inverse of the time required to cross the spatial domain with the
typical flow velocity. Since the later $\left( U_{0}/L\right) $ involves
macroscopic quantities this ratio can be large. It is multiplied by the
ratio of the spatial length to the density gradient length (these quantities
can be comparable and the ratio not too different of unity).

We change the notations eliminating the primes. The equation becomes 
\begin{equation*}
\beta \frac{\partial \phi }{\partial y}+\left( \frac{\partial }{\partial t}+U%
\frac{\partial }{\partial y}\right) \nabla _{\perp }^{2}\phi -\frac{d^{2}U}{%
dx^{2}}\frac{\partial \phi }{\partial y}+\varepsilon \left[ \left( -\mathbf{%
\nabla }_{\perp }\phi \times \widehat{\mathbf{n}}\right) \cdot \mathbf{%
\nabla }_{\perp }\right] \nabla _{\perp }^{2}\phi =0
\end{equation*}

This is the \emph{barotropic} equation, known in the physics of the
atmosphere.

\section{Multiple time and space scale analysis of the ion mode equation
reduced to the barotropic form}

The quasigeostrophic barotropic atmospheric model \ leads to the \textbf{%
barotropic equation} (see Horton \cite{Horton1} where the quasigeostrophic
approximation is explained in relation with the Ertel's theorem). This
equation has been studied in the atmosphere science by means of the \emph{%
multiple space and time scales method}. Our aim is similar, \emph{i.e.} to
obtain an equation for the envelope of the fluctuating field $\phi $ in
order to study the possible poloidally nonuniform profiles of the turbulence
averaged amplitude. The connection between the original equation and the
final slow-time and large-spatial scales equations is not simple : much
numerical work is required in order to connect the input physical parameters
of the plasma with the formal coefficients in the final equations. The
detailed analytical work is presented in Ref(\cite{Tan}) and the necessary
formulas are reproduced here for convenience. To simplify the expressions
the following notation is introduced $\left[ \phi ,\nabla ^{2}\phi \right]
\equiv \left[ \left( -\mathbf{\nabla }_{\perp }\phi \times \widehat{\mathbf{n%
}}\right) \cdot \mathbf{\nabla }_{\perp }\right] \nabla _{\perp }^{2}\phi $
and the equation can be written 
\begin{equation*}
\left( \frac{\partial }{\partial t}+U\left( x\right) \frac{\partial }{%
\partial y}\right) \nabla ^{2}\phi +\left( \beta -\frac{d^{2}U\left(
x\right) }{dx^{2}}\right) \frac{\partial \phi }{\partial y}=-\varepsilon %
\left[ \phi ,\nabla ^{2}\phi \right]
\end{equation*}

The analysis on multiple scales starts by introducing the variables on
scales separated by the parameter $\varepsilon $: 
\begin{equation*}
T_{1}=\varepsilon t,\;\;T_{2}=\varepsilon ^{2}t
\end{equation*}
\begin{equation*}
Y_{1}=\varepsilon y,\;\;Y_{2}=\varepsilon ^{2}y
\end{equation*}
This gives for the derivatives 
\begin{equation}
\frac{\partial }{\partial t}\rightarrow \frac{\partial }{\partial t}%
+\varepsilon \frac{\partial }{\partial T_{1}}+\varepsilon ^{2}\frac{\partial 
}{\partial T_{2}}  \label{dertserie}
\end{equation}
\begin{equation}
\frac{\partial }{\partial y}\rightarrow \frac{\partial }{\partial y}%
+\varepsilon \frac{\partial }{\partial Y_{1}}+\varepsilon ^{2}\frac{\partial 
}{\partial Y_{2}}  \label{derxserie}
\end{equation}

The solution is adopted to be of the form 
\begin{equation}
\phi =\phi ^{\left( 1\right) }+\varepsilon \phi ^{\left( 2\right)
}+\varepsilon ^{2}\phi ^{\left( 3\right) }+\cdots  \label{psiserie}
\end{equation}
We denote the linear part of the operator in the equation by 
\begin{equation*}
L\equiv \left( \frac{\partial }{\partial t}+U\left( x\right) \frac{\partial 
}{\partial y}\right) \nabla ^{2}+\left( \beta -\frac{d^{2}U\left( x\right) }{%
dx^{2}}\right) \frac{\partial }{\partial y}
\end{equation*}
Substituting the Eq.(\ref{psiserie}) and the forms of the derivation
operators Eqs.(\ref{dertserie}, \ref{derxserie}) in the equation of motion
we obtain from the equality between the coefficients of the powers of the
variable $\varepsilon $: 
\begin{equation}
L\phi ^{\left( 1\right) }=0  \label{linear}
\end{equation}
\begin{eqnarray}
L\phi ^{\left( 2\right) } &=&  \label{linear2} \\
&=&\left( \frac{\partial }{\partial T_{1}}+U\frac{\partial }{\partial Y_{1}}%
\right) \nabla ^{2}\phi ^{\left( 1\right) }+\left( \beta -U^{\prime \prime
}\right) \frac{\partial \phi ^{\left( 1\right) }}{\partial Y_{1}}  \notag \\
&&-2\left( \frac{\partial }{\partial t}+U\frac{\partial }{\partial y}\right) 
\frac{\partial ^{2}\phi ^{\left( 1\right) }}{\partial y\partial Y_{1}} 
\notag \\
&&-\left[ \phi ^{\left( 1\right) },\nabla ^{2}\phi ^{\left( 1\right) }\right]
\notag
\end{eqnarray}
\begin{eqnarray}
L\phi ^{\left( 3\right) } &=&  \label{lpsi3} \\
&=&\left( \frac{\partial }{\partial T_{2}}+U\frac{\partial }{\partial Y_{2}}%
\right) \nabla ^{2}\phi ^{\left( 1\right) }-\left( \beta -U^{\prime \prime
}\right) \frac{\partial \phi ^{\left( 1\right) }}{\partial Y_{2}}  \notag \\
&&-2\left( \frac{\partial }{\partial t}+U\frac{\partial }{\partial y}\right) 
\frac{\partial ^{2}\phi ^{\left( 1\right) }}{\partial y\partial Y_{2}} 
\notag \\
&&-\left( \frac{\partial }{\partial t}+U\frac{\partial }{\partial y}\right) 
\frac{\partial ^{2}\phi ^{\left( 1\right) }}{\partial Y_{1}^{2}}  \notag \\
&&-2\left( \frac{\partial }{\partial T_{1}}+U\frac{\partial }{\partial Y_{1}}%
\right) \frac{\partial ^{2}\phi ^{\left( 1\right) }}{\partial y\partial Y_{1}%
}  \notag \\
&&-\left( \frac{\partial }{\partial T_{1}}+U\frac{\partial }{\partial Y_{1}}%
\right) \nabla ^{2}\phi ^{\left( 2\right) }  \notag \\
&&-2\left( \frac{\partial }{\partial t}+U\frac{\partial }{\partial y}\right) 
\frac{\partial ^{2}\phi ^{\left( 2\right) }}{\partial y\partial Y_{1}}%
-\left( \beta -U^{\prime \prime }\right) \frac{\partial \phi ^{\left(
2\right) }}{\partial Y_{1}}  \notag \\
&&-\frac{\partial \phi ^{\left( 2\right) }}{\partial Y_{1}}\frac{\partial }{%
\partial x}\nabla ^{2}\phi ^{\left( 1\right) }+\frac{\partial \phi ^{\left(
1\right) }}{\partial x}\frac{\partial }{\partial Y_{1}}\nabla ^{2}\phi
^{\left( 1\right) }  \notag \\
&&-2\left( \frac{\partial \phi ^{\left( 1\right) }}{\partial y}\frac{%
\partial ^{3}\phi ^{\left( 1\right) }}{\partial y\partial x\partial Y_{1}}-%
\frac{\partial \phi ^{\left( 1\right) }}{\partial x}\frac{\partial ^{3}\phi
^{\left( 1\right) }}{\partial y^{2}\partial Y_{1}}\right)  \notag \\
&&-\left[ \phi ^{\left( 1\right) },\nabla ^{2}\phi ^{\left( 2\right) }\right]
-\left[ \phi ^{\left( 2\right) },\nabla ^{2}\phi ^{\left( 1\right) }\right] 
\notag
\end{eqnarray}

The solution is represented on the space and time scales.

\subsection{The enveloppe equation}

The solution of the linear part of the operator, Eq.(\ref{linear}) is
adopted as the superposition of two propagating waves 
\begin{eqnarray}
\phi ^{\left( 1\right) } &=&A_{1}\left( T_{1},T_{2},Y_{1},Y_{2}\right)
\varphi _{1}\left( x\right) \exp \left( ik_{1}y-i\omega _{1}t\right)
\label{psiat1} \\
&&+A_{2}\left( T_{1},T_{2},Y_{1},Y_{2}\right) \varphi _{2}\left( x\right)
\exp \left( ik_{2}y-i\omega _{2}t\right)  \notag \\
&&+c.c.  \notag
\end{eqnarray}
The two functions $\varphi _{n}$, $n=1,2$, are solutions of the equation 
\begin{equation}
\frac{d^{2}\varphi _{n}}{dx^{2}}-\left( k_{n}^{2}-\frac{\beta -U^{\prime
\prime }}{U-c_{n}}\right) \varphi _{n}=0  \label{phin}
\end{equation}
with 
\begin{equation*}
\varphi _{n}\left( 0\right) =\varphi _{n}\left( L\right) =0
\end{equation*}
In this formula, $\left( 0,L\right) $ are the limits of the spatial domain
in the minor radius direction where the flow $U\left( x\right) \neq 0$ ; $%
c_{n}=\omega _{n}/k_{n}$ are the phase velocities. To find the solutions of
this equation we have to use numerical methods. This is a Schrodinger-type
equation where the potential depends on the energy, (since the phase
velocity $c_{n}$ depends on the wavenumber $k_{n}$). We must use numerical
methods suitable for Sturm-Liouville equations and in general the solution
exists for only particular values of the parameters. We choose to fix the
frequencies $\omega _{n}$ and consider $k_{n}$ as the parameter to be
determined. The problem is complicated by the possible occurence of
singularities in the potential due to resonances between the phase velocity
and the fixed zonal flow. We will avoid any resonance since the physical
content of the processes consisting of the direct energy transfer between
the fluctuations and the flow represents a different physical problem and
requires separate consideration. In the present work we examine the effect
of the turbulence on the poloidal rotation as being due to the nonuniform
diffusion rates and Stringer mechanism.

The parameters in the Eqs.(\ref{phin}) are $\beta $, $\varepsilon $ and the
sheared flow velocity $U(x)$. We take for $U\left( x\right) $ a symmetric
Gaussian-like profile\ shifted in amplitude and retain its maximum,
half-width and asymptotic value (\emph{i.e.} at $x=0$ or $L$) as parameters.
The two frequencies have fixed values during the eigenvalue search but they
must be considered free parameters as well. This builds up a large space of
parameters which should be sampled. For some point in this space the
eigenvalue problems Eqs.(\ref{phin}) are solved and the phase velocities
compared with the profile of $U\left( x\right) $. The presence of resonances
renders the set of parameters useless. Since (as will result from the
formuals below) there are also other possible resonances, the determination
of a useful solution $\left( k_{1},k_{2}\right) $ is difficult and requires
many trials.

The numerical methods that have been used in solving Eqs.(\ref{phin}) belong
to two classes: boundary value integration methods and respectively shooting
methods. When the sign of the ``potential'' in the equation is negative
everywhere on $(0,L)$ , which occurs for short wavelengths and smooth fluid
velocity profile the boundary conditions cannot be fulfilled except for
asymtotically, \emph{i.e.} assuming arbitrary small nonzero values at these
limits, since the solution decays exponentially. However most of the
boundary value integrators we have tried gave amplitudes of magnitudes
comparable to the boundary value, which essentially are homotopic to the
trivial vanishing solution (the equation is homogeneous) and render useless
this approach. We have to use a shooting method accepting the difficulty
that the initial value of the parameter to be determined (eigenvalue) must
be placed not far from the final value. We have used the specialized package
SLEIGN2 interactively. For the investigation of ranges of parameters most of
the calculations have been performed using the routine D02KEF of NAG. For
identical conditions the two numerical codes gave the same results within
the accepted tolerance.

Using the routine \textbf{D02KEF} we divide the integration range $\left(
0,L\right) $ into four intervals corresponding roughly to a different
behaviour of the potential. This should simplify the task of integrator. A
single matching point is assumed, usually at the centre. The tolerence is $%
10^{-8}$ which however does not exclude fake convergence in case of an
initialisation of the eigenvalue very far from the true solution..

\begin{figure}[tbp]
\centerline{
 \psfig{file=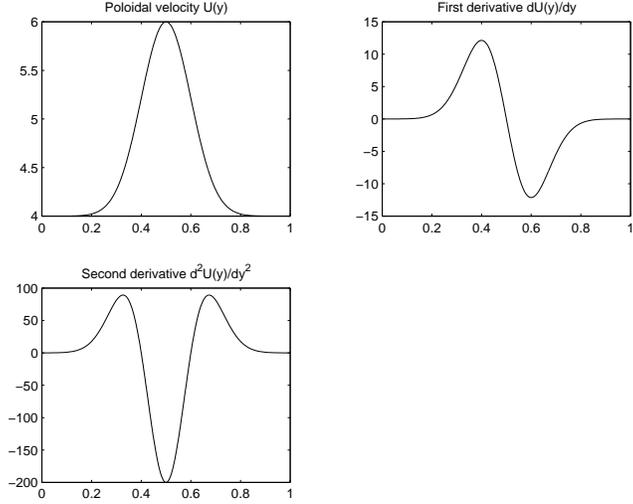,width=0.7\textwidth}}
\caption{Profile of the seed sheared poloidal velocity. Also are plotted the
first and the second derivatives.}
\label{figu}
\end{figure}

\begin{figure}[tbp]
\centerline{
 \psfig{file=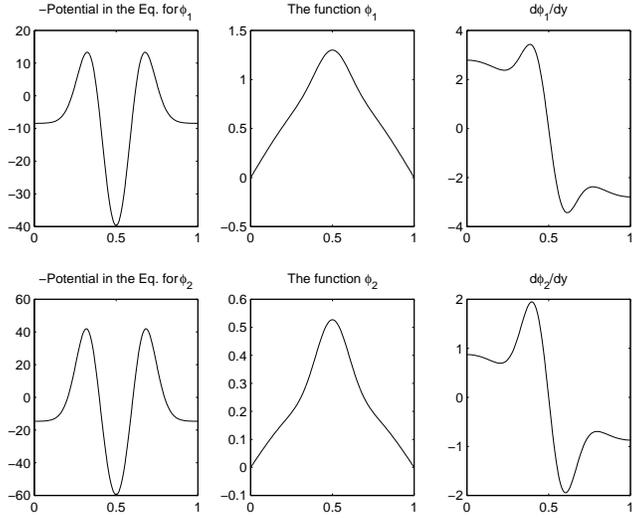,width=0.7\textwidth}}
\caption{The potentials in the Schrodinger-type equations and the
eigenfunctions $\protect\varphi _{1,2} $. The derivatives of the
eigenfunctions are aslo plotted.}
\label{figphi}
\end{figure}

We notice from Eqs.(\ref{phin}) that the potential can easily be dominated
by high shear $U^{\prime \prime }$ especially for smaller wavenumbers $k_{n}$%
. Actually we are more interested to investigate regimes where only mild
effects of the seed sheared rotation arise during the process of
self-modulation since, as explained before, the possible evolution to the
nonuniformity in the poloidal direction will be the source of the torque on
the plasma. Avoiding regimes of $U^{\prime \prime }$-dominated potentials we
assume low shear and take the Gaussian profile with $U_{\max }-U_{\min
}=2\times 10^{4}$ ($m/s$) represented in Figure (\ref{figu}). 
The following set of physical parameters has been used: minor radius $a=1\;m$%
, magnetic field $B_{T}=3.5\;T$, electron temperature $T_{e}=1\;KeV$, ion
temperature $T_{i}=1\;KeV$, density $n=10^{20}\;m^{-3}$, $L_{n}=10\;m$. The
parameters in the barotropic equation are then: $\ \beta =33.526$ and $%
\varepsilon =0.28571 $. The two frequencies are $\omega _{1}=85\times
10^{2}\;s^{-1}$ and $\omega _{2}=85\times 10^{4}\;s^{-1}$. It is assumed
that the space region of significant magnitude of $U\left( x\right) $ is $%
L=0.1\;m$. The following two eigenvalues are obtained, normalised to $L^{-1}$%
: $k_{1}=0.5525 $ and $k_{2}=3.2029$. The potentials in the Schrodinger-like
equations and the eigenfunctions are shown in Figure (\ref{figphi}). 

\begin{figure}[tbp]
\centerline{
 \psfig{file=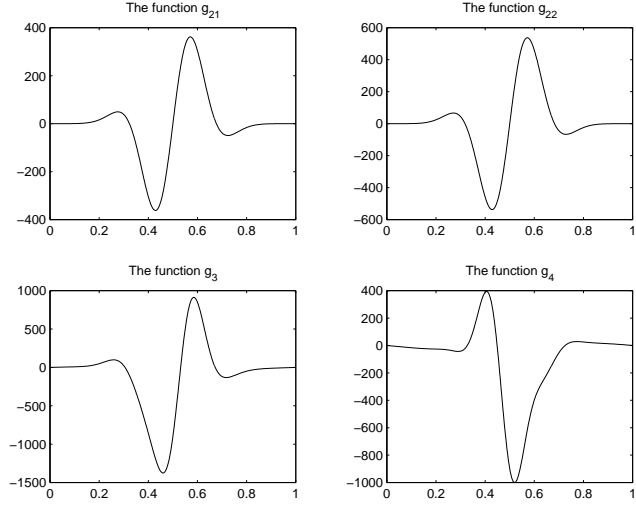,width=0.7\textwidth}}
\caption{Graphs of the functions $g$'s.}
\label{figg}
\end{figure}

The two eigenfunctions $\varphi _{1,2}$ have an important role in the
following steps of the calculation. Substituting the Eq.(\ref{psiat1}) in
the barotropic equation and expanding the operators, we obtain: 
\begin{eqnarray}
L\phi ^{\left( 2\right) } &=&-\sum_{n=1}^{2}G_{1n}\exp \left(
ik_{n}y-i\omega _{n}t\right)  \label{lpsi2} \\
&&-\sum_{n=1}^{2}ig_{2n}A_{n}^{2}\exp \left[ i2\left( k_{n}y-\omega
_{n}t\right) \right]  \notag \\
&&-g_{3}A_{1}A_{2}\exp \left[ i\left( k_{12}y-\omega _{12}t\right) \right] 
\notag \\
&&-ig_{4}A_{1}A_{2}^{\ast }\exp \left[ i\left( \alpha _{12}y-\sigma
_{12}t\right) \right]  \notag \\
&&+c.c.  \notag
\end{eqnarray}

where 
\begin{eqnarray*}
k_{12} &=&k_{1}+k_{2} \\
\omega _{12} &=&\omega _{1}+\omega _{2}
\end{eqnarray*}
We have $k_{12}=3.755$, $\omega =8.585$. 
\begin{eqnarray*}
\alpha _{12} &=&k_{1}-k_{2} \\
\sigma _{12} &=&\omega _{1}-\omega _{2}
\end{eqnarray*}
or: $\alpha _{12}=-2.650$, $\sigma _{12}=-8.415$. 
\begin{eqnarray*}
G_{1n} &=&\left( \frac{d^{2}\varphi _{n}}{dx^{2}}-k_{n}^{2}\varphi
_{n}\right) \frac{\partial A_{n}}{\partial T_{1}}+ \\
&&+\left[ U\left( \frac{d^{2}\varphi _{n}}{dx^{2}}-k_{n}^{2}\varphi
_{n}\right) +\left( \beta -U^{\prime \prime }\right) \varphi
_{n}-2k_{n}\left( U-c_{n}\right) \right] \frac{\partial A_{n}}{\partial Y_{1}%
}
\end{eqnarray*}
\begin{equation*}
g_{2n}=k_{n}\left( \varphi _{n}\frac{d}{dx}-\frac{d\varphi _{n}}{dx}\right) 
\frac{d^{2}\varphi _{n}}{dx^{2}}
\end{equation*}
\begin{eqnarray*}
g_{3} &=&\left( k_{1}\varphi _{1}\frac{d}{dx}-k_{2}\frac{d\varphi _{1}}{dx}%
\right) \left( \frac{d^{2}\varphi _{2}}{dx^{2}}-k_{2}^{2}\varphi _{2}\right)
\\
&&+\left( k_{2}\varphi _{2}\frac{d}{dx}-k_{1}\frac{d\varphi _{2}}{dx}\right)
\left( \frac{d^{2}\varphi _{1}}{dx^{2}}-k_{1}^{2}\varphi _{1}\right)
\end{eqnarray*}
\begin{eqnarray*}
g_{4} &=&\left( k_{1}\varphi _{1}\frac{d}{dx}+k_{2}\frac{d\varphi _{1}}{dx}%
\right) \left( \frac{d^{2}\varphi _{2}}{dx^{2}}-k_{2}^{2}\varphi _{2}\right)
\\
&&+\left( k_{2}\varphi _{2}\frac{d}{dx}+k_{1}\frac{d\varphi _{2}}{dx}\right)
\left( \frac{d^{2}\varphi _{1}}{dx^{2}}-k_{1}^{2}\varphi _{1}\right)
\end{eqnarray*}
These functions can be obtained after calculating the eigenfunctions $%
\varphi _{1,2}$ and are represented in Figure (\ref{figg}). 

Various solutions to the equation (\ref{lpsi2}) are possible, taking the
space-time dependence similar to one of the terms which compose the right
hand side.

The second term suggests solutions of the type 
\begin{equation*}
\phi _{2n}^{\left( 2\right) }=W_{2n}A_{n}^{2}\exp \left[ i2\left(
k_{n}y-\omega _{n}t\right) \right] +c.c.
\end{equation*}

The third term suggets solutions of the type 
\begin{equation*}
\phi _{3}^{\left( 2\right) }=W_{3}A_{1}A_{2}\exp \left[ i\left(
k_{12}y-\omega _{12}t\right) \right] +c.c.
\end{equation*}

The fourth term gives 
\begin{equation*}
\phi _{4}^{\left( 2\right) }=W_{4}A_{1}A_{2}^{\ast }\exp \left[ i\left(
\alpha _{12}y-\sigma _{12}t\right) \right] +c.c.
\end{equation*}

In the formulas above the functions $W_{2n}\left( x\right) $ are solutions
of the equations: 
\begin{eqnarray*}
\frac{d^{2}W_{2n}}{dx^{2}}-\left( 4k_{n}^{2}-\frac{\beta -U^{\prime \prime }%
}{U-c_{n}}\right) W_{2n} &=&\frac{g_{2n}}{2\left( Uk_{n}-\omega _{n}\right) }
\\
\text{with\ \ }W_{2n}\left( 0\right) &=&W_{2n}\left( L\right) =0
\end{eqnarray*}
\begin{eqnarray*}
\frac{d^{2}W_{3}}{dx^{2}}-\left( k_{12}^{2}-\frac{k_{12}\left( \beta
-U^{\prime \prime }\right) }{Uk_{12}-\omega _{12}}\right) W_{3} &=&\frac{%
g_{3}}{Uk_{12}-\omega _{12}} \\
\text{with\ \ }W_{3}\left( 0\right) &=&W_{3}\left( L\right) =0
\end{eqnarray*}
\begin{eqnarray*}
\frac{d^{2}W_{4}}{dx^{2}}-\left( \alpha _{12}^{2}-\frac{\alpha _{12}\left(
\beta -U^{\prime \prime }\right) }{U\alpha _{12}-\sigma _{12}}\right) W_{4}
&=&\frac{g_{4}}{U\alpha _{12}-\sigma _{12}} \\
\text{with\ \ }W_{4}\left( 0\right) &=&W_{4}\left( L\right) =0
\end{eqnarray*}
These equations are integrated numerically with a boundary value integrator,
a finite difference method. The results are shown in Figures (\ref{figw1}), (%
\ref{figw2}) and (\ref{figw3}). In the following figures are explained also
the functions $W_{1n}$ which will be introduced in Eq.(\ref{eqw1n}). 
\begin{figure}[tbp]
\centerline{
 \psfig{file=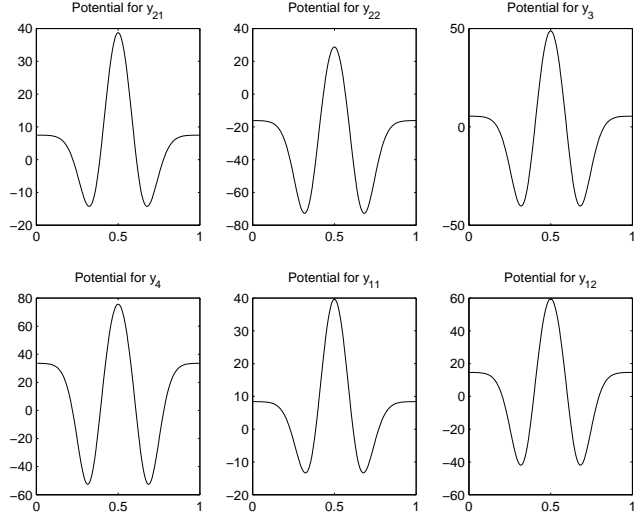,width=0.7\textwidth}}
\caption{Graphs of the potentials occuring in the equations for the
functions $w$'s.}
\label{figw1}
\end{figure}
\begin{figure}[tbp]
\centerline{
 \psfig{file=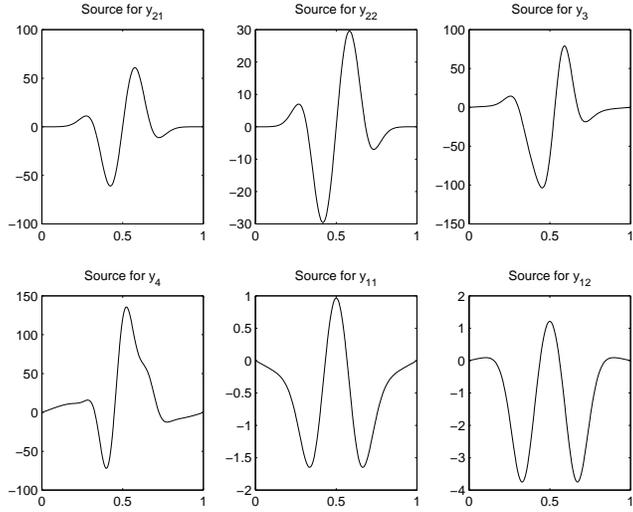,width=0.7\textwidth}}
\caption{Graphs of the source terms in the equations for the functions $w$%
's. }
\label{figw2}
\end{figure}
\begin{figure}[tbp]
\centerline{
 \psfig{file=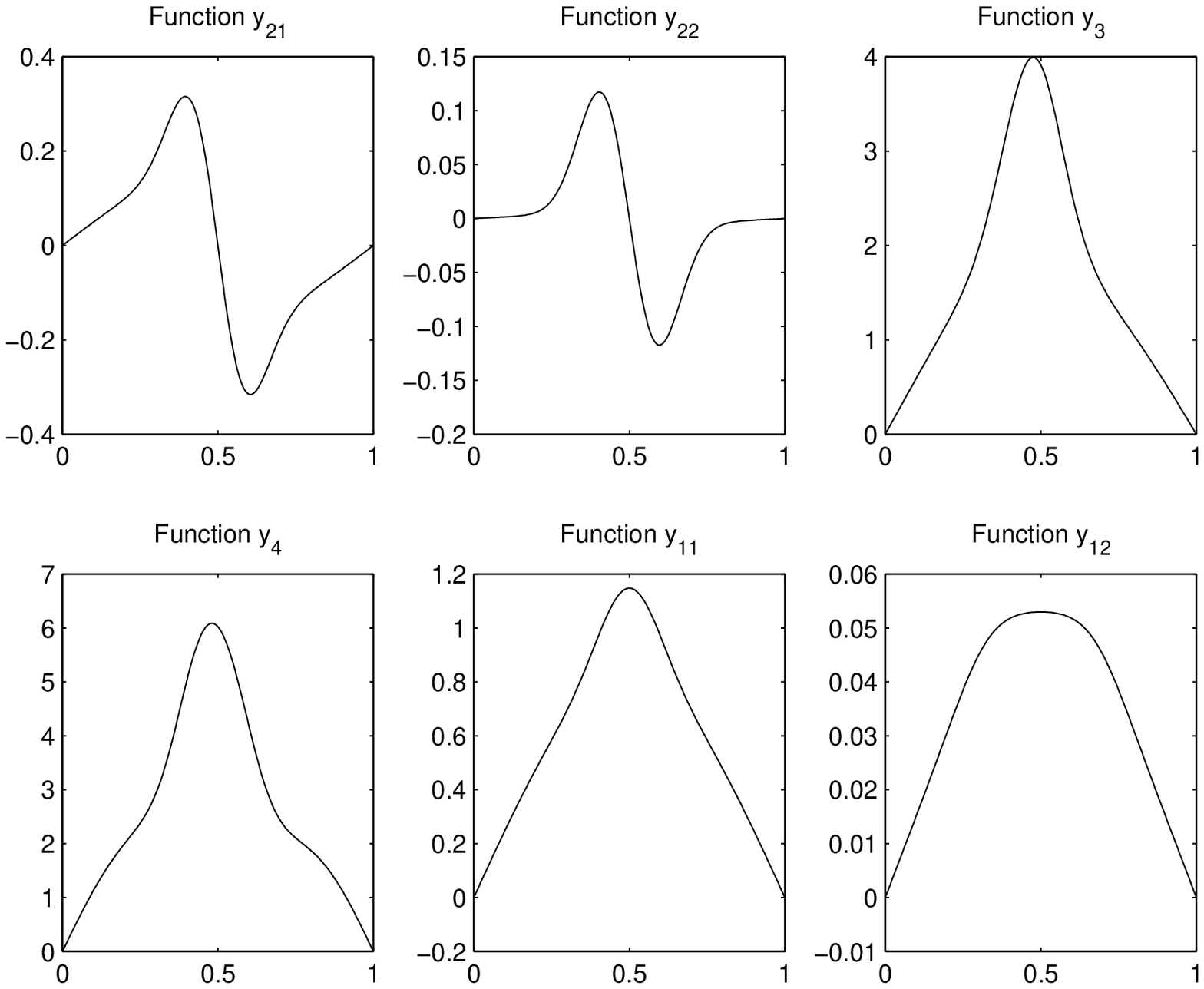,width=0.7\textwidth}}
\caption{Graphs of the functions $w$'s.}
\label{figw3}
\end{figure}

The first term gives a solution of the following form 
\begin{equation*}
\phi _{1n}^{\left( 2\right) }=\varphi _{1n}^{\left( 2\right) }\exp \left(
ik_{n}y-i\omega _{n}t\right) +c.c.
\end{equation*}
where the function $\varphi _{1n}^{\left( 2\right) }$ satisfies the equation 
\begin{eqnarray*}
\frac{d^{2}\varphi _{1n}^{\left( 2\right) }}{dx^{2}}-\left( k_{n}^{2}-\frac{%
\beta -U^{\prime \prime }}{U-c_{n}}\right) \varphi _{1n}^{\left( 2\right) }
&=& \\
&=&\frac{i}{k_{n}\left( U-c_{n}\right) }\left\{ \left( \frac{d^{2}\varphi
_{n}}{dx^{2}}-k_{n}^{2}\varphi _{n}\right) \frac{\partial A_{n}}{\partial
T_{1}}\right. \\
&&+\left[ U\left( \frac{d^{2}\varphi _{n}}{dx^{2}}-k_{n}^{2}\varphi
_{n}\right) +\left( \beta -U^{\prime \prime }\right) \varphi _{n}\right. \\
&&\left. \left. -2k_{n}^{2}\left( U-c_{n}\right) \right] \frac{\partial A_{n}%
}{\partial Y_{1}}\right\}
\end{eqnarray*}

This equation serves to generate a condition on the two amplitudes $A_{1,2}$.

When the left hand side of the equation is multiplied by the function $%
\varphi _{n}$ and integrated between the limits on the $x$ domain, it gives
zero, due to the boundary conditions. The same must then be true for the
right hand side. This gives a ``solubility condition'', on the space time
scales slower at order $1$: 
\begin{equation*}
\frac{\partial A_{n}}{\partial T_{1}}+c_{gn}\frac{\partial A_{n}}{\partial
Y_{1}}=0
\end{equation*}
where 
\begin{equation*}
c_{gn}=c_{n}+2\int_{0}^{L}k_{n}^{2}\varphi _{n}^{2}dx\diagup \int_{0}^{L}%
\frac{\beta -U^{\prime \prime }}{\left( U-c_{n}\right) ^{2}}\varphi
_{n}^{2}dx
\end{equation*}
Expressed in units of length $L$ and units of time $t_{0}=10^{-4}$, the 
\emph{group velocities} are : $c_{g1}=0.30328$ and $c_{g2}=3.9908$.

Then it is clear that the amplitude $A_{n}$ propagates at the speed $c_{gn}$
: 
\begin{equation*}
A_{n}=A_{n}\left( Y_{1}-c_{gn}T_{1},Y_{2},T_{2}\right)
\end{equation*}
Now we introduce a new function 
\begin{equation*}
\varphi _{1n}^{\left( 2\right) }=iW_{1n}\frac{\partial A_{n}}{\partial Y_{1}}
\end{equation*}
The equation satisfied by $W_{1n}$ is 
\begin{eqnarray}
&&\frac{d^{2}W_{1n}}{dx^{2}}-\left( k_{n}^{2}-\frac{\beta -U^{\prime \prime }%
}{u-c_{n}}\right) W_{1n}  \label{eqw1n} \\
&=&\frac{1}{k_{n}\left( U-c_{n}\right) }\left[ \left( c_{n}-c_{gn}\right)
\left( \frac{d^{2}\varphi _{n}}{dx^{2}}-k_{n}^{2}\varphi _{n}\right)
-2k_{n}^{2}\left( U-c_{n}\right) \varphi _{n}\right]  \notag
\end{eqnarray}
with the bounday conditions 
\begin{equation*}
W_{1n}\left( 0\right) =W_{1n}\left( L\right) =0
\end{equation*}

At this order $\left( 2\right) $ there are the following solutions 
\begin{eqnarray}
\phi ^{\left( 2\right) } &=&\phi _{11}^{\left( 2\right) }+\phi _{12}^{\left(
2\right) }+\phi _{21}^{\left( 2\right) }+\phi _{22}^{\left( 2\right) }
\label{psiat2} \\
&&+\phi _{3}^{\left( 2\right) }+\phi _{4}^{\left( 2\right) }+  \notag \\
&&+\Phi \left( x,T_{1},Y_{1}\right)  \notag
\end{eqnarray}
The last function will be determined by the condition of solubility at the
next order on the space-time scales ($O\left( \varepsilon ^{2}\right) $). To
obtain the evolution equations for $\Phi $ and $A_{n}$, we use the
expressions for $\phi ^{\left( 1\right) }$ Eq.(\ref{psiat1}) and $\phi
^{\left( 2\right) }$, Eq.(\ref{psiat2}) in the right hand side of the
equation (\ref{lpsi3}), and all inhomogeneous terms are obtained. In
additions there are terms which are not dependent on $y$ and $t$, and 
\textbf{these terms should be zero}. This leads to the solubility condition
relating the function $\Phi $ (correction to the mean flow due to the
presence of the wave packets) to the amplitude $A_{n}$. 
\begin{eqnarray}
&&\left( \frac{\partial }{\partial T_{1}}+U\frac{\partial }{\partial Y_{1}}%
\right) \frac{\partial ^{2}\Phi }{\partial x^{2}}+\left( \beta -U^{\prime
\prime }\right) \frac{\partial \Phi }{\partial Y_{1}}  \label{relphia} \\
&=&\sum_{n=1}^{2}\left\{ \frac{d^{2}}{dx^{2}}\left[ k_{n}\left( W_{1n}\frac{%
d\varphi _{n}}{dx}-\varphi _{n}\frac{dW_{1n}}{dx}\right) \right]
-4k_{n}^{2}\varphi _{n}\frac{d\varphi _{n}}{dx}\right.  \notag \\
&&\left. +\left( \varphi _{n}\frac{d}{dx}-\frac{d\varphi _{n}}{dx}\right)
\left( \frac{d^{2}\varphi _{n}}{dx^{2}}-k_{n}^{2}\varphi _{n}\right)
\right\} \frac{\partial }{\partial Y_{1}}\left| A_{n}\right| ^{2}  \notag
\end{eqnarray}
This relation will be used later.

Returning to the equation (\ref{lpsi3}), we examine the content of the
inhomogeneous part (the right hand side). There are terms which have a
space-time dependence 
\begin{equation}
\exp \left[ ik_{n}y-i\omega _{n}t\right]  \label{expkn}
\end{equation}
and in order to have solutions of this type, we need to impose solubility
conditions, otherwise there will arise secular terms at the order $\phi
^{\left( 3\right) }$. To find these restrictions (solubility conditions) we
multiply the equation with the function conjugated to (\ref{expkn}) 
\begin{equation*}
\exp \left[ -ik_{n}y+i\omega _{n}t\right]
\end{equation*}
and integrate time on the interval $\left( 0,2\pi /\omega _{n}\right) $ and
space: $y$ on the interval $\left( 0,2\pi /k_{n}\right) $ and $x$ on the
interval $\left( 0,L\right) $. On the left hand side we obtain zero, so this
is what we must obtain also from the right hand side. The conditions which
results are: 
\begin{equation*}
\left( \frac{\partial }{\partial T_{2}}+c_{g1}\frac{\partial }{\partial Y_{2}%
}\right) A_{1}-i\alpha _{1}\frac{\partial ^{2}A_{1}}{\partial Y_{1}^{2}}%
-i\left( \rho _{1}\left| A_{1}\right| ^{2}+\gamma _{12}\left| A_{2}\right|
^{2}+\lambda _{1}\right) A_{1}=0
\end{equation*}
\begin{equation*}
\left( \frac{\partial }{\partial T_{2}}+c_{g2}\frac{\partial }{\partial Y_{2}%
}\right) A_{2}-i\alpha _{2}\frac{\partial ^{2}A_{2}}{\partial Y_{1}^{2}}%
-i\left( \rho _{2}\left| A_{2}\right| ^{2}+\gamma _{21}\left| A_{1}\right|
^{2}+\lambda _{2}\right) A_{2}=0
\end{equation*}
The notations which have been introduced are: $\alpha _{n}=\frac{I_{n0}}{\Pi
_{n}}$, $\rho _{n}=\frac{I_{nn}}{\Pi _{n}}$, $\gamma _{12}=\frac{I_{12}}{\Pi
_{1}}$, $\gamma _{21}=\frac{I_{21}}{\Pi _{2}}$, $\lambda _{n}=\frac{I_{n3}}{%
\Pi _{n}}$ and 
\begin{equation*}
\Pi _{n}=-\int_{0}^{L}\frac{\varphi _{n}}{U-c_{n}}f_{n}dx\;,\;I_{n0}=%
\int_{0}^{L}\frac{\varphi _{n}}{U-c_{n}}f_{n0}dx
\end{equation*}
\begin{equation*}
I_{nn}=\int_{0}^{L}\frac{\varphi _{n}}{U-c_{n}}f_{nn}dx\;,\;I_{12}=%
\int_{0}^{L}\frac{\varphi _{1}}{U-c_{1}}f_{12}dx
\end{equation*}
\begin{equation*}
I_{21}=\int_{0}^{L}\frac{\varphi _{2}}{U-c_{2}}f_{21}dx\;,\;I_{n3}=%
\int_{0}^{L}\frac{\varphi _{n}}{U-c_{n}}f_{n3}dx
\end{equation*}

\begin{figure}[tbp]
\centerline{
 \psfig{file=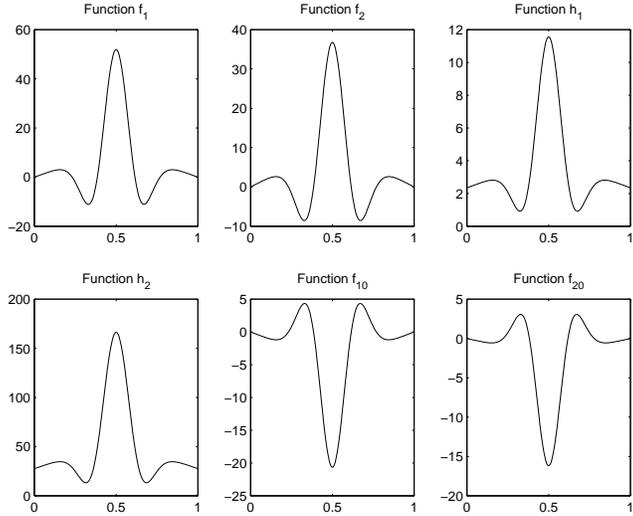,width=0.7\textwidth}}
\caption{Graphs of the first group of functions $f$.}
\label{figf1}
\end{figure}

\begin{figure}[tbp]
\centerline{
 \psfig{file=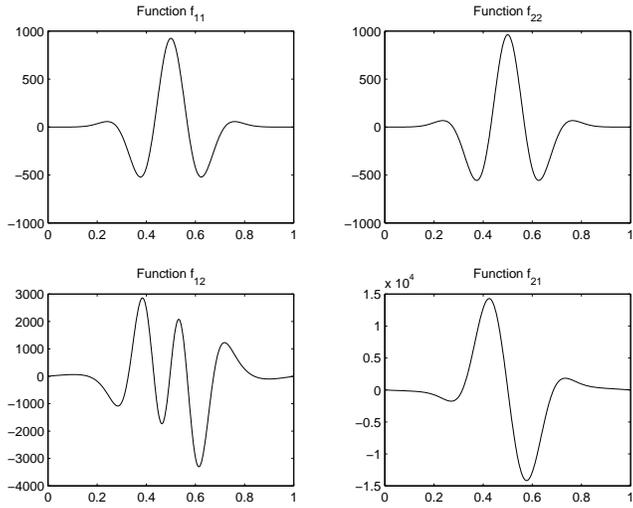,width=0.7\textwidth}}
\caption{Graphs of the second group of functions $f$.}
\label{figf2}
\end{figure}

The equations satisfied by all these functions are given given in Ref.(\cite
{Tan}). 
\begin{equation*}
f_{n}=-\left( \frac{d^{2}\varphi _{n}}{dx^{2}}-k_{n}^{2}\varphi _{n}\right)
\end{equation*}
\begin{equation*}
h_{n}=-U\left( \frac{d^{2}\varphi _{n}}{dx^{2}}-k_{n}^{2}\varphi _{n}\right)
-\left( \beta -U^{\prime \prime }\right) \varphi _{n}+2k_{n}^{2}\left(
U-c_{n}\right)
\end{equation*}
\begin{eqnarray*}
f_{n0} &=&-2k_{n}\left( U-c_{gn}\right) \varphi _{n}-k_{n}\left(
U-c_{n}\right) \\
&&-\left( U-c_{gn}\right) \left( \frac{d^{2}W_{1n}}{dy^{2}}%
-k_{n}^{2}W_{1n}\right) \\
&&+2k_{n}^{2}\left( U-c_{n}\right) W_{1n}-\left( \beta -U^{\prime \prime
}\right) W_{1n}
\end{eqnarray*}
\begin{eqnarray*}
f_{nn} &=&-\left( 2k_{n}W_{2n}\frac{d}{dx}+k_{n}\frac{dW_{2n}}{dx}\right)
\left( \frac{d^{2}\varphi _{n}}{dx^{2}}-k_{n}^{2}\varphi _{n}\right) \\
&&+\left( k_{n}\varphi _{n}\frac{d}{dx}+2k_{n}\frac{d\varphi _{n}}{dx}%
\right) \left( \frac{d^{2}W_{2n}}{dx^{2}}-4k_{n}^{2}W_{2n}\right)
\end{eqnarray*}
\begin{eqnarray*}
f_{12} &=&-\left( k_{12}W_{3}\frac{d}{dx}+k_{2}\frac{dW_{3}}{dx}\right)
\left( \frac{d^{2}\varphi _{2}}{dx^{2}}-k_{2}^{2}\varphi _{2}\right) \\
&&+\left( k_{2}\varphi _{2}\frac{d}{dx}+k_{12}\frac{d\varphi _{2}}{dx}%
\right) \left( \frac{d^{2}W_{3}}{dx^{2}}-k_{12}^{2}W_{3}\right) \\
&&+\left( k_{2}\varphi _{2}\frac{d}{dx}-\alpha _{12}\frac{d\varphi _{2}}{dx}%
\right) \left( \frac{d^{2}W_{4}}{dx^{2}}-\alpha _{12}^{2}W_{4}\right) \\
&&-\left( \alpha _{12}W_{4}\frac{d}{dx}-k_{2}\frac{dW_{4}}{dx}\right) \left( 
\frac{d^{2}\varphi _{2}}{dx^{2}}-k_{2}^{2}\varphi _{2}\right)
\end{eqnarray*}
\begin{eqnarray*}
f_{21} &=&-\left( k_{12}W_{3}\frac{d}{dx}+k_{1}\frac{dW_{3}}{dx}\right)
\left( \frac{d^{2}\varphi _{1}}{dx^{2}}-k_{1}^{2}\varphi _{1}\right) \\
&&-\left( k_{1}\varphi _{1}\frac{d}{dx}+\alpha _{12}\frac{d\varphi _{1}}{dx}%
\right) \left( \frac{d^{2}W_{4}}{dx^{2}}-\alpha _{12}^{2}W_{4}\right) \\
&&+\left( k_{1}\varphi _{1}\frac{d}{dx}+k_{12}\frac{d\varphi _{2}}{dx}%
\right) \left( \frac{d^{2}W_{3}}{dx^{2}}-k_{12}^{2}W_{3}\right) \\
&&+\left( \alpha _{12}W_{4}\frac{d}{dx}+k_{1}\frac{dW_{4}}{dx}\right) \left( 
\frac{d^{2}\varphi _{1}}{dx^{2}}-k_{1}^{2}\varphi _{1}\right)
\end{eqnarray*}
\begin{equation*}
f_{n3}=-k_{n}\left( \varphi _{n}\frac{\partial ^{3}\Phi }{\partial x^{3}}%
-\left( \frac{d^{2}\varphi _{n}}{dx^{2}}-k_{n}^{2}\varphi _{n}\right) \frac{%
\partial \Phi }{\partial x}\right)
\end{equation*}
We use a boundary value integrator for the inhomogeneous equations and (the
NAG routine \textbf{D02JAF}). The results are given in the Figures (\ref
{figf1}) and (\ref{figf2}). 

After numerical integrations the follwing values are obtained for the
constants: $\alpha _{1}=0.39976$, $\alpha _{2}=0.45246$, $\rho _{1}=-0.10182$%
, $\rho _{2}=1.8115$, $\gamma _{12}=0.091474$, $\gamma _{21}=-5.3768$.

The following transformation (Jeffrey and Kawahara) is performed in Ref.( 
\cite{Tan}) : 
\begin{eqnarray*}
T &=&T_{2} \\
X &=&\frac{1}{\varepsilon }\left( Y_{2}-c_{g1}T_{2}\right) =Y_{1}-c_{g1}T_{1}
\end{eqnarray*}
Then the equation of connection between $\Phi $ and $A_{n}$ becomes 
\begin{eqnarray}
&&\frac{\partial }{\partial Y}\left[ \left( U-c_{g1}\right) \frac{\partial
^{2}\Phi }{\partial x^{2}}+\left( \beta -U^{\prime \prime }\right) \Phi %
\right]  \label{eqphia} \\
&=&\sum_{n=1}^{2}\left[ k_{n}\frac{d^{2}}{dx^{2}}\left( W_{1n}\frac{d\varphi
_{n}}{dx}-\varphi _{n}\frac{dW_{1n}}{dx}\right) -4k_{n}\varphi _{n}\frac{%
d\varphi _{n}}{dx}\right.  \notag \\
&&\left. -\left( \varphi _{n}\frac{d}{dx}-\frac{d\varphi _{n}}{dx}\right)
\left( \frac{d^{2}\varphi _{n}}{dx^{2}}-k_{n}^{2}\varphi _{n}\right) \right] 
\frac{\partial }{\partial Y}\left| A_{n}\right| ^{2}  \notag
\end{eqnarray}

The solution of the equation (\ref{eqphia}) is considered to be of the form 
\begin{equation}
\Phi =\sum_{n=1}^{2}H_{n}\left( x\right) \left| A_{n}\right| ^{2}
\label{phimarea}
\end{equation}
and the equation for $H_{n}\left( x\right) $ results: 
\begin{eqnarray*}
&&\left( U-c_{g1}\right) \frac{\partial ^{2}H_{n}}{\partial x^{2}}+\left(
\beta -U^{\prime \prime }\right) H_{n} \\
&=&k_{n}\frac{d^{2}}{dx^{2}}\left( W_{1n}\frac{d\varphi _{n}}{dx}-\varphi
_{n}\frac{dW_{1n}}{dx}\right) -4k_{n}\varphi _{n}\frac{d\varphi _{n}}{dx} \\
&&-\left( \varphi _{n}\frac{d}{dx}-\frac{d\varphi _{n}}{dx}\right) \left( 
\frac{d^{2}\varphi _{n}}{dx^{2}}-k_{n}^{2}\varphi _{n}\right)
\end{eqnarray*}

\begin{figure}[tbp]
\centerline{
 \psfig{file=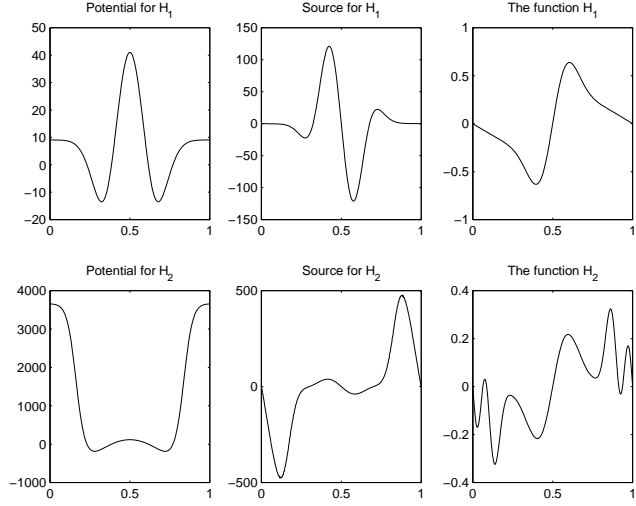,width=0.7\textwidth}}
\caption{Graphs of the functions $H$.}
\label{figH}
\end{figure}

The change of variables is performed in the equations for the amplitudes $%
A_{n}$ and, again, the Eq.(\ref{phimarea}) is used: 
\begin{equation*}
i\frac{\partial A_{1}}{\partial T}+\alpha _{1}\frac{\partial ^{2}A_{1}}{%
\partial Y^{2}}+\left( \sigma _{1}\left| A_{1}\right| ^{2}+\nu _{12}\left|
A_{2}\right| ^{2}\right) A_{1}=0
\end{equation*}
\begin{equation*}
i\left( \frac{\partial }{\partial T}-\delta \frac{\partial }{\partial Y}%
\right) A_{2}+\alpha _{2}\frac{\partial ^{2}A_{2}}{\partial Y^{2}}+\left(
\sigma _{2}\left| A_{2}\right| ^{2}+\nu _{21}\left| A_{1}\right| ^{2}\right)
A_{2}=0
\end{equation*}
where 
\begin{equation*}
\delta =\frac{1}{\varepsilon }\left( c_{g1}-c_{g2}\right)
\end{equation*}
\begin{equation*}
\sigma _{n}=\rho _{n}-\frac{1}{\Pi _{n}}\int_{0}^{L}\frac{k_{n}\varphi _{n}}{%
U-c_{n}}\left[ \varphi _{n}\frac{d^{3}H_{n}}{dx^{3}}-\left( \frac{%
d^{2}\varphi _{n}}{dx^{2}}-k_{n}^{2}\varphi _{n}\right) \frac{dH_{n}}{dx}%
\right] dx
\end{equation*}
\begin{equation*}
\nu _{12}=\gamma _{12}-\frac{1}{\Pi _{1}}\int_{0}^{L}\frac{k_{1}\varphi _{1}%
}{U-c_{1}}\left[ \varphi _{1}\frac{d^{3}H_{2}}{dx^{3}}-\left( \frac{%
d^{2}\varphi _{1}}{dx^{2}}-k_{1}^{2}\varphi _{1}\right) \frac{dH_{2}}{dx}%
\right] dx
\end{equation*}
\begin{equation*}
\nu _{21}=\gamma _{21}-\frac{1}{\Pi _{2}}\int_{0}^{L}\frac{k_{2}\varphi _{2}%
}{U-c_{2}}\left[ \varphi _{2}\frac{d^{3}H_{1}}{dx^{3}}-\left( \frac{%
d^{2}\varphi _{2}}{dx^{2}}-k_{2}^{2}\varphi _{2}\right) \frac{dH_{1}}{dx}%
\right] dx
\end{equation*}

The following transformation is convenient: 
\begin{equation*}
A_{2}=B\exp \left[ i\frac{\delta }{2\alpha _{2}}\left( Y+\frac{\delta }{2}%
T\right) \right]
\end{equation*}
This leads to the following equations: 
\begin{equation}
i\frac{\partial A_{1}}{\partial T}+\alpha _{1}\frac{\partial ^{2}A_{1}}{%
\partial Y^{2}}+\left( \sigma _{1}\left| A_{1}\right| ^{2}+\nu _{12}\left|
B\right| ^{2}\right) A_{1}=0  \label{cse1}
\end{equation}
\begin{equation}
i\frac{\partial B}{\partial T}+\alpha _{2}\frac{\partial ^{2}B}{\partial
Y^{2}}+\left( \sigma _{2}\left| B\right| ^{2}+\nu _{21}\left| A_{1}\right|
^{2}\right) B=0  \label{cse2}
\end{equation}

The constants are: $\sigma _{1}=-1.6324$, $\sigma _{2}=1.9331$, $\nu
_{12}=-0.53387$, $\nu _{21}=-2.9014$. \ The constants $\alpha _{1,2}$ are
called the ``dispersion'', $\sigma _{1,2}$ are called the nonlinearity
(``Landau'') constants ; $\nu _{12,21}$ represent the coupling of the two
amplitudes.

Numerical experience shows that the dispersion constants are sensitive to
the precision of the integration scheme. We use for \textbf{D02KEF} the
value for tolerence $10^{-8}$.

The system of coupled Cubic Nonlinear Schrodinger Equations has been
integrated numerically in relation with many applications \cite{Class0}, 
\cite{Class1}. One important conclusion is that, in the case of a single
equation, the sign of the product of the dispersion and of the nonlinearity
parameters establishes the possibility of a soliton formation.

\subsection{Numerical search for classes of admissible envelope solutions}

In principle an interactive search for eigenvalues (either with SLEIGN or 
\textbf{D02KEF}) is able to obtain the eigenvalues $k_{1},k_{2}$ eventually
after series of less-inspired initializations. However in many situations
resonances occur and the sources and/or the potentials in the differential
equations for the functions $W_{n}$, $f_{n}$ become discontinuous. Since
these eigenvalues cannot be used we have to devise an automatic search over
the space of parameters with checks of continuity of the functions and
automatic discarding of wrong results. This is accomplished by the routine 
\textbf{P01ABF} of NAG.

The search reveals that the short and respectively long \emph{radial}
wavelength part of the ion wave spectrum have different behavior under the
modulation instability. The most affected is the long wavelength part (small 
$k_{x}$) with $\omega $ of the order of the ion-diamagnetic frequency. This
corresponds to the dominant part of the ion spectrum, because, as can be
seen in numerous numerical simulations, the dominant structures are
elongated in the radial direction, with various degrees of tilting compared
to the radius. It is also important to note that finding the eigenvalues in
the scan of the parameter space weas possible in the condition that the
eigenfunctions $\varphi _{1,2}\left( x\right) $ have low number of
oscillations in the radial direction (one or two). Again this corresponds to
the important part of the ITG turbulence spectrum.

The variation of the parameters of the system of NSE with the physical
parameters of the problem can be studied by the automatic search of
eigenvalues, as described before. Below are plotted four graphs showing this
dependence. 
\begin{figure}[tbp]
\centerline{
 \psfig{file=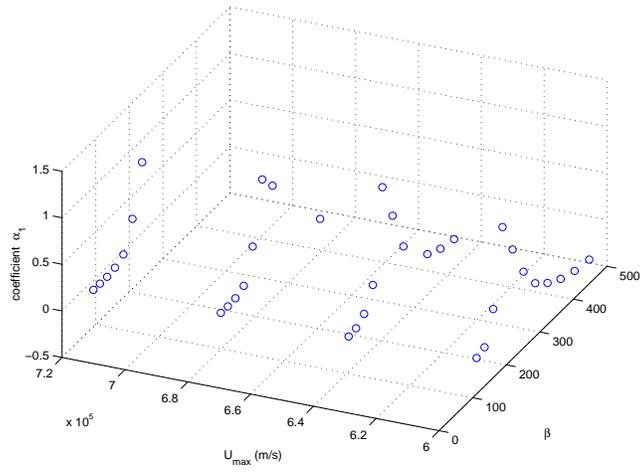,width=0.7\textwidth}}
\caption{Parameter dependence of the coefficient $\protect\alpha _{1}$.}
\end{figure}
\begin{figure}[tbp]
\centerline{
 \psfig{file=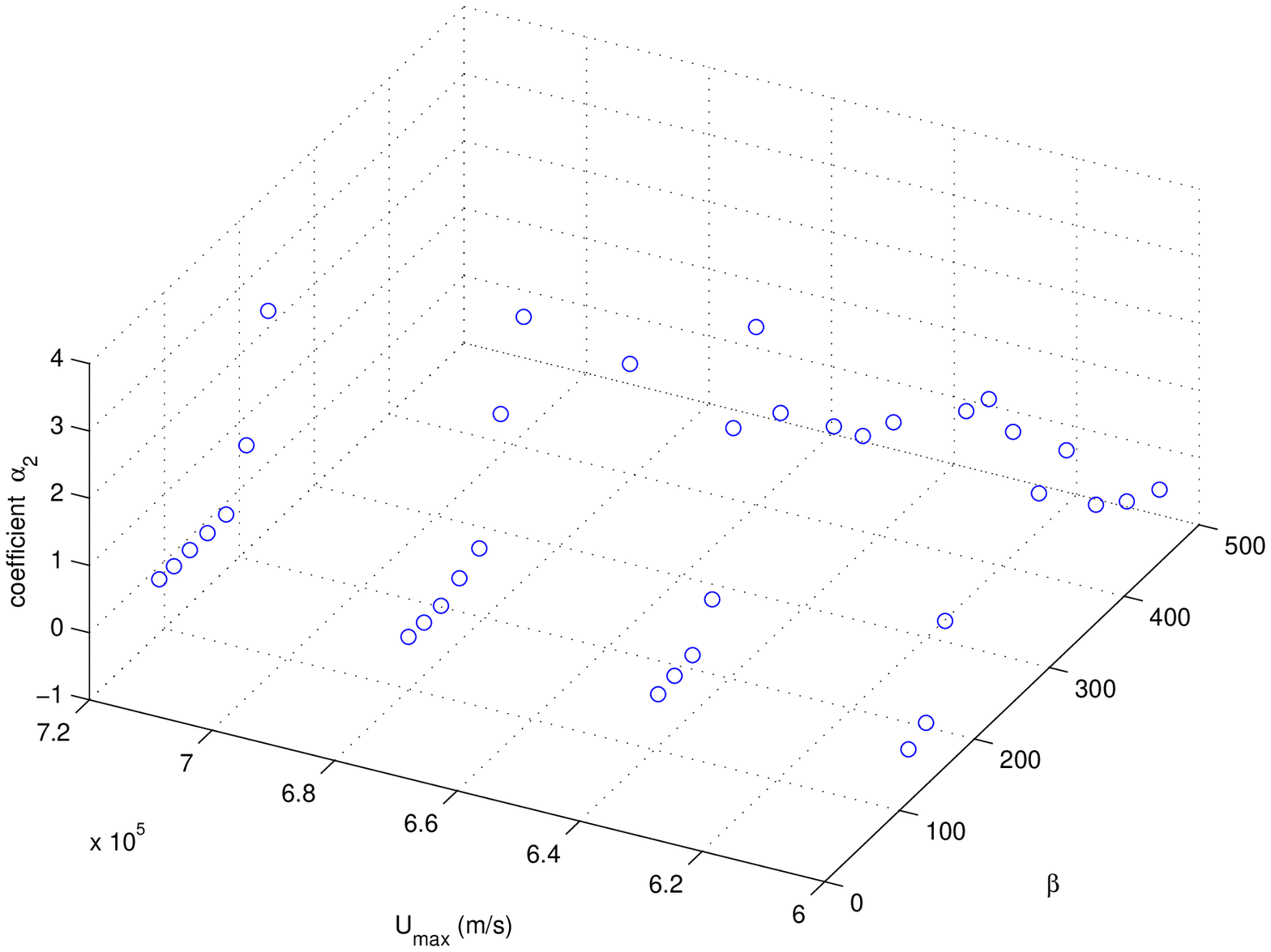,width=0.7\textwidth}}
\caption{Parameter dependence of the coefficient $\protect\alpha _{2}$.}
\end{figure}
\begin{figure}[tbp]
\centerline{
 \psfig{file=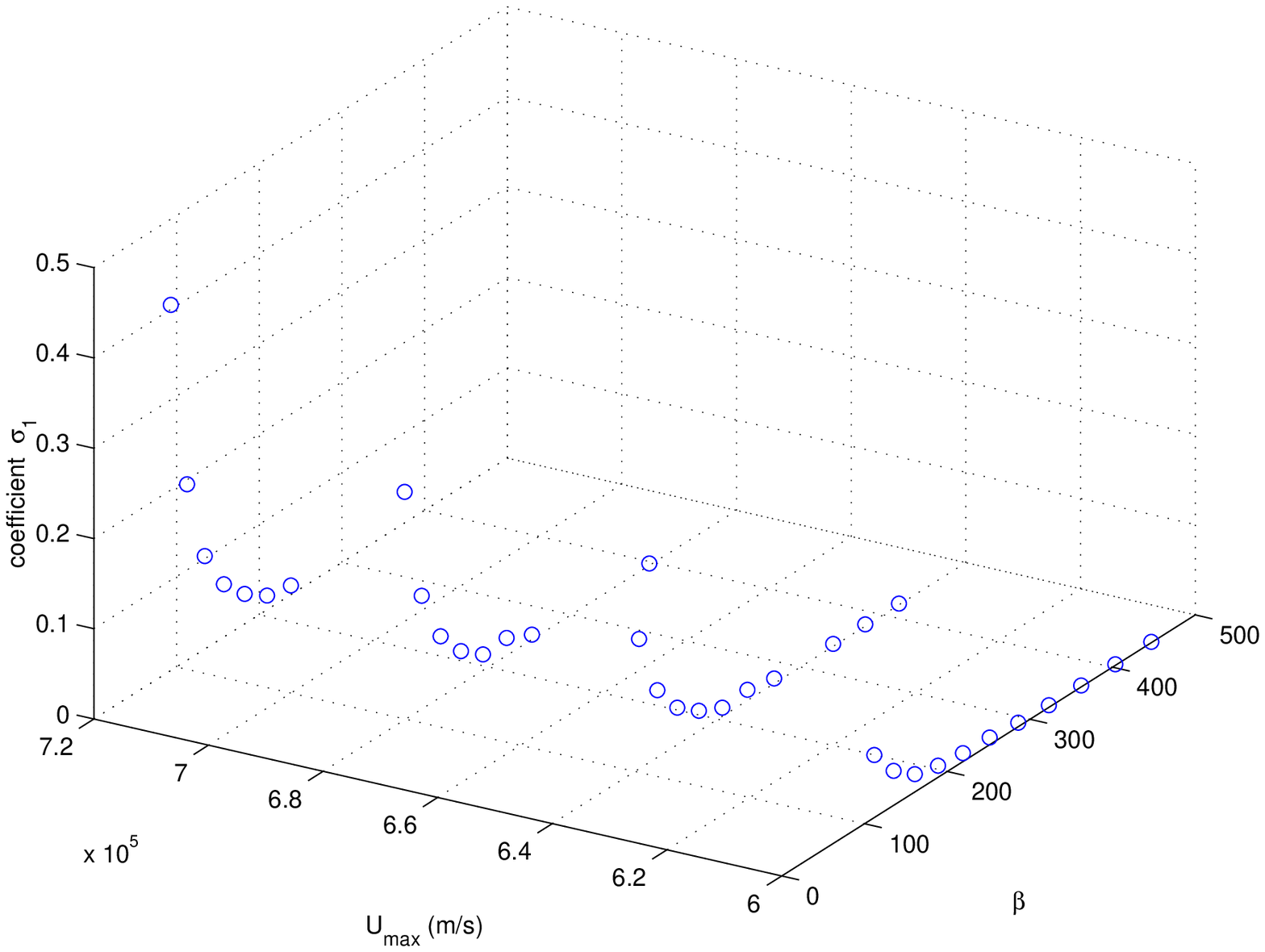,width=0.7\textwidth}}
\caption{Parameter dependence of the coefficient $\protect\sigma _{1}$.}
\end{figure}
\begin{figure}[tbp]
\centerline{
 \psfig{file=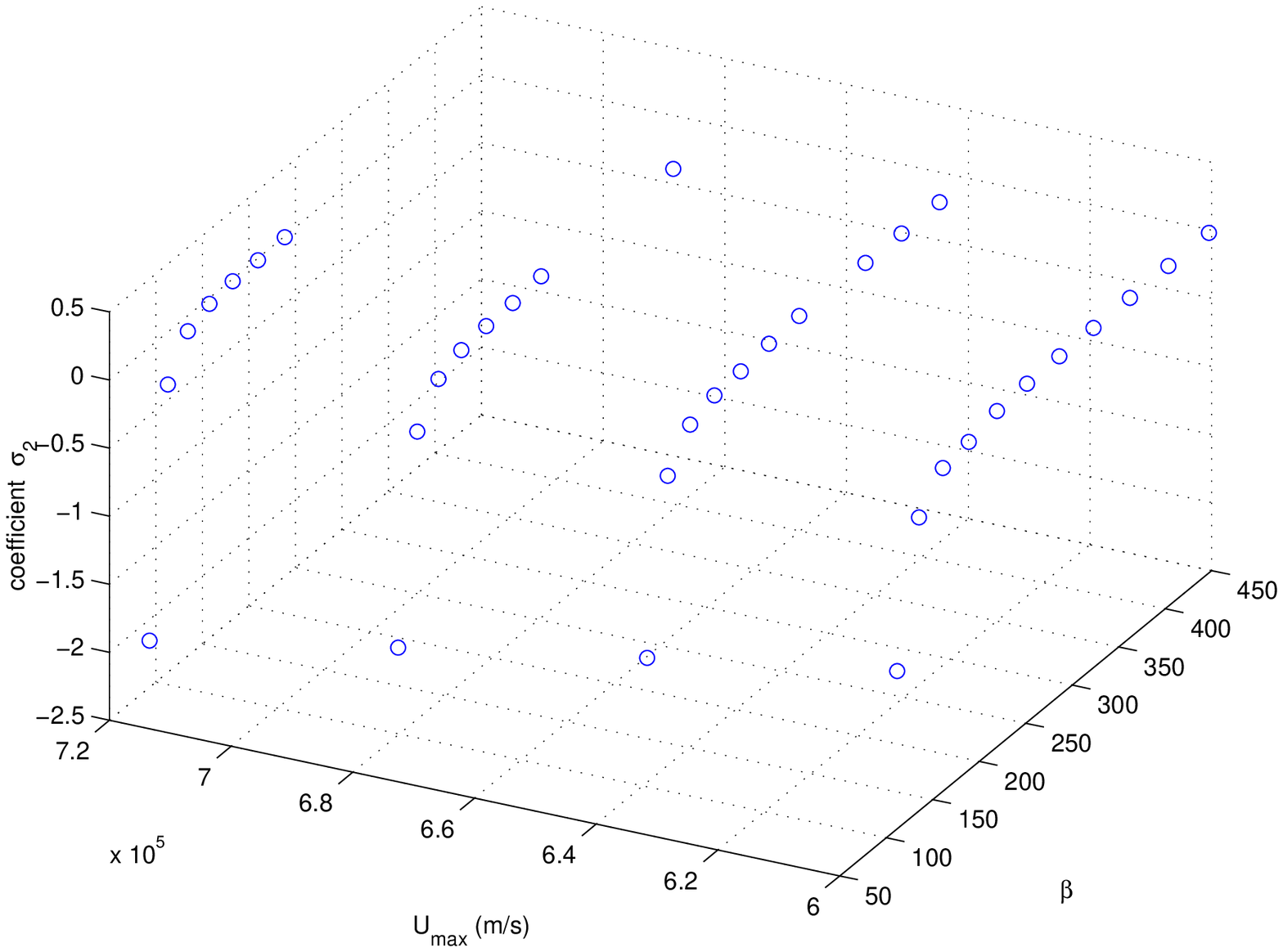,width=0.7\textwidth}}
\caption{Parameter dependence of the coefficient $\protect\sigma _{2}$.}
\end{figure}

\section{Exact solutions of the Nonlinear Schrodinger Equation and the
nonlinear stability problem}

\subsection{Introduction}

The Nonlinear Schrodinger Equation can be solved\emph{\ exactly} on an
infinite spatial domain using the Inverse Scattering Transform. The first
step is to represent the nonlinear equation as a condition of compatibility
of two \emph{linear} equations. This is achieved by finding a pair of linear
operators (Lax operators) and noticing that for the first of them the
eigenfunction problem is a Schrodinger equation leading to a quantum
scattering problem. The unknown function of the NSE is the potential of the
Schrodinger equation. Using the initial condition of the NSE the scattering
data are determined at $t=0$. The scattering data have simple time
dependence and so they can be evloved in time to the desired moment.
Returning from this new set of scattering data (to the potential that has
produced it) is achieved by solving an integral equation
(Gelfand-Levitan-Marchenko) and provides the solution of the NSE. In general
this solution consists of solitons and ``radiation''.

The Inverse Scattering Transform on periodic domain \cite{Novikov}, \cite
{Tracy}, \cite{Osborne} is a more powerful procedure since it reveals and
takes advantage of the deep geometric and topological nature of the problem.
The admissible solutions of the Lax eigenvalue problem can now only be
periodic functions. Since after one period the change of any solution can
only be linear (via the \emph{monodromy} matrix) the periodic or
antiperiodic functions must be found as eigenfunctions of this monodromy
operator (matrix), under the constrain that the eigenvalues are complex of
unit absolute value. This singles out a set of complex values of the
spectral parameter $\lambda $ (the formal eigenvalue in the Lax equation),
the \emph{main spectrum}. From this set, the particular values that make the
two eigenfunctions to coincide are called \emph{non-degenerate}. Here the
squared Wronskian (which is a space and time invariant) has zeros on the
complex $\lambda $ plane. Then the non-degenerate $\lambda $'s are singular
points for the Wronskian. Removing the square-root indeterminancy one has to
connect pairs of non-degenerate eigenvalues by cuts in the $\lambda $ plane.
The Wronskian becomes a hyperelliptic Riemann surface. The evolution of the
unknown NSE solutions (called hyperelliptic functions $\mu _{j}\left(
x,t\right) $) on this surface is as complicated as the original NSE
equation. However, via the \emph{Abel map} the Riemann surface is mapped
onto a torus and on this torus the motion magically becomes \emph{linear}.
After obtaining the space-time dependence on the torus we need to come back
to the original framework. This is called the Jacobi inversion and can be
done exactly in terms of Riemann \emph{theta} functions, giving at the end
the exact solution of the NSE.

Not only the geometric approach is more clear but it also allows a treatment
of the stability problem for the solutions of the NSE since it allows to
trace the changes of the main spectrum after a perturbation of the initial
condition. The following subsections provide a more detailed discussion of
the IST of a periodic domain. The next section will discuss the stability.
This information is available from the abundant literature on the IST and is
only mentioned here in order to understand the mechanism governing the
stability of the envelope of the ion wave turbulence. For this reason we
will focus on the determination of the \emph{main spectrum} and its role in
the construction of the solution. The effective steps to be undertaken to
obtain the solution will only be briefly described. We strongly recommend
the lecture of Ref.(\cite{Tracy}) on which this presentation is based.

\subsection{The Lax operators and the main spectrum}

The IST method starts by introducing a pair of linear operators, called the 
\textbf{Lax pair} (see \cite{Tracy}) which allow to express the nonlinear
equation as a compatibility condition for a system of two linear equations.
For the cubic Schr\"{o}dinger equation 
\begin{equation}
i\frac{\partial u}{\partial t}+\frac{\partial ^{2}u}{\partial x^{2}}+2\left|
u\right| ^{2}u=0  \label{CSE}
\end{equation}
the \textbf{Lax pair} of operators is 
\begin{equation}
L\equiv \left( 
\begin{array}{cc}
i\partial _{x} & u\left( x,t\right) \\ 
-u^{\ast }\left( x,t\right) & -i\partial _{x}
\end{array}
\right)  \label{Laxl}
\end{equation}
\begin{equation}
A\equiv \left( 
\begin{array}{cc}
i\left| u\right| ^{2}-2i\lambda ^{2} & -u_{x}+2i\lambda u \\ 
u_{x}^{\ast }+2i\lambda u^{\ast } & -i\left| u\right| ^{2}+2i\lambda ^{2}
\end{array}
\right)  \label{Laxa}
\end{equation}
The action of the operators on a two-component (column) wave function $\phi
=\left( 
\begin{array}{c}
\phi _{1} \\ 
\phi _{2}
\end{array}
\right) $ \ is given by the equations 
\begin{equation}
L\phi =\lambda \phi  \label{leigen}
\end{equation}
\begin{equation}
\frac{\partial \phi }{\partial t}=A\phi  \label{equa}
\end{equation}
and the condition of compatibility of these two equations $\phi _{xt}=\phi
_{tx}$ is precisely the cubic Nonlinear Schr\"{o}dinger Equation.

\subsection{Solving the eigenvalue equation for the operator $L$}

As in the ``infinite-domain'' IST, we have to solve the eigenvalue equation
Eq.(\ref{leigen}) using the initial condition for the function\emph{\ }$%
u\left( x,t\right) $, $u\left( x,0\right) $. However, for the particular
case of \textbf{periodic }solutions of the NSE, we must have $u\left(
x+d,0\right) =u\left( x,0\right) $ where $d$ is the length of the spatial
period (\emph{i.e.} actually $L$ in the previous notation; however we use $d$
in this and the next section, keeping $L$ for the \emph{Lax} operator).
Fixing an arbitrary base point $x=x_{0}$ one considers the two independent
solutions of the equation (\ref{leigen}) which take the following
``initial'' values at $x=x_{0}$: 
\begin{equation}
\phi \left( x_{0}\right) =\left( 
\begin{array}{c}
1 \\ 
0
\end{array}
\right) \text{\ \ and \ \ }\widetilde{\phi }\left( x_{0}\right) =\left( 
\begin{array}{c}
0 \\ 
1
\end{array}
\right)  \label{inival}
\end{equation}
The matrix $\Phi \left( x,x_{0};\lambda \right) $ of solutions is given by 
\begin{equation}
\Phi \left( x,x_{0};\lambda \right) \equiv \left( 
\begin{array}{cc}
\phi _{1}\left( x,x_{0};\lambda \right) & \widetilde{\phi }_{1}\left(
x,x_{0};\lambda \right) \\ 
\phi _{2}\left( x,x_{0};\lambda \right) & \widetilde{\phi }_{2}\left(
x,x_{0};\lambda \right)
\end{array}
\right)  \label{fundmatrix}
\end{equation}
\emph{i.e.} this solution satisfies the equation 
\begin{equation}
L\Phi =\lambda \Phi \text{\ \ and\ \ }\Phi \left( x_{0},x_{0};\lambda
\right) =\left( 
\begin{array}{cc}
1 & 0 \\ 
0 & 1
\end{array}
\right)  \label{eqfimare}
\end{equation}

It can be seen that the Wronskian of $\left( \phi ,\widetilde{\phi }\right) $
is the \emph{determinant}\ of the matrix of fundamental solutions 
\begin{equation*}
W\left( \phi ,\widetilde{\phi }\right) =\det \left( \Phi \right)
\end{equation*}
It can be shown that $\frac{\partial W}{\partial x}=0$ \emph{i.e.} the
Wronskian is constant, which gives 
\begin{equation*}
\det \left( \Phi \left( x\right) \right) =\det \left( \Phi \left(
x_{0}\right) \right) =1
\end{equation*}
Using the initial condition for $\Phi $ one can solve the equation (\ref
{eqfimare}) and obtain $\Phi \left( x,x_{0};\lambda \right) $; in particular
we can obtain it at $x_{0}+d$, \emph{i.e.} after a period : 
\begin{equation}
\Phi \left( x_{0}+d,x_{0};\lambda \right) =\left( 
\begin{array}{cc}
\phi _{1}\left( x_{0}+d,x_{0};\lambda \right) & \widetilde{\phi }_{1}\left(
x_{0}+d,x_{0};\lambda \right) \\ 
\phi _{2}\left( x_{0}+d,x_{0};\lambda \right) & \widetilde{\phi }_{2}\left(
x_{0}+d,x_{0};\lambda \right)
\end{array}
\right)  \label{phix0}
\end{equation}
This is called \textbf{the monodromy matrix} and is noted 
\begin{equation}
M\left( x_{0};\lambda \right) \equiv \Phi \left( x_{0}+d,x_{0};\lambda
\right)  \label{monmatrix}
\end{equation}
The monodromy matrix is the matrix of the fundamental solutions calculated
for one spatial period. In general the \emph{monodromy matrix} is the
element of the monodromy group associated to a loop and to its homotopy
class, for a marked point on a manifold (here the circle $S^{1}\equiv
\lbrack 0,d$). The linear monodromy group acts by replacing the elements of
the vector column by \emph{linear} combinations of the initial ones. For
changes of the marked point $x_{0}$ the matrix has two invariants : \textbf{%
the trace} and \textbf{the determinant}: 
\begin{equation*}
\left[ Tr\,M\right] \left( x_{0};\lambda \right) =\left[ Tr\,M\right] \left(
\lambda \right) \equiv \Delta \left( \lambda \right)
\end{equation*}
\begin{equation*}
\det M=1
\end{equation*}
The function $\Delta \left( \lambda \right) $ is called \textbf{the
discriminant} and is independent of $x_{0}$.

\paragraph{Looking for Bloch (Floquet) solutions}

The values the discriminant $\Delta \left( \lambda \right) $ takes on the
complex $\lambda $-plane control the monodromy and by consequence select the
values of $\lambda $ for which admissible eigenfunction of the Lax problem
exist. In other words $\Delta $ governs the \emph{spectral properties of the
operator }$L$. To find under what conditions periodic solutions are
possible, we construct the Bloch (or Floquet) solutions of the equation $%
L\phi =\lambda \phi $. The Bloch function has the property 
\begin{equation}
\psi \left( x+d;\lambda \right) =e^{ip\left( \lambda \right) }\psi \left(
x;\lambda \right)  \label{bloch1}
\end{equation}
where $p\left( \lambda \right) $ is the \textbf{Floquet exponent}. Like any
other solution of the Lax eigenproblem $\psi $ can be expressed as a linear
combination of the two fundamental solutions $\phi $ and $\widetilde{\phi }$%
\begin{equation*}
\psi \left( x;\lambda \right) =A\phi \left( x;\lambda \right) +B\widetilde{%
\phi }\left( x;\lambda \right)
\end{equation*}
For the particular point $x_{0}$ we have, taking into account the
``initial'' conditions (\ref{inival}) 
\begin{equation*}
\psi \left( x_{0};\lambda \right) =A\phi \left( x_{0};\lambda \right) +B%
\widetilde{\phi }\left( x_{0};\lambda \right) =\left( 
\begin{array}{c}
A \\ 
B
\end{array}
\right)
\end{equation*}
After one period $d$ the function $\psi $ is linearly modified by the
monodromy matrix and, according to Eq.(\ref{bloch1}) is multiplied by a
complex number of unit absolute value. We write $m\left( \lambda \right) $
for this number and look for those $\lambda $'s where this is complex of
modulus one. The regions of the $\lambda $ plane where $m\left( \lambda
\right) $ has modulus different of one correspond to unstable functions $%
\psi $. We write 
\begin{equation}
\psi \left( x_{0}+d;\lambda \right) =m\left( \lambda \right) \,\psi \left(
x_{0};\lambda \right)  \label{mdel}
\end{equation}
\begin{equation}
M\left( 
\begin{array}{c}
A \\ 
B
\end{array}
\right) =m\left( \lambda \right) \left( 
\begin{array}{c}
A \\ 
B
\end{array}
\right)  \label{eigenmono}
\end{equation}
The Bloch functions are invariant directions for the \emph{monodromy operator%
} $M$ acting in the base formed by the fundamental solutions of the Lax
eigenproblem. The monodromy operator preserves these directions and
multiplies the Bloch vector with $m\left( \lambda \right) =\exp \left(
ip\left( \lambda \right) \right) $. 
\begin{equation*}
\det \left( M-m\right) =m^{2}-\left( Tr\,M\right) m+\det M=m^{2}-\Delta
\left( \lambda \right) m+1=0
\end{equation*}
This gives 
\begin{equation*}
m^{\pm }\left( \lambda \right) =\frac{\Delta \left( \lambda \right) \pm
\left( \Delta ^{2}\left( \lambda \right) -4\right) ^{1/2}}{2}
\end{equation*}

In general $\Delta \left( \lambda \right) $ is an analytic function on the
plane of the complex variable $\lambda $. The equation $\func{Im}\left[
\Delta \left( \lambda \right) \right] =0$ is a single relation with two
unknowns, $\left( \func{Re}\lambda ,\func{Im}\lambda \right) $ and
calculating one of them leaves the dependence on the other as a free
parameter. This gives a curve in the plane and the real-$\lambda $ axis is
such a curve: for $\func{Im}\lambda =0$ we have $\func{Im}\Delta \left(
\lambda \right) =0$.

We look at the variation of $\Delta \left( \lambda \right) $ on the complex
plane to find the effect on $m^{\pm }\left( \lambda \right) $.

\begin{itemize}
\item  Where the Floquet multiplier is a complex number of magnitude unity,
the functions are only effected by a phase factor after one period. The
Bloch functions (eigenfunctions of $L$) are stable under translations with $%
d $. The \emph{real} $\lambda $ axis is a region of stability. The region
where $\Delta $ is real and 
\begin{equation*}
\Delta \left( \lambda \right) <4
\end{equation*}
is the \textbf{band of stability}, the modulus of $m$ being $1$.

\item  When $\lambda $ is such that the discriminant ($\Delta \left( \lambda
\right) \equiv Tr\left( M\right) $ ) is equal to $4$, the eigenvalues of the
monodromy matrix, $m\left( \lambda \right) $, are $\pm 1$, and the Bloch
functions are \emph{periodic} or \emph{antiperiodic}. \textbf{This is the
Main Spectrum} in $\lambda $ and is noted $\left\{ \lambda _{j}\right\} $.
The main spectrum consists of a set of discrete values $\lambda _{j}$.

\item  for those $\lambda $ which gives $\left| m\left( \lambda \right)
\right| \neq 1$ the Bloch eigenfunctions are unstable.
\end{itemize}

\smallskip Consider the case where $\Delta \left( \lambda \right) $ is 
\textbf{complex}. For all values of the parameter $\lambda $ on the complex
plane, [with the exception of the Main Spectrum where $\Delta \left( \lambda
\right) =\pm 2$ ], the eigenvalues of the monodromy matrix are distinct
which means that the eigenfunctions Bloch are distinct and \textbf{%
Independent} (The \textbf{Wronskian} is different of zero). 
\begin{eqnarray*}
\psi ^{+}\left( x+d\right) &=&m^{+}\psi ^{+}\left( x\right) \\
\psi ^{-}\left( x+d\right) &=&m^{-}\psi ^{-}\left( x\right)
\end{eqnarray*}

However we can only be intersted in points of the \emph{main spectrum}, as
they can provide admissible (periodic) eigenfunctions of the Lax operator.
(1) There are points $\lambda _{j}$ in the main spectrum where the two
eigenfunctions Bloch are distinct and independent: these points are called%
\textbf{\ }\emph{degenerate} (with the meaning that for two independent
eigenfunctions there is only one eigenvalue). (2) There are points $\lambda
_{j}$ \ where the two eigenfunctions Bloch are NOT independent: these points
are called \emph{nondegenerate}. For such values of $\lambda $ the Wronskian
is zero.

The important class of initial conditions $u\left( x,t=0\right) $ with the
property that there is only a \emph{finite number of non-degenerate
eigenvalues} $\lambda _{j}$ is called \textbf{finite-band potential}.

\subsection{The ``squared'' eigenfunctions}

Starting from the initial condition $u\left( x,0\right) $ and leting it
evolve in time according to the NSE, $u\left( x,0\right) \overset{NSE}{%
\rightarrow }u\left( x,t\right) $, the \textbf{discriminant} remains
invariant. All the spectral structure obtained from the discriminant $\Delta
\left( \lambda \right) $, \emph{i.e.} the main spectrum, the stability
bands, are invariant. The eigenvalues of the monodromy matrix $\lambda _{j}$
are invariant.\textbf{\ }

\subsubsection{The two Bloch functions for the operator $L$}

Consider the two Bloch eigenfunctions of the operator $L$: 
\begin{equation*}
\psi ^{+}=\left( 
\begin{array}{c}
\psi _{1}^{+} \\ 
\psi _{2}^{+}
\end{array}
\right) \text{\ \ and\ \ }\psi ^{-}=\left( 
\begin{array}{c}
\psi _{1}^{-} \\ 
\psi _{2}^{-}
\end{array}
\right)
\end{equation*}
and define the following functions 
\begin{equation}
f\left( x,t;\lambda \right) \equiv -\frac{i}{2}\left( \psi _{1}^{+}\psi
_{2}^{-}+\psi _{2}^{+}\psi _{1}^{-}\right)  \label{f}
\end{equation}
\begin{equation}
g\left( x,t;\lambda \right) \equiv \psi _{1}^{+}\psi _{1}^{-}  \label{g}
\end{equation}
\begin{equation}
h\left( x,t;\lambda \right) \equiv -\psi _{2}^{+}\psi _{2}^{-}  \label{h}
\end{equation}
These functions have the property that they are periodic, $f\left(
x+d,t;\lambda \right) =f\left( x,t;\lambda \right) $for all values of $%
\lambda $. \ Since $f$ is composed of $\psi ^{+}$ and $\psi ^{-}$ and these
are Bloch solutions for the eigenvalue equation $L\varphi =\lambda \varphi $
we have 
\begin{equation}
\frac{\partial f}{\partial x}=u^{\ast }g-uh  \label{fx}
\end{equation}
\begin{equation}
\frac{\partial g}{\partial x}=-2i\lambda g-2uf  \label{gx}
\end{equation}
\begin{equation}
\frac{\partial h}{\partial x}=2i\lambda h+2u^{\ast }f  \label{hx}
\end{equation}
and similarly we write the time derivatives taking account of the
compatibility condition which involves the operator $A$. 
\begin{equation}
\frac{\partial f}{\partial t}=-\left( u_{x}^{\ast }+2i\lambda u^{\ast
}\right) g+i\left( -u_{x}+2i\lambda u\right) h  \label{ft}
\end{equation}
\begin{equation}
\frac{\partial g}{\partial t}=2i\left( \left| u\right| ^{2}-2\lambda
^{2}\right) g+2i\left( -u_{x}+2i\lambda u\right) f  \label{gt}
\end{equation}
\begin{equation}
\frac{\partial h}{\partial t}=-2i\left( \left| u\right| ^{2}-2\lambda
^{2}\right) h-2i\left( -u_{x}^{\ast }+2i\lambda u^{\ast }\right) f
\label{ht}
\end{equation}

It is shown in Ref.(\cite{Tracy}) that the condition on the initial function 
$u\left( x,0\right) $ for there to be only a finite number of nondegenerate
points in the main spectrum is equivalent to the requirement that $f,g$ and $%
h$ be finite-order polynomials in the parameter $\lambda $, which we take of
degree $N+1$: $f\left( x,t;\lambda \right) =\sum_{j=0}^{N+1}f_{j}\left(
x,t\right) \lambda ^{j}$, $g\left( x,t;\lambda \right)
=\sum_{j=0}^{N+1}g_{j}\left( x,t\right) \lambda ^{j}$, $h\left( x,t;\lambda
\right) =\sum_{j=0}^{N+1}h_{j}\left( x,t\right) \lambda ^{j}$. From the Eqs.(%
\ref{fx} - \ref{hx}) it results however that $g$ and $h$ have degree $N$ in $%
\lambda $.

One can check that the following combination is invariant in time and space
and we note that it is actually the square of the Wronskian: 
\begin{equation}
\frac{\partial }{\partial x}\left( f^{2}-gh\right) =\frac{\partial }{%
\partial t}\left( f^{2}-gh\right) =0  \label{intxt}
\end{equation}
\begin{equation*}
f^{2}-gh=-\frac{1}{4}\left[ W\left( \psi ^{+},\psi ^{-}\right) \right] ^{2}
\end{equation*}
The Wronskian only depends on the spactral parameter $\lambda $. We also
know that for a subset of the \emph{main spectrum}, the nondegenerate
eigenvalues, the Wronskian is zero. Then the number of the nondegenerate
eigenvalues is $2N+2$ (the degree of $f^{2}-gh$ as a polynomial in $\lambda $%
) and we express the function $f^{2}-gh$ as a polinomial with constant
coefficients (not depending on $x$ and $t$). 
\begin{equation}
-\frac{1}{4}\left[ W\left( \psi ^{+},\psi ^{-}\right) \right]
^{2}=f^{2}-gh\equiv P\left( \lambda \right) =\sum_{k=1}^{2N+2}P_{k}\lambda
^{k}=\prod_{j=0}^{2N+2}\left( \lambda -\lambda _{j}\right)  \label{f2gh}
\end{equation}

\subsubsection{Product expansion of $g$ and its zeros $\protect\mu %
_{j}\left( x,t\right) $. Introduction of the \emph{hyperelliptic} functions $%
\protect\mu _{j}$}

The functions $g$ and $h$ are both of order $N$ in $\lambda $. For $g$ we
will note the $N$ roots by $\mu _{j}\left( x,t\right) $. \ The coefficient
of $\lambda ^{2N+2}$ in $f^{2}-gh$ is $f_{N+1}^{2}$. Due to Eq.(\ref{intxt})
it is a constant which can be taken $1$. Now since $f_{N+1}=1$ the
coefficient of $\lambda ^{N}$ in $g$ is $iu\left( x,t\right) $. Written as a
product 
\begin{equation}
g=iu\left( x,t\right) \prod_{j=1}^{N}\left[ \lambda -\mu _{j}\left(
x,t\right) \right]  \label{gprod}
\end{equation}
By similar arguments we find that the coefficient of $\lambda ^{N}$ in $h$
is $iu^{\ast }\left( x,t\right) $ and the zeros of $h$ are $\mu _{j}^{\ast
}\left( x,t\right) $. The functions $\mu _{j}\left( x,t\right) $ are called
the \emph{hyperelliptic functions}. It will be proved below that finding the 
\emph{hyperelliptic} functions leads immediately to the solution $u\left(
x,t\right) $.

Now we calculate the expression in Eq.(\ref{f2gh}) for $\lambda $ = a zero
of the function $g$, \emph{i.e.} $\lambda $ is equal to \emph{hyperelliptic
function} $\mu _{j}\left( x,t\right) $: 
\begin{equation*}
f^{2}\left( x,t;\lambda =\mu _{m}\right) -gh=f^{2}\left( x,t;\mu _{m}\right)
=P\left( \lambda =\mu _{m}\right)
\end{equation*}
or 
\begin{equation}
f\left( x,t;\mu _{m}\right) =\sigma _{m}\sqrt{P\left( \mu _{m}\right) }
\label{fatmiu}
\end{equation}
Here the factor $\sigma _{m}$ is a sheet index that indicates which sheet of
the Riemann surface associated with $\sqrt{P\left( \lambda \right) }$ the
complex $\mu _{m}$ lies on.

Let us calculate from Eq.(\ref{gx}) and Eq.(\ref{gprod}) the derivative at $%
x $ of $\mu _{m}$: 
\begin{eqnarray*}
\frac{\partial g\left( x,t;\lambda \right) }{\partial x} &=&i\frac{\partial
u\left( x,t;\lambda \right) }{\partial x}\prod_{j=1}^{N}\left( \lambda -\mu
_{j}\left( x,t\right) \right) \\
&&-iu\left( x,t;\lambda \right) \sum_{j=1}^{N}\frac{\partial \mu _{j}\left(
x,t\right) }{\partial x}\prod_{k=1,k\neq j}^{N}\left( \lambda -\mu
_{k}\left( x,t\right) \right)
\end{eqnarray*}
Replace here a zero of $g$: $\lambda =\mu _{m}$%
\begin{equation*}
\frac{\partial g\left( x,t;\mu _{m}\right) }{\partial x}=-iu\left( x,t;\mu
_{m}\right) \frac{\partial \mu _{m}\left( x,t\right) }{\partial x}%
\prod_{k=1,k\neq m}^{N}\left( \mu _{m}-\mu _{k}\left( x,t\right) \right)
\end{equation*}
On the other hand we have, from Eq.(\ref{gx}) 
\begin{eqnarray*}
\frac{\partial g}{\partial x} &=&-2i\mu _{m}g\left( x,t;\lambda =\mu
_{m}\right) -2u\left( x,t;\lambda =\mu _{m}\right) f\left( x,t;\lambda =\mu
_{m}\right) \\
&=&-2uf
\end{eqnarray*}
then, since we have calculated $f\left( x,t;\lambda =\mu _{m}\right) $, Eq(%
\ref{fatmiu}), we obtain 
\begin{equation}
\frac{\partial \mu _{m}\left( x,t\right) }{\partial x}=\frac{-2i\sigma _{m}%
\sqrt{P\left( \mu _{m}\right) }}{\prod_{k=1,k\neq m}^{N}\left( \mu _{m}-\mu
_{k}\left( x,t\right) \right) }  \label{dxdmiu}
\end{equation}
for $m=1,2,...,N$.

The coefficients of the term with $\lambda ^{N}$ in the equation (\ref{gx})
are matched and we obtain $iu_{x}=-2ig_{N-1}-2uf_{N}$ .Using Eq.(\ref{gprod}%
) we find 
\begin{equation*}
\partial _{x}\ln u=2i\left( \sum_{j=1}^{N}\mu _{j}+f_{N}\right)
\end{equation*}
From the Eq.(\ref{f2gh}) we obtain the coefficient of the term $\lambda ^{N}$%
, \emph{i.e.} $f_{N}$%
\begin{equation*}
f_{N}=-\frac{1}{2}\sum_{k=1}^{2N+2}\lambda _{k}
\end{equation*}
Then from the preceding equation it results 
\begin{equation}
\partial _{x}\ln u=2i\left( \sum_{j=1}^{N}\mu _{j}-\frac{1}{2}%
\sum_{k=1}^{2N+2}\lambda _{k}\right)  \label{dxlnu}
\end{equation}

The same procedure is applied to the equation for $g_{t}$, (\ref{gt}) and we
obtain 
\begin{equation}
\frac{\partial \mu _{j}\left( x,t\right) }{\partial t}=-2\left( \sum_{m\neq
j}\mu _{m}-\frac{1}{2}\sum_{k=1}^{2N+2}\lambda _{k}\right) \frac{\partial
\mu _{j}\left( x,t\right) }{\partial x}  \label{dtdmiu}
\end{equation}
\begin{eqnarray}
\partial _{t}\ln u &=&2i\left[ \sum_{j>k}\lambda _{j}\lambda _{k}-\frac{3}{4}%
\left( \sum_{k=1}^{2N+2}\lambda _{k}\right) ^{2}\right]  \label{dtlnu} \\
&&-4i\left[ \left( -\frac{1}{2}\sum_{k=1}^{2N+2}\lambda _{k}\right) \left(
\sum_{j=1}^{N}\mu _{j}\right) +\sum_{j>k}\mu _{j}\mu _{k}\right]  \notag
\end{eqnarray}

We can see that the problem has been reformulated: from the equation for the
function $u\left( x,t\right) $ the Lax-operators formulation leads to the
problem for Bloch functions, $\psi ^{+}$ and $\psi ^{-}$ and their
Wronskian; then, using the ``squared'' eigenfunctions $f$, $g$ and $h$ we
arrive at a formulation for the \emph{hyperelliptic }functions $\mu
_{j}\left( x,t\right) $ . Finding $\mu _{j}\left( x,t\right) $ leads
immediately to $u\left( x,t\right) $.

These operations have a significative geometric counterpart: every point of
the complex plane of the spectral parameter $\lambda $ is mapped, via Eq.(%
\ref{f2gh}) on the two-sheeted Riemann surface $\sqrt{-\frac{1}{4}W\left(
x,t;\lambda \right) }=\sqrt{P\left( \lambda \right) }$ with singularities at
the \emph{nondegenerate} points of the main spectrum, $\lambda _{j}$.
Removing the indeterminancy (by cutting and glueing the \ two sheets) we
obtain a \textbf{hyperelliptic Riemann surface} of genus $g=N$. It results
that we have to consider the variables $\mu _{j}\left( x,t\right) $, $j=1,N$
, as points on this surface but the motion of $\mu _{j}\left( x,t\right) $
is not simpler than the equation itself.

Now the next steps would be: (1) using the initial condition $\mu _{j}\left(
0,0\right) $ on the function $\mu _{j}\left( x,t\right) $ we can solve the
two equations (\ref{dxdmiu}) and (\ref{dtdmiu}) and find $\mu _{j}\left(
x,t\right) $; (2) then using the initial condition $\left| u\left(
0,0\right) \right| $ for the function $u\left( x,t\right) $ we can solve the
equations (\ref{dxlnu}) and (\ref{dtlnu}). The procedure will be:

\begin{itemize}
\item  Take the parameters $\left\{ \lambda _{j}|\;j=1,2,...,2N+2\right\} $
as known; these parameters are \underline{\textbf{non-degenerate}}\textbf{\
eigenvalues} from the main spectrum, they can be found from the equation $%
\Delta \left( \lambda \right) =\pm 2$ and $\Delta $ is determined from $%
u\left( x,0\right) $, the initial condition.

\item  Choose initial conditions $\mu _j\left( 0,0\right) $ and $\left|
u\left( 0,0\right) \right| $ (see below)

\item  solve the equations for $\mu _{j}\left( x,t\right) $ ; this means to
find the \emph{hyperelliptic functions}. Then solve the equations for $%
u\left( x,t\right) $.
\end{itemize}

\subsection{Solution: the two-sheeted Riemann surface (hyperelliptic genus-$%
N $ Riemann surface)}

The variable $\mu _{j}$'s are points lying on the two-sheeted Riemann
surface associated with 
\begin{equation*}
\sqrt{P\left( \lambda \right) }=\left( \prod_{k=1}^{2N+2}\left( \lambda
-\lambda _{k}\right) \right) ^{1/2}\equiv R\left( \lambda \right)
\end{equation*}
which has \textbf{branch cuts at each of the nondegenerate points} $\lambda
_{k}$. To eliminate the indeterminacy related to the square-root
singularities (located at $\lambda _{j}$'s) the surfaces must be cut and
reglued, obtaining a complex manifold of complex dimension one, a
hyperelliptic Riemann surface. We need some constructions on this surface.

\subsubsection{Holomorphic differential forms and cycles on the Riemann
surface}

On this hyperelliptic Riemann surface (denoted $M$) it is possible to define 
$N$ linearly independent holomorphic (regular) differentials. The following
is the canonical basis of differentials one-forms defined on the manifold $M$%
: 
\begin{equation*}
dU_{1}\equiv \frac{d\lambda }{R\left( \lambda \right) }
\end{equation*}
\begin{equation*}
dU_{2}\equiv \frac{\lambda \,d\lambda }{R\left( \lambda \right) },\;\;...
\end{equation*}
\begin{equation*}
dU_{j}\equiv \frac{\lambda ^{j}\,d\lambda }{R\left( \lambda \right) },\;...
\end{equation*}
\begin{equation*}
dU_{N}\equiv \frac{\lambda ^{N-1}\,d\lambda }{R\left( \lambda \right) }
\end{equation*}

On the surface $M$ there are $2N$ topologically distinct closed curves (%
\textbf{cycles}). On a torus ($N=1$, called \emph{elliptic surface}) there
are two curves which cannot be deformed one into another. A more general
Riemann surface, with $N>1$ can be shown to be equivalent to a sphere with $%
N $ handles and is called \emph{hyperelliptic Riemann surface}. The number $%
N $ is called the \textit{genus} of the surface. We have 
\begin{equation*}
N=\text{genus of the surface}=\text{number of independent holomorphic
differential forms}
\end{equation*}
There are $2N$ cycles which are split into two classes: $a_{j}$\ \ cycles
and $b_{j}\;\;$cycles. Each of these cycles has a specified direction (an
arrow) attached to it. To construct the cycles, the rules to be applied are:

\begin{enumerate}
\item  $a_j$ cycles do not cross any other $a_j$ cycle; $b_j$ cycles do not
cross any other $b_j$ cycle;

\item  the cycle $a_k$ intersects $b_k$ only once and intersects no other $b$
cycle;

\item  the intersection is such that, at the point of intersection, the
vector tangent to the cycle $a_{k}$, the vector tangent to the cycle $b_{k}$
and the normal to the tangent plane of $M$ represent a system of three
vectors compatible with the \emph{orientation} of $M$.
\end{enumerate}

This can be represented as: 
\begin{equation*}
a\circ a=0\;,\;b\circ b=0\;,\;a_{j}\circ b_{k}=\delta _{jk}
\end{equation*}

\subsubsection{Periods and matrices of periods}

With the set of holomorphic differential forms and the set of cycles we can
define the \textbf{matrices of periods}, $A$ and $B$%
\begin{equation*}
A_{kj}\equiv \int_{a_{j}}dU_{k}
\end{equation*}
\begin{equation*}
B_{kj}=\int_{b_{j}}dU_{k}
\end{equation*}
and the matrices $A$ and $B$ are \textbf{invertible}. A change of basis of
holomorphic differential forms is a linear transformation represented by the
matrix $C$: 
\begin{equation*}
d\psi _{j}\equiv \sum_{k=1}^{N}C_{jk}\,dU_{k}
\end{equation*}
We can choose the matrix $C$ as 
\begin{equation*}
C=A^{-1}
\end{equation*}
Then the periods in the new basis becomes 
\begin{equation*}
\int_{a_{n}}d\psi
_{j}=\sum_{k=1}^{N}C_{jk}\int_{a_{n}}dU_{k}=\sum_{k=1}^{N}C_{jk}A_{kn}=%
\delta _{jn}
\end{equation*}
\begin{equation*}
\int_{b_{n}}d\psi _{j}=\tau _{jn}
\end{equation*}
where 
\begin{equation*}
\tau =A^{-1}B
\end{equation*}
It can be shown that the matrix $\tau $ is symmetric $\tau _{jk}=\tau _{kj}$
and has a positive-definite imaginary part\textbf{\ }$\func{Im}\tau >0$.

\subsubsection{The Abel map}

The Abel map is defined from the Riemann surface $M$ to the space $C^{N}$
and associates to a hyperelliptic Riemann surface (a sphere with $N$ hadles)
a $N$-torus. Using the differentials $d\psi _{j}$ one constructs a change of
variables by the following procedure:

\begin{itemize}
\item  choose a base point on $M$ and call it $p_0$;

\item  define the variables $W_{j}\left( x,t\right) $ as integrals on the
Riemann $\lambda $-surface of the \emph{differential forms} from the
arbitrary point $p_{0}$ to the point which is the \textbf{hyperelliptic
function} $\mu _{k}\left( x,t\right) $: 
\begin{eqnarray*}
W_{j}\left( x,t\right) &=&\sum_{k=1}^{N}\int_{p_{0}}^{\mu _{k}\left(
x,t\right) }\,d\psi _{j} \\
&=&\sum_{k=1}^{N}\sum_{m=1}^{N}C_{jm}\int_{p_{0}}^{\mu _{k}\left( x,t\right)
}\frac{\lambda ^{m-1}d\lambda }{R\left( \lambda \right) }
\end{eqnarray*}
These variables have the remarkable property that their $x$ and $t$
dependence is trivial 
\begin{equation*}
\frac{d}{dx}W_{j}\left( x,t\right) =\sum_{k=1}^{N}\sum_{m=1}^{N}C_{jm}\frac{%
\mu _{k}^{m-1}\frac{d\mu _{k}}{dx}}{\sigma _{k}R\left( \mu _{k}\right) }
\end{equation*}
and using 
\begin{equation*}
\frac{d\mu _{k}\left( x,t\right) }{dx}=-2i\sigma _{k}\frac{R\left( \mu
_{k}\right) }{\prod_{n\neq k}\left( \mu _{k}-\mu _{n}\right) }
\end{equation*}
it results 
\begin{equation*}
\frac{d}{dx}W_{j}\left( x,t\right) =\sum_{m=1}^{N}C_{jm}\sum_{k=1}^{N}\frac{%
-2i\mu _{k}^{m-1}}{\prod_{n\neq k}\left( \mu _{k}-\mu _{n}\right) }
\end{equation*}
To calculate the sum 
\begin{equation*}
\sum_{k=1}^{N}\frac{\mu _{k}^{m-1}}{\prod_{n\neq k}\left( \mu _{k}-\mu
_{n}\right) }
\end{equation*}
one can use the contour integral 
\begin{equation*}
I_{m}\equiv \frac{1}{2\pi i}\int_{C}\frac{\lambda ^{m-1}d\lambda }{%
\prod_{n\neq k}\left( \mu _{k}-\mu _{n}\right) }
\end{equation*}
where the contour $C$ encloses all of the poles $\mu _{n}$ counterclockwise.
By the residue theorem 
\begin{equation*}
\sum_{k=1}^{N}\frac{\mu _{k}^{m-1}}{\prod_{n\neq k}\left( \mu _{k}-\mu
_{n}\right) }=I_{m}
\end{equation*}
One can evaluate this integral by compactifying the $\lambda $ plane into a
sphere and noticing that the contour encloses the pole at $z=\infty $.
Evaluating the residuu at $z=\infty $%
\begin{equation*}
\sum_{k=1}^{N}\frac{\mu _{k}^{m-1}}{\prod_{n\neq k}\left( \mu _{k}-\mu
_{n}\right) }=\delta _{m,N}
\end{equation*}
and from this we obtain 
\begin{equation*}
\frac{d}{dx}W_{j}\left( x,t\right) =-2iC_{j,N}\equiv \frac{1}{2\pi }k_{j}
\end{equation*}
A similar calculation gives 
\begin{eqnarray*}
\frac{d}{dt}W_{j}\left( x,t\right) &=&-4i\left[ C_{j,N-1}+\left( \frac{1}{2}%
\sum_{k=1}^{2N+2}\lambda _{k}\right) C_{j,N}\right] \\
&=&\frac{1}{2\pi }\Omega _{j}
\end{eqnarray*}
It results from these calculations 
\begin{equation*}
W_{j}\left( x,t\right) =\frac{1}{2\pi }\left( k_{j}x+\Omega
_{j}t+d_{j}\right)
\end{equation*}
where $d_{j}$ is a phase which is determined by the initial condition on $%
\mu _{k}$.
\end{itemize}

The Abel map has \emph{linearized} the motion of the points $\mu _{k}\left(
x,t\right) $. Now, in order to determine the function $u\left( x,t\right) $
we must invert the Abel map, returning from the variables $W$ to $\mu $.
This is the \emph{Jacobi inversion problem}\textbf{.}

\subsubsection{The Jacobi inversion problem}

The return from the variables $W_{j}\left( x,t\right) $ of the $N$-torus
back to the variables $\mu _{k}\left( x,t\right) $ of the Riemann surface $M$
\ (notice we will only need particular combinations of the functions $\mu $)
can be done in an exact way using the Riemann $\theta $ function.

The argument of $\theta $ is the N-tuple of complex numbers $\overline{z}\in
C^{N}$. 
\begin{equation*}
\theta \left( z|\tau \right) \equiv \sum_{m_{1}=-\infty }^{\infty }\cdot
\cdot \cdot \sum_{m_{N}=-\infty }^{\infty }\;\exp \left( 2\pi i\overline{m}%
\cdot \overline{z}+\pi i\overline{m}\cdot \tau \cdot \overline{m}\right)
\end{equation*}
with the notations $\overline{m}\cdot \overline{z}=\sum_{k=1}^{N}m_{k}z_{k}$
and $\overline{m}\cdot \tau \cdot \overline{m}=\sum_{k,j=1}^{N}\tau
_{jk}\,m_{j}\,m_{k}$. Adding to the vector $z$ a vector with a single
nonzero element in the position $k$ we obtain from the definition: 
\begin{equation*}
\theta \left( z+e_{k}|\tau \right) =\theta \left( z|\tau \right)
\end{equation*}
which means that the $\theta $ -function has $N$ real periods\textbf{. }%
Adding the full $k$-th column of the matrix $\tau $, we obtain: 
\begin{equation}
\theta \left( z+\tau _{k}|\tau \right) =\exp \left( -2\pi iz_{k}-\pi i\tau
_{kk}\right) \theta \left( z|\tau \right)  \label{thetaaddtau}
\end{equation}
where $\tau _{kk}$ is the diagonal element of the $\underline{\tau }$ matrix.

There are two quantities which must be determined in order to have the
explicit form of the solution. The space and respectively time equations for 
$u\left( x,t\right) $ depend on the quantity $\sum_{j>k}\mu _{j}\mu _{k}=%
\frac{1}{2}\left[ \left( \sum_{m=1}^{N}\mu _{m}\right)
^{2}-\sum_{m=1}^{N}\mu _{m}^{2}\right] $. So we must determine the following
combinations of the variables $\mu _{j}$: 
\begin{equation*}
\sum_{m=1}^{N}\mu _{m}\left( x,t\right) \;\;\text{and}\;\;\sum_{m=1}^{N}\mu
_{m}^{2}\left( x,t\right)
\end{equation*}

In order to calculate these two quantities, we shall start by introducing a
series of functions , $\psi _{j}\left( p\right) $ defined on the Riemann
surface $M$: 
\begin{equation*}
\psi _{j}\left( p\right) =\int_{p_{0}}^{p}d\psi _{j}
\end{equation*}
where $\ p\;\;$is a point on$\;M$ , $d\psi _{j}$\ \ is the $j$ -th
holomorphic differential form defined on $M$ and $p_{0}$ is an arbitrary
fixed point on $M$. The functions $\psi _{j}\left( p\right) $ are multiple
valued since the contour is defined up to addition of any combination of the
cycles on $M$.

Now consider the function, $F\left( p\right) $: 
\begin{equation*}
F\left( p\right) \equiv \theta \left( \psi \left( p\right) -K|\tau \right)
\end{equation*}
where $\theta $ is the $N$-dimensional Riemann \emph{theta} function
associated with the surface $M$; $\psi \left( p\right) $ is the $N$%
-dimensional column of complex functions $\psi _{j}\left( p\right) $; $K$ is
an $N$-dimensional column of complex numbers independent of the point $p$.
The function $F\left( p\right) $ is multivalued on the Riemann surface $M$
for the same reason as $\psi $: moving the point $p$ on the surface such as
to turn around one of the $b$ cycle and returning to the initial position,
the functions $\psi _{j}$ will add elements of the $\tau $ matrix. This will
make the function $F$ to change as imposed by the properties of the $\theta $
function, shown in Eq.(\ref{thetaaddtau}).

In order to render the function $F$ single-valued, we replace its domain $M$
by a new surface, obtained from $M$ by dissecting it in a canonical fashion,
along a basis of cycles. This new surface, $M^{\ast }$ is simply connected
and has the bord composed of a number of arcs equal to $4N$.

This operation is necessary in order to render the function $F\left(
p\right) $ entire and allow us define the inegral 
\begin{equation}
I_{0}=\frac{1}{2\pi i}\int_{\partial M^{\ast }}d\ln F\left( p\right)
\label{lnfp}
\end{equation}
around the contour of the surface $M^{\ast }$. Applying Cauchy theorem, the
integral is the \emph{number of zeroes} of $F\left( p\right) $ on the
surface $M^{\ast }$. This number is $N$ (Riemann).

We now impose the condition that the $N$\ zeros of $F\left( p\right) $
coincide with the $N$ points $\left( \mu _{j}\left( x,t\right) ,\sigma
_{j}\right) $. This fixes the values of the $N$\ complex numbers $%
K_{j}\;\left( j=1,...,N\right) $. By doing so the contour integral (\ref
{lnfp}) calculated with the reziduum theorem will involve the variables $\mu
_{j}$.

\paragraph{Calculation of the two combination of $\protect\mu $'s}

We must remember that the complex $\lambda $ plane is covered by the
two-sheeted Riemann surface whose compact version is $M$. Further this is
mapped by Abel map onto the Jacobi $N$-torus. To a variable on the Riemann
surface $M$ (formally also on $M^{\ast }$), say $p$, corresponds a certain $%
\lambda \left( p\right) $. One defines the following integrals, which are
proved to be real constants : 
\begin{equation*}
I_{1}=\frac{1}{2\pi i}\int_{\partial M^{\ast }}\lambda \left( p\right) d\ln %
\left[ F\left( p\right) \right] \equiv A_{1}
\end{equation*}
\begin{equation*}
I_{2}=\frac{1}{2\pi i}\int_{\partial M^{\ast }}\lambda ^{2}\left( p\right)
d\ln \left[ F\left( p\right) \right] \equiv A_{2}
\end{equation*}
They can be evaluated by the residuu theorem. By the choice of the constants 
$K$'s, the zeroes of the of the function $F\left( p\right) $ are located at $%
\mu _{j}$. Then the residues are just the integrand ($\lambda $) calculated
in $\mu $ plus the residuu at the infinite, $\pm \infty $: 
\begin{equation*}
I_{1}=\sum_{m=1}^{N}\mu _{m}+\underset{\lambda \rightarrow \infty ^{+}}{%
\func{Re}s}\left[ \lambda \left( p\right) d\ln F\left( p\right) \right] +%
\underset{\lambda \rightarrow \infty ^{-}}{\func{Re}s}\left[ \lambda \left(
p\right) d\ln F\left( p\right) \right]
\end{equation*}
\begin{equation*}
I_{2}=\sum_{m=1}^{N}\mu _{m}^{2}+\underset{\lambda \rightarrow \infty ^{+}}{%
\func{Re}s}\left[ \lambda ^{2}\left( p\right) d\ln F\left( p\right) \right] +%
\underset{\lambda \rightarrow \infty ^{-}}{\func{Re}s}\left[ \lambda
^{2}\left( p\right) d\ln F\left( p\right) \right]
\end{equation*}
The reason to write \textbf{two residues} at infinity is that $\lambda
=\infty $ is not a branching point which means that there are \textbf{two
points} on the manifold $M$ corresponding to $\lambda =\infty $, one on each
sheet of the surface. We note $r^{\pm }\;\;$the value of $\psi \left(
p\right) $\ when $p$ \ is the point on $M^{\ast }$ corresponding to $\lambda
\rightarrow \infty ^{\pm }$.

The result is 
\begin{equation*}
A_{1}=\sum_{j=1}^{N}\mu _{j}+\frac{i}{2}\frac{\partial }{\partial x}\ln %
\left[ \frac{F\left( r^{-}-K\right) }{F\left( r^{+}-K\right) }\right]
\end{equation*}
or 
\begin{equation*}
A_{1}=\sum_{j=1}^{N}\mu _{j}+\frac{i}{2}\frac{\partial }{\partial x}\ln %
\left[ \frac{F\left( r^{-}-K\right) }{F\left( r^{+}-K\right) }\right]
\end{equation*}
with $\theta ^{\pm }=F\left( r^{\pm }-K\right) $. The expression of $I_{2}$%
\begin{equation*}
A_{2}=\sum_{j=1}^{N}\mu _{j}^{2}-\frac{1}{4}\frac{\partial }{\partial x}\ln %
\left[ F\left( r^{+}-K\right) F\left( r^{-}-K\right) \right] +\frac{i}{4}%
\frac{\partial }{\partial t}\ln \left[ \frac{F\left( r^{-}-K\right) }{%
F\left( r^{+}-K\right) }\right]
\end{equation*}

\paragraph{Return to the equations for the function $u$}

With these expressions we come back to the two equations for the two partial
derivatives of $u$. 
\begin{equation*}
\frac{\partial }{\partial x}\ln u=\frac{\partial }{\partial x}\ln \frac{%
F\left( r^{+}-K\right) }{F\left( r^{-}-K\right) }+2iA_{1}-i\sum_{m=1}^{2N+2}%
\lambda _{m}
\end{equation*}
\begin{equation*}
\frac{\partial }{\partial t}\ln u=\frac{\partial }{\partial t}\ln \frac{%
F\left( r^{+}-K\right) }{F\left( r^{-}-K\right) }+i\text{const}
\end{equation*}

\paragraph{The solution}

Let us note 
\begin{equation*}
\omega _{0}=Q
\end{equation*}
\begin{equation*}
k_{0}=2A_{1}-\sum_{m=1}^{2N+2}\lambda _{m}
\end{equation*}
which are ``external'' frequency and wavelength.

The solution is 
\begin{equation*}
u\left( x,t\right) =u\left( 0,0\right) \exp \left( ik_{0}-i\omega
_{0}t\right) \frac{\theta (W^{-}|\tau )}{\theta \left( W^{+}|\tau \right) }
\end{equation*}
where 
\begin{equation*}
W_{j}^{\pm }=\frac{1}{2\pi }\left( k_{j}x+\Omega _{j}t+\delta _{j}^{\pm
}\right)
\end{equation*}

The phases $\delta _{j}^{\pm }$ are the part of $r^{\pm }-K_{j}$ which is
independent of $\left( x,t\right) $.

\section{Stability of the envelope solutions}

In this Section we return to the equations obtained by multiple space and
time scales analysis. Then the meaning of the variables $\phi $, $y$, $t$
is: first order (envelope) amplitude of the potential fluctuations in the
ion turbulence and respectively large space and slow time variables. We
focus on a single NSE in order to examine the properties of the stability of
the solutions. The equation has the generic form 
\begin{equation}
i\frac{\partial \phi }{\partial t}+\alpha \frac{\partial ^{2}\phi }{\partial
y^{2}}+2\sigma \left| \phi \right| ^{2}\phi =0  \label{eqalfsig}
\end{equation}
and represents the equation for the envelope of a fast oscillation arising
as solution of the barotropic equation. On infinite spatial domain this
equation has soliton as solutions. On a periodic spatial domain (which is
our case) the solutions can be cnoidal waves and they may be modulationally
unstable. The instability can be examined in the neighbourhood of a solution
by imposing a slight deviation and studying its time behaviour by classical
perturbation expansion. This analysis and comparison with the
geometric-algebraic results have been carried out in Ref.(\cite{Tracy}).

Consider $\phi \left( y,t\right) $ is a solution and perturb it at the
initial moment $t=0$ as 
\begin{equation*}
\widetilde{\phi }\left( y,t\right) =\phi \left( y,0\right) +\varepsilon
h\left( y\right)
\end{equation*}
The question is to find the relative behaviour of the two functions $\phi
\left( y,t\right) $ and $\widetilde{\phi }\left( y,t\right) $. It they
depart exponentially for time close to the initial moment, then $\phi \left(
y,t\right) $ is an unstable solution. Suppose the initial solution $\phi
\left( y,t\right) $ is the envelope of an exact plane wave \emph{i.e.} it is
constant in space. As a function of time it must have the form $\phi \left(
y,t\right) =p\exp \left( 2iqt\right) $ and we find 
\begin{equation}
\phi \left( y,t\right) =p\exp \left( 2i\sigma p^{2}t\right)  \label{envplane}
\end{equation}
To this solution we add a perturbation 
\begin{equation*}
\widetilde{\phi }\left( y,t\right) =p\exp \left( 2i\sigma p^{2}t\right)
\left\{ 1+\varepsilon \left[ A_{1}\exp \left( iky-i\Omega t\right)
+A_{2}\exp \left( -iky+i\Omega t\right) \right] \right\}
\end{equation*}
where $A_{1}$ and $A_{2}$ are \emph{real} coefficients. Linearising the
equation for small $\varepsilon $ we obtain: 
\begin{eqnarray*}
&&\left\{ \left( 2p^{3}\sigma +p\Omega -p\alpha k^{2}\right) \exp \left(
iky-i\Omega t\right) +2p^{3}\sigma \exp \left( -iky+i\Omega t\right)
\right\} A_{1}+ \\
&&\left\{ \left( 2p^{3}\sigma -p\Omega ^{\ast }-p\alpha k^{2}\right) \exp
\left( -iky+i\Omega t\right) +2p^{3}\sigma \exp \left( iky-i\Omega t\right)
\right\} A_{2} \\
&=&0
\end{eqnarray*}
Taking the complex conjugate we obtain another equation for the two \emph{%
real} coefficients $A_{1}$ and $A_{2}$. The condition of compatibility of
this system of equations (the determinant equals zero) can only be fulfilled
if the factor multiplying the function of $y$ and $t$ is allways zero. This
gives us the dispersion relation 
\begin{equation}
\Omega =\pm \sqrt{\alpha }k\left( \alpha k^{2}-4p^{2}\sigma \right) ^{1/2}
\label{disprel}
\end{equation}
which shows that for 
\begin{equation*}
\left| k\right| <2\left| p\right| \sqrt{\frac{\sigma }{\alpha }}
\end{equation*}
(\emph{i.e.} for long wavelengths) the two solutions $\phi \left( y,t\right) 
$ and $\widetilde{\phi }\left( y,t\right) $ diverge exponentially in time
and the function $\phi \left( y,t\right) $ is modulationally unstable. From
Eq.(\ref{disprel}) we notice that when $\sigma $ (the Landau coefficient)$<0$
and $\alpha $ (the dispersion coefficient)$>0$ the frequency $\Omega $ is
real and there is no linear instability. This remark is useful in connection
with the numerical calculations described above, which connect the physical
parameters with the coefficients $\left( \alpha ,\sigma \right) $ of the
NSE. Sampling the space of the physical parameters we must retain those
points where the linear instability is possible. In the above formula $%
\sigma $ , $\alpha $ and $p$ are dimensionless constants and the variables
of the equation are normalised : time to $t_{0}$ and length to $L$. The
value of $p$ is connected with the frequency of oscillation of the uniform
poloidal envelope $\omega _{unif}^{phys}$ ($s^{-1}$) by 
\begin{equation}
p=\sqrt{\frac{t_{0}}{2\sigma }\omega _{unif}^{phys}}  \label{eqp}
\end{equation}
and the condition of linear instability is for the poloidal wavenumber $%
\overline{m}$%
\begin{equation}
\overline{m}<\sqrt{2}\frac{a}{L}\sqrt{\frac{\sigma }{\alpha }}\left(
t_{0}\omega _{unif}^{phys}\right) ^{1/2}  \label{mbar}
\end{equation}
In the case of the following parameters: $\alpha =0.399$, $\sigma =1.452$, $%
d=2\pi a/L=2\pi \left( 1/0.3\right) =\allowbreak 20.\,\allowbreak 944$, $%
\omega _{unif}^{phys}=10^{5}$ (sec$^{-1}$) we have 
\begin{equation*}
\overline{m}\lesssim 8
\end{equation*}

For the Nonlinear Schrodinger Equation it is possible to follow the
perturbed solution much further in time in an exact manner. In Ref.(\cite
{Tracy}) it is shown that from a solution it is possible to return toward
the initial time and to recover the results of the linear analysis. This is
possible because the equation can be exactly solved for initial conditions
starting arbitrary close to the envelope of the plane wave. The case of the
plane wave (uniform envelope) is particular due to the simple form of the
main spectrum, as will be discussed below.

\subsection{Modulation instability of the envelope of a plane wave}

\subsubsection{General method}

The geometric-algebraic setting of Inverse Scattering Transform is the
appropriate framework for understanding the nature of the stability of the
exact solutions of the NSE. This is because the solution is strongly
dependent on the topology of the hyperelliptic Riemann curve which in turn
is dependent on the initial function. A perturbation of the initial
condition may change the main spectrum turning degenerate eigenvalues into
new nondegenerate pairs of eigenvalues. This means that the hyperelliptic
curve will have a different genus $N^{\prime }>N$ and new cycles and
differential forms should be defined. The solution depends precisely on this
topology.

We order to investigate the stability of an exact $N$-band solution of the
NSE we have to develop in detail the construction of solution. In the
particular case where the solution is the plane wave this analysis is
simpler since all the spectral properties can be found analytically. It has
been proved \cite{Tracy} that under the perturbation of the initial
condition the degeneracies in the main spectrum will be removed and new
degrees of freedom will become active. In the new solution some of the new
degrees of freedom have little effect and can be neglected.

\subsubsection{Calculation of the main spectrum associated to the uniform
amplitude solution}

The unperturbed solution is the envelope of the plane wave, Eq.(\ref
{envplane}). To deteremine the main spectrum, the solution is considered
initial condition (at $t=0$) and introduced in the Lax operator. The
eigenvalue problem of the Lax operator has the form 
\begin{equation*}
\left( 
\begin{array}{cc}
i\sqrt{\frac{\sigma }{\alpha }}\frac{\partial }{\partial y} & 1 \\ 
-1 & -i\sqrt{\frac{\sigma }{\alpha }}\frac{\partial }{\partial y}
\end{array}
\right) \Phi =\lambda \Phi
\end{equation*}
where 
\begin{equation*}
\Phi \equiv \left( 
\begin{array}{c}
\phi _{1} \\ 
\phi _{2}
\end{array}
\right)
\end{equation*}
The equation is: 
\begin{equation*}
\frac{\partial ^{2}\phi _{1}}{\partial y^{2}}+\frac{\alpha }{\sigma }\left(
1+\lambda ^{2}\right) \phi _{1}=0
\end{equation*}
The general solution is written 
\begin{equation*}
\phi _{1}=A\cos \left( \kappa y\right) +B\sin \left( \kappa y\right)
\end{equation*}
\begin{equation*}
\kappa =\sqrt{\frac{\alpha }{\sigma }}\sqrt{1+\lambda ^{2}}
\end{equation*}
The second component will be also a combination of harmonic functions: 
\begin{equation*}
\phi _{2}=C\cos \left( \kappa y\right) +D\sin \left( \kappa y\right)
\end{equation*}
The equation can have two independent one-column solutions. The one
corresponding to the boundary conditions 
\begin{equation*}
\Phi _{1}\left( y=0;\lambda \right) \equiv \left( 
\begin{array}{c}
\phi _{1} \\ 
\phi _{2}
\end{array}
\right) _{y=0}=\left( 
\begin{array}{c}
1 \\ 
0
\end{array}
\right)
\end{equation*}
is determined by $A=1$, $B=\sqrt{\alpha /\sigma }\,\lambda /\left( i\kappa
\right) $ , $C=0$ and $D=\sqrt{\alpha /\sigma }\,i/\kappa $. The solution is 
\begin{equation}
\Phi _{1}\left( y;\lambda \right) =\left( 
\begin{array}{c}
\cos \kappa y-\lambda i\sqrt{\frac{\alpha }{\sigma }}\frac{\sin \kappa y}{%
\kappa } \\ 
i\sqrt{\frac{\alpha }{\sigma }}\frac{\sin \kappa y}{\kappa }
\end{array}
\right)  \label{Phibig1}
\end{equation}
The second fundamental solution corresponds to the initial condition; 
\begin{equation*}
\Phi _{2}\left( y=0;\lambda \right) =\left( 
\begin{array}{c}
0 \\ 
1
\end{array}
\right)
\end{equation*}
and it results: 
\begin{equation}
\Phi _{2}\left( y;\lambda \right) =\left( 
\begin{array}{c}
i\sqrt{\frac{\alpha }{\sigma }}\frac{\sin \kappa y}{\kappa } \\ 
\cos \kappa y+\lambda i\sqrt{\frac{\alpha }{\sigma }}\frac{\sin \kappa y}{%
\kappa }
\end{array}
\right)  \label{Phibig2}
\end{equation}

The fundamental solution matrix $\Phi $ is a $2\times 2$ matrix having as
column the two above solutions. From the fundamental solution matrix we
calculate the monodromy matrix 
\begin{equation*}
M\left( \lambda \right) \equiv \Phi \left( y=d;\lambda \right)
\end{equation*}

The discriminant is the trace of the monodromy matrix. 
\begin{equation}
\Delta \left( \lambda \right) =2\cos \left( d\sqrt{\frac{\alpha }{\sigma }}%
\sqrt{1+\lambda ^{2}}\right)  \label{discr2}
\end{equation}

We can proceed to the calculation of the \emph{spectrum}.

\subsection{The spectrum of the plane wave solution}

\subsubsection{The spectrum of the unperturbed initial condition (case of
plane wave)}

We take a set of constants as derived from the numerical calculations that
connects the physical conditions to the coefficients of the NSE, $d=2\pi
\left( \frac{a}{L}\right) ,\;\alpha =0.399,\;\sigma =1.452$. With these
constants we calculate the spectrum on the complex $\lambda $-plane.
Although we can have analytical solutions the graphs below show the
difficulty of resolving the details of the variation of the functions. To be
noted that the function is intersected with a plane arbitrarly placed at the
magnitude $TrM=5$, for easier 3d-representation. 
\begin{figure}[tbp]
\centerline{
 \psfig{file=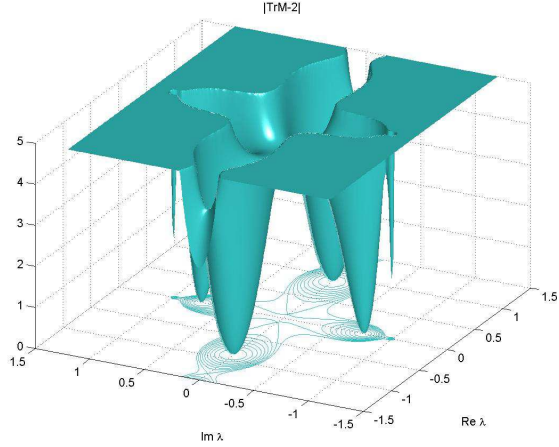,width=0.7\textwidth}}
\caption{Absolute value of $Tr(M)-2.$ The zeroes on the $Im \protect\lambda $
axis are difficult to resolve.}
\end{figure}

This graph shows the presence of roots on the imaginary axis of $\lambda $
of the equation $\left| TrM\left( \lambda \right) -2\right| =0$ with very
narrow variation of the function. Actually it is easily to see that for the
parameters adopted here, there are \emph{three} roots of this equation,
while in the graph we can hardly see two of them. A graphical better way of
finding these roots is to look for the intersections of the contours of the
two equations: 
\begin{eqnarray*}
\func{Re}\left( TrM\right) -2 &=&0 \\
\func{Im}\left( TrM\right) &=&0
\end{eqnarray*}
\begin{figure}[tbp]
\centerline{
 \psfig{file=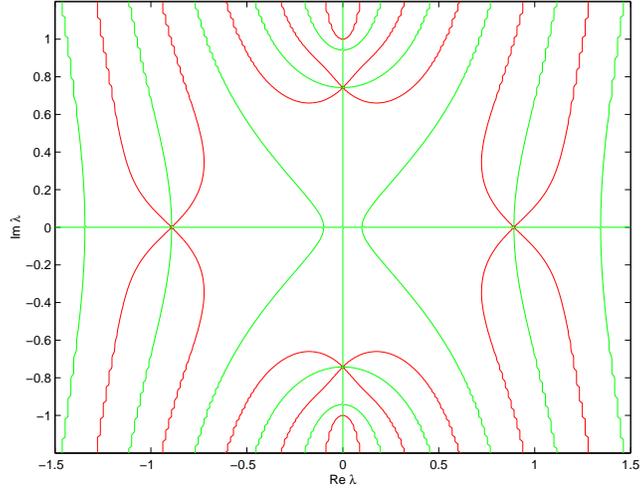,width=0.7\textwidth}}
\caption{Contours of the solutions of: $TrM-2=0$ and $ImTrM=0$.}
\end{figure}
where, again, the other two roots cannot be easily resolved. However the
analytical expression allows to find the positions of the roots both on the $%
\func{Re}\lambda $ axis 
\begin{figure}[tbp]
\centerline{
 \psfig{file=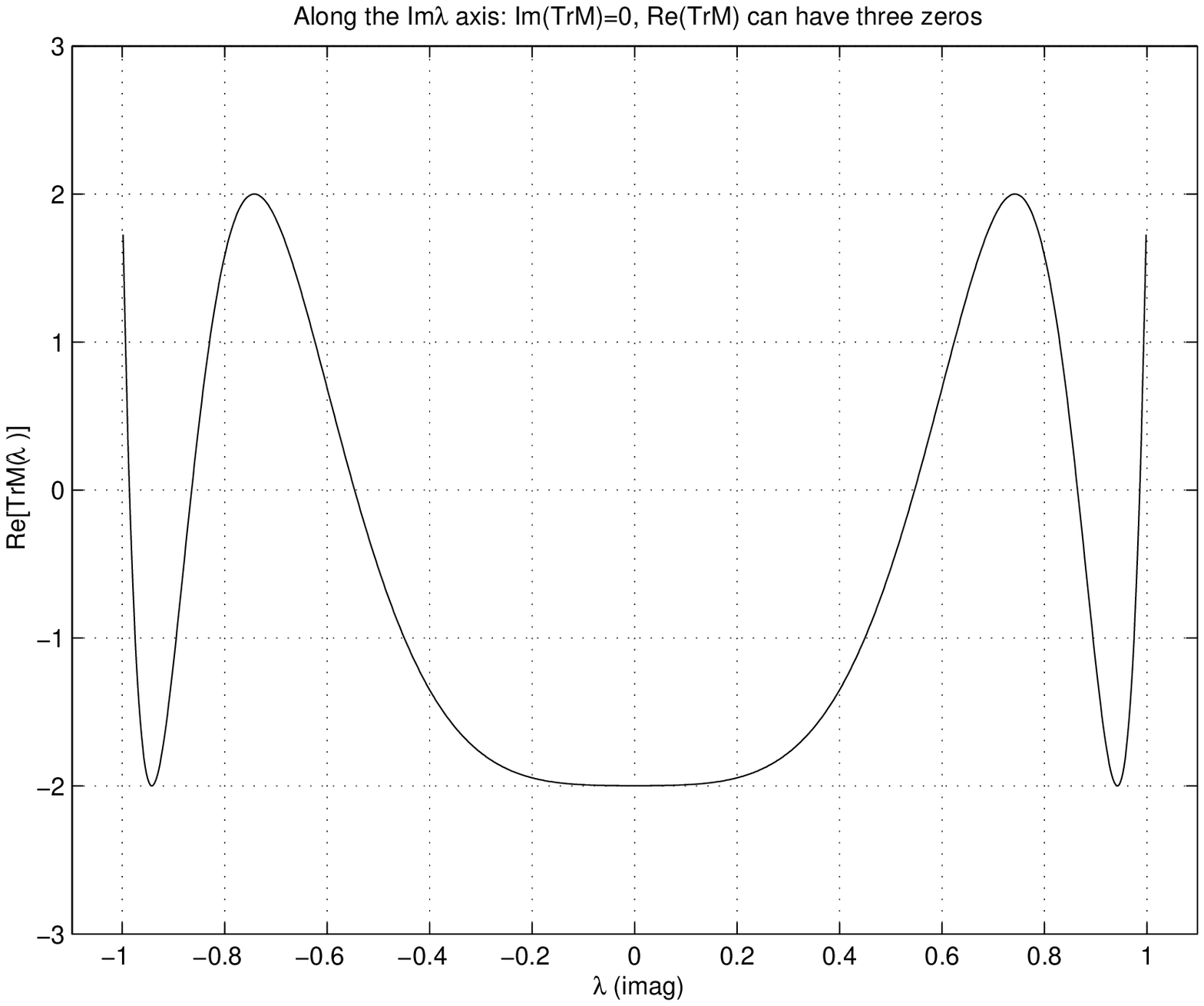,width=0.7\textwidth}}
\caption{}
\end{figure}
and on the $\func{Im}\lambda $ axis: 
\begin{figure}[tbp]
\centerline{
 \psfig{file=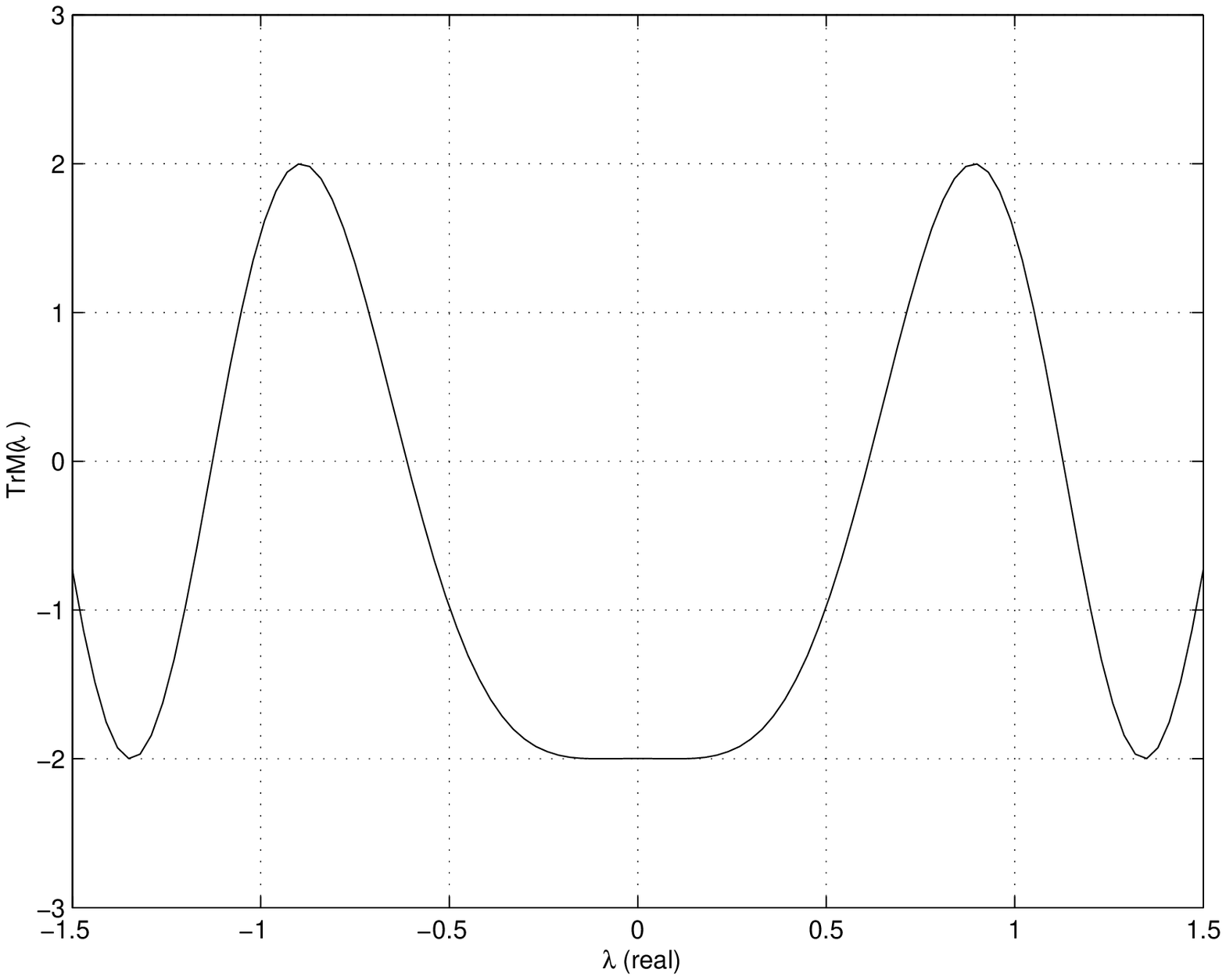,width=0.7\textwidth}}
\caption{}
\end{figure}

By contrast, the variation along the real $\lambda $ axis easily evidences
the zeroes of the corresponding to the \emph{main spectrum}. 
\begin{figure}[tbp]
\centerline{
 \psfig{file=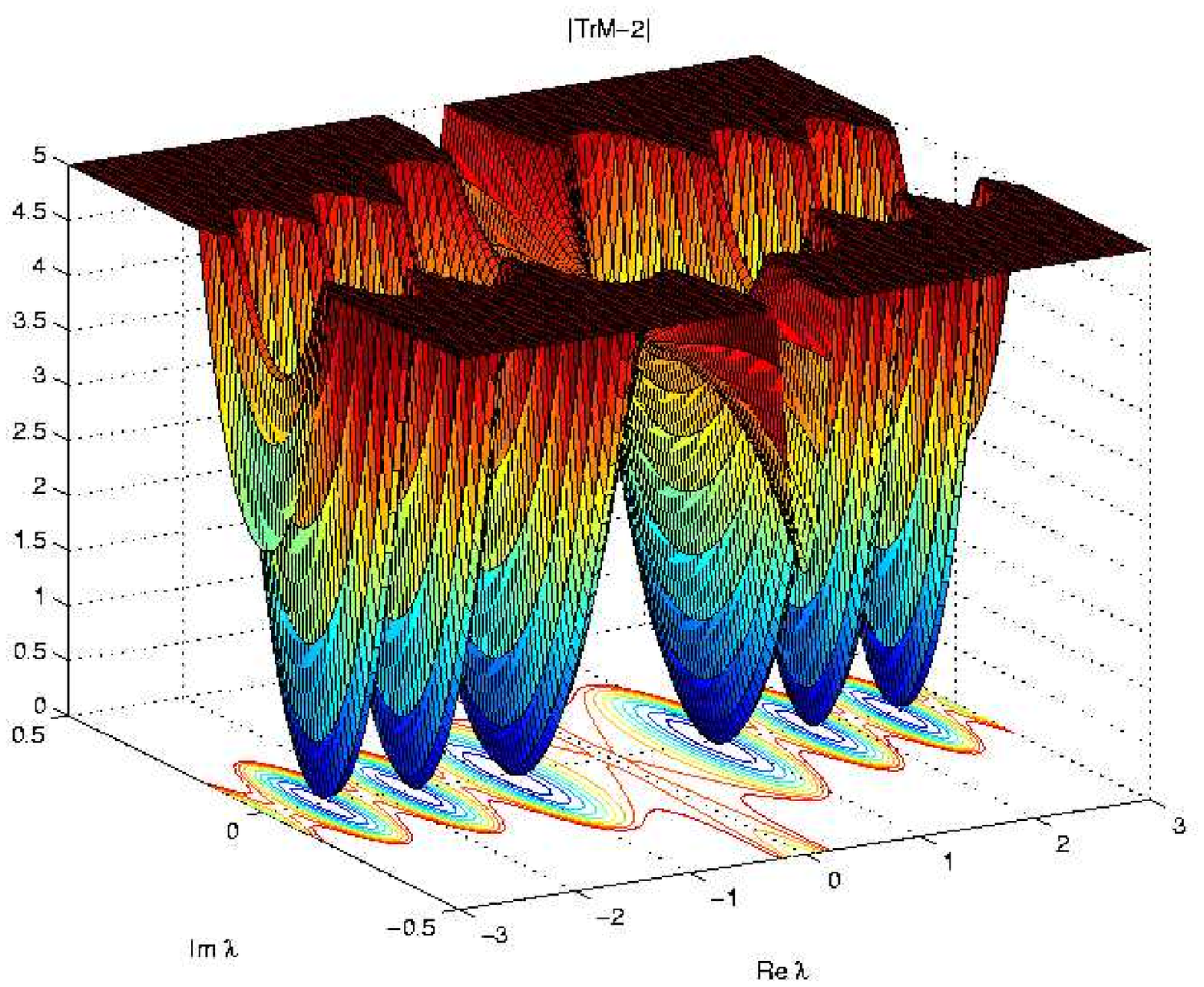,width=0.7\textwidth}}
\caption{The graph of $Tr(M)-2$ showing the zeroes on the $Re \protect\lambda
$ axis.}
\end{figure}

The stability or instability of the Bloch fuctions (\emph{i.e.} the values
of the quantity $m\left( \lambda \right) $ which is the eigenvalue of the
monodromy matrix) is governed by the \emph{discriminant} $\Delta \left(
\lambda \right) $. We have $-2\leqslant \Delta \left( \lambda \right)
\leqslant 2$ for all real $\lambda $ and for $\lambda =i\alpha $ with $%
-1\leqslant \alpha \leqslant 1$. The \emph{main spectrum} corresponds to the
values of $\lambda $ for which $\Delta \left( \lambda \right) =\pm 2$. For 
\emph{real} $\lambda $, $\lambda \equiv \lambda _{n,\func{real}}^{0}$ this
means 
\begin{equation*}
\lambda _{n,\func{real}}^{0}=\pm \left[ \left( \frac{n\pi }{d\sqrt{\alpha
/\sigma }}\right) ^{2}-1\right] ^{1/2}
\end{equation*}
for $n\geq 3.$ For pure imaginary $\lambda $, $\lambda \equiv i\lambda
_{n,imag}^{0}$ 
\begin{equation*}
\lambda _{n,imag}^{0}=\pm \left[ 1-\left( \frac{n\pi }{d\sqrt{\alpha /\sigma 
}}\right) ^{2}\right] ^{1/2}
\end{equation*}
for $n=0$, $1$, $2$ (with the values of the parameters $\alpha $, $\sigma $
and $d$). These values are: $i\lambda _{0,imag}^{0}=\pm i$, $i\lambda
_{1,imag}^{0}=\pm i0.9422$ and $i\lambda _{2,imag}^{0}=\pm i0.7424$. Since $%
d $ is finite only a finite number of degenerate pairs will appear on the
imaginary axis. At $n=0$ there are two eigenvalues $\lambda _{0}=\pm i$ .
These eigenvalues are \emph{nondegenerate} in the sense that there is only
one Bloch function for each of these values of\textbf{\ }$\lambda $. This
can be shown directly, constructing the Bloch functions form the matrix of
fundamental solutions (\ref{Phibig1}) and (\ref{Phibig2}) with $\kappa
\rightarrow 0$ (because $\Delta \left( \lambda \right) =2$ implies $\kappa
=0 $) and showing that only \textbf{one Bloch function} can be obtained.

\subsubsection{The spectrum of the perturbed initial condition (case of
plane wave)}

Starting from an initial condition slightly different of the \emph{plane
wave's envelope} 
\begin{equation}
u\left( y,0\right) \rightarrow u\left( y,0\right) +\varepsilon h\left(
y\right)  \label{pertini}
\end{equation}
we shall have to repeat all steps in the determination of the spectrum. This
must be done in a comparative manner, identifying the degenerate eigenvalues 
$\lambda _{j}$ of the main spectrum which becomes nondegenerate after
perturbation. The time behaviour of the exact solution developing from the
perturbed initial condition is dominated by the imaginary part of the
variables $W$ which are obtained on the basis of the new (perturbed)
quantities $\mu _{j}$ and this can be calculated.

Using the new initial condition Eq.(\ref{pertini}) we shall calculate the
new main spectrum, i.e. the new values of the parameter $\lambda $ for which
the discriminant is $\pm 2$. All the spectrum of the simple initial
condition $u\left( y,0\right) $ is perturbed by order-$\varepsilon $
quantities. The main effect of this perturbation is that the degeneracies
are broken. In order to construct the exact solution corresponding to the
new position of the nondegenerate $\lambda $ values, we need to specify the
cycles on the two-sheeted Riemann surface. The steps are as usual and
consists of first choosing the cycles on the Riemann surface and the base of
the holomorphic differentials; then, compute the matrices of periods (the
matrix of $a$ periods, and the matrix of the $b$ periods); compute the $\tau 
$ matrix; construct the Abel map, which maps the points of the compact
two-sheeted Riemann surface onto the Jacobi torus; find the wavenumbers and
the frequencies in the Jacobi manifold; make the Jacobi inversion, using the
theta functions;

\paragraph{Finding the base of the holomorphic differentials}

One of the eigenvalue pair of the unperturbed system is $\left( -i,i\right) $%
. Consider now that under the perturbation a degenerate eigenvalue $\lambda
_{j}^{0}$ splits into the conjugated pair $\left( \lambda
_{k}^{0}-\varepsilon _{k},\lambda _{k}^{0}+\varepsilon _{k}\right) $. In
Ref.(\cite{Tracy}) it was constructed a new basis as a linear combination of
the canonical basis: 
\begin{equation*}
dU_{j}=\frac{1}{2\pi i}\sqrt{1+\left( \lambda _{j}^{0}\right) ^{2}}\frac{%
\prod_{m\neq j}\left( \lambda -\lambda _{m}^{0}\right) d\lambda }{R\left(
\lambda \right) },\;\;\text{for}\;\;j=1,2,...,N
\end{equation*}
where 
\begin{equation*}
R\left( \lambda \right) =\sqrt{1+\lambda ^{2}}\prod_{k=1}^{N}\left[ \left(
\lambda -\lambda _{k}^{0}-\varepsilon _{k}\right) \left( \lambda -\lambda
_{k}^{0}+\varepsilon _{k}\right) \right] ^{1/2}
\end{equation*}

To understand the choice of basis it is useful to look at the limit $%
\varepsilon \rightarrow 0$ where the basis of holomorphic differential
becomes 
\begin{equation*}
\underset{\varepsilon \rightarrow 0}{\lim }dU_{j}=\frac{1}{2\pi i}\sqrt{%
1+\left( \lambda _{j}^{0}\right) ^{2}}\frac{1}{\sqrt{1+\lambda ^{2}}}\frac{%
d\lambda }{\left( \lambda -\lambda _{j}^{0}\right) }
\end{equation*}
This expression shows that, at the limit $\varepsilon \rightarrow 0$, each
differential ``sees'' only one pole, at $\lambda _{j}$ (which is a \emph{%
double point} at this limit), and the square-root \emph{branch} points $\pm
i $.

The loop (cycle) $a_{j}$ encircles the pole $\lambda _{j}^{0}$ and the
period is 
\begin{equation*}
\underset{\varepsilon \rightarrow 0}{\lim }A_{kj}=\underset{\varepsilon
\rightarrow 0}{\lim }\int_{a_{j}}dU_{k}=\int_{a_{j}}\underset{\varepsilon
\rightarrow 0}{\lim }dU_{k}=\delta _{kj}
\end{equation*}
This is because the cycle $a_{j}$ surrounds the pole $\lambda _{j}^{0}$ or
it does not surrounds any other pole.

We find that, under the \emph{limit operation} $\varepsilon \rightarrow 0$,
the matrix $\mathbf{A}$ goes over to the identity matrix to the first order
in $\varepsilon $, $O\left( \varepsilon \right) $ and that the matrix $%
\mathbf{B}$ becomes 
\begin{equation*}
\underset{\varepsilon \rightarrow 0}{\lim }B_{jk}=\tau _{jk}+O\left(
\varepsilon \right)
\end{equation*}

To see what happens with the $\mathbf{B}$ matrix in this limit, we must
examine the \emph{off-diagonal} terms and the diagonal terms.

The \emph{off-diagonal} terms are well-behaved and can be integrated
analytically: 
\begin{equation*}
\underset{\varepsilon \rightarrow 0}{\lim }B_{jk}=\int_{b_{j}}\underset{%
\varepsilon \rightarrow 0}{\lim }dU_{k}\;\;\text{for}\;\;j\neq k
\end{equation*}
and 
\begin{equation*}
\tau _{kj}=\frac{1}{2\pi i}\sqrt{1+\left( \lambda _{j}^{0}\right) ^{2}}%
\int_{b_{j}}\frac{1}{\sqrt{1+\lambda ^{2}}}\frac{d\lambda }{\left( \lambda
-\lambda _{k}^{0}\right) }+O\left( \varepsilon \right)
\end{equation*}
The contour $b_{j}$ passes through the middle of the segment connecting the
two new $\lambda _{j}$ and goes around the pole at $\lambda =-i$, for
example. If this contour does not surround the pole at $\lambda _{k}^{0}$
then the integral is elementary. If the contour intersecting the cut from $%
-i $ to infinity encircles the pole at $\lambda _{k}^{0}$ then it may be
chnaged such that to intersect the other cut, connecting the branch point $%
+i $ to infinity, and so it does not encircle any pole. In conclusion since
the off-diagonal elements of the matrix $\mathbf{\tau }$ depend (to $0\left(
\varepsilon \right) $) only of the positions of the original double points,
which are determined only by the parameter $d$, it results that: \textbf{the
off-diagonal terms of the }$\mathbf{\tau }$ \textbf{matrix does not carry
any information on the }\emph{perturbed initial condition}. \textbf{Only the
diagonal terms of the }$\tau $\textbf{\ matrix are affected by the initial
condition.}

The \emph{diagonal} terms of the $\mathbf{\tau }$ - matrix are \emph{singular%
} at the limit $\varepsilon \rightarrow 0$. Consider 
\begin{equation*}
\tau _{kk}=\frac{1}{2\pi i}\sqrt{1+\left( \lambda _{j}^{0}\right) ^{2}}%
\int_{b_{k}}\frac{1}{\sqrt{1+\lambda ^{2}}}\frac{d\lambda }{\sqrt{\left(
\lambda -\lambda _{k}^{0}\right) ^{2}-\varepsilon _{k}^{2}}}+O\left(
\varepsilon \right)
\end{equation*}

Suppose that 
\begin{equation*}
\lambda _{k}^{0}\;\;\text{is on the \textbf{imaginary axis}}
\end{equation*}
Then we have 
\begin{equation*}
\tau _{kk}=\frac{1}{2\pi i}\sqrt{1+\left( \lambda _{j}^{0}\right) ^{2}}%
\left( 2\int_{-i}^{\lambda _{k}^{0}}+\frac{1}{2}\int_{a_{k}}\right) \frac{1}{%
\sqrt{1+\lambda ^{2}}}\frac{d\lambda }{\sqrt{\left( \lambda -\lambda
_{k}^{0}\right) ^{2}-\varepsilon _{k}^{2}}}+O\left( \varepsilon \right)
\end{equation*}
\begin{equation*}
\tau _{kk}=\frac{1}{2}+\frac{1}{\pi i}\sqrt{1+\left( \lambda _{j}^{0}\right)
^{2}}\int_{-i}^{\lambda _{k}^{0}}\frac{1}{\sqrt{1+\lambda ^{2}}}\frac{%
d\lambda }{\sqrt{\left( \lambda -\lambda _{k}^{0}\right) ^{2}-\varepsilon
_{k}^{2}}}+O\left( \varepsilon \right)
\end{equation*}
The integral can be made with \emph{elliptic functions} but an approximation
is (near $\varepsilon _{k}=0$) 
\begin{equation*}
\tau _{kk}=\frac{1}{2}-\frac{i}{\pi }\ln \left| \varepsilon _{k}\right|
+O\left( 1\right) +O\left( \varepsilon _{k}\right)
\end{equation*}

Consider now 
\begin{equation*}
\lambda _{k}^{0}\text{\ \ is on \textbf{real axis}}
\end{equation*}
then the integral defining the diagonal elements of the matrix $\tau $ (the
integral along the $b_{k}$ cycles) is 
\begin{equation*}
\int_{-i}^{\lambda _{k}^{0}}=\int_{-i}^{0}+\int_{0}^{\lambda _{k}^{0}}
\end{equation*}
where the second integral is along the \textbf{real axis}.

The first integral is done with the residuu theorem after taking safely the
limit $\varepsilon _{k}=0$ and gives a finite imaginary contribution which
is well-behaved for this limit. The second can be done and the result is 
\begin{equation*}
\tau _{kk}=-\frac{i}{\pi }\ln \left| \varepsilon _{k}\right| +iO\left(
1\right) +O\left( \varepsilon _{k}\right)
\end{equation*}

We then conclude that for $\varepsilon _{k}\rightarrow 0$ the $b$ - period
matrix $\tau $ becomes logarithmically singular on the main diagonal.

The main conclusion is: (1) The real-$\lambda $ axis degenaracies have $%
\Omega _{j}$ real so they are linearly stable. The imaginary axis
degeneracies have $\Omega _{j}$ imaginary so they are linearly unstable.

This conclusion can have important practical effects: in our case the
degenerate eigenvalues on the \emph{imaginary} $\lambda $-axis split and
induce instability of the plane wave solution, \emph{i.e.} of the envelope.
At the same time the solutio evolves toward one of the stable (soliton-like)
solutions. Still much work is required to examine the stability of that
solution to additional perturbation. For example, the main spectrum of the
solution of the type $\sec h(y)$ can be obtained from the discriminant whose
form is very complicated, with many purely imaginary (dangerous) eigenvalues.%
%
%
%
%
%
%
%
\begin{figure}[tbp]
\centerline{
 \psfig{file=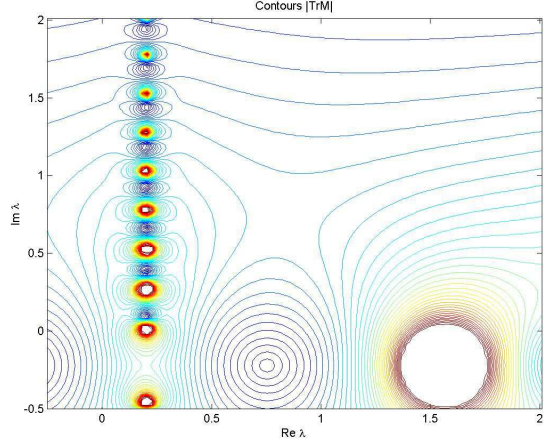,width=0.7\textwidth}}
\caption{The discriminant $TrM\left( \protect\lambda \right) $ of the $\sec
h\left( y\right) $ solution, to compare with the plane wave solution's
discriminant.}
\end{figure}

\section{The growth of the perturbed solution as a source of poloidal
asymmetry and spontaneous plasma spin-up}

\subsection{The minimal condition for instability of the poloidal rotation}

We have found that the space-uniform solution (poloidal symmetry) is
unstable and the turbulence will self-modulate according to one of the
possible soliton-like solution. In a time picture, we can say that the
poloidal symmetry will be left and the turbulence amplitude will develop a
poloidal asymmetry which grows exponentially in time. The onset of the
poloidal asymmetry will induce an asymmetric rate of transport on the
poloidal direction \cite{cehii}. When the particle diffusion is poloidally
asymmetric and the local particle confinement time is shorter than the
damping time of the poloidal rotation, the poloidal rotation is unstable.
The plasma can spontaneously spin-up. The growth rate of the asymmetry
(which is given by the rate of departure from the uniform solution toward a
soliton-like solution) determines the growth rate of the torque applied to
plasma in the poloidal direction but the relation is not direct, depending
essentially of the momentum balance in the poloidal direction. Any torque
acting to generate a poloidal rotation must overcome the neoclassical
magnetic pumping, which is an efficient damping mechanism. A balance of
these forces depends on the comparison between the rates of (1) asymmetry
growth, the radial variation of the asymmetry-induced terms, on one hand,
and (2) rate of damping through magnetic pumping, on the other hand.

For the simplest case of the space-uniform envelope solution, we have the
instability has the \emph{linear} growth rate 
\begin{equation*}
i\gamma ^{env}=i\func{Im}\Omega =i\left| \sqrt{\alpha }k\left( \alpha
k^{2}-4p^{2}\sigma \right) ^{1/2}\right| 
\end{equation*}
where $k<2p\sqrt{\sigma /\alpha }$ is here the wavenumber of the
perturbation of the envelope. In the case that has served as example in the
preceeding Section, we have $p=\sqrt{t_{0}\omega _{unif}^{phys}/\left(
2\sigma \right) }\approx 1.86$ and $k<7.08$. Taking for example $k=3$ we
obtain $\gamma ^{env}=7.67$ or $\gamma _{phys}^{env}=7.67\times 10^{4}$ ($%
s^{-1}$).

\subsection{The soliton developing from the initial perturbation}

Since we know that the uniform state can be unstable we have to investigate
the problem of the evolution of the system after a perturbation is applied
and look for stable solutions. The solitons of this system can be realisable
in real conditions since they are intrinsically more stable. We will look
for soliton-like solution for the uncoupled equations, taking a single NSE
like (\ref{eqalfsig}). The time dependence can be
\begin{equation*}
\phi =\exp \left( i\mu ^{2}\alpha t\right) w\left( x\right) 
\end{equation*}
giving the reduced equation
\begin{equation*}
\frac{\partial ^{2}w}{\partial x^{2}}-\mu ^{2}w+\frac{\sigma }{\alpha }%
w^{3}=0
\end{equation*}
where $\mu $ is a real constant. This equation has the following solution
\begin{equation*}
w=\sqrt{\frac{2\alpha }{\sigma }}\mu \sec h\left( \mu x\right) 
\end{equation*}
The constant $\mu $ can be related to the modulation wavenumber whose upper
bound is (\ref{mbar}). Taking $\mu \simeq 3$  and $\alpha =0.399$ we obtain
the frequency of steadily oscillating soliton $\omega _{osc}^{env}\equiv \mu
^{2}\alpha \simeq 1.2$, which is slower than the growth rate of the
deformation, $\gamma ^{lin}=7.67$ obtained above (all in absolute units).
Since the rate of change of the poloidal velocity (which will be calculated
below) has comparable magnitude with $\omega _{osc}^{env}$ we conclude that
the rotation has the time to develop before the oscillation reverses the
sense of the applied torque. The maximum amplitude of the deformation is $%
\sqrt{2\alpha /\sigma }\mu =2.224$. Large smooth deformations (small $\mu $)
with less localised  profile will have even slower steady oscillatory motion
and however significant amplitude, and we can expect they are favored in
real situations.

\subsection{The poloidal asymmetry of the diffusion fluxes}

Our approach, which is adapted to the investigation of the stability of the
turbulence envelope, can only provide a \emph{model} of the turbulent field,
according to Eq.(\ref{psiat1}). This is composed of the primary modes
propagating in the poloidal direction with amplitude modulated by the slowly
varying envelopes $A_{1,2}$ obeying the system of coupled Nonlinear
Schr\"{o}dinger Equations (\ref{cse1}) and (\ref{cse2}). In the radial
direction the field has a variation given by the functions $\varphi
_{1,2}\left( x\right) $ obtained by solving the equations (\ref{phin}). We
will need to reconstitute the turbulent field in order to calculate the
modified diffusion coefficient. We choose the following set of parameters: 
\begin{eqnarray*}
a &=&1\;\left( m\right) \\
B_{T} &=&3.5\;\left( T\right) \\
T_{e} &=&1\;\left( KeV\right) \;,\;T_{i}=1\;\left( KeV\right)
\end{eqnarray*}
\begin{eqnarray*}
n &=&10^{20}\;\left( m^{-3}\right) \\
L_{n} &=&0.1\;\left( m\right)
\end{eqnarray*}
\begin{equation*}
U_{0}=6\times 10^{5}\;\left( m/s\right)
\end{equation*}
The extension on the radial direction of the mode is taken $L=0.1\;\left(
m\right) $ and this is also the normalising length. Other quantities are:
normalising time $t_{0}=0.16\times 10^{-6}\;\left( s\right) $ and $\Omega
_{i}=0.335\times 10^{9}\;\left( s^{-1}\right) $, the ion-cyclotron
frequency. From this it results the two parameters of the barotropic form of
the ion mode, $\beta =111.75$ and $\varepsilon =0.476\times 10^{-2}$. We
take the following frequencies for the primary poloidal modes: $\omega
_{1}=0.1451\times 10^{6}\;\left( s^{-1}\right) $ and $\omega
_{2}=0.1931\times 10^{6}\;\left( s^{-1}\right) $. From the solution of the
equations (\ref{phin}) the two eigenvalues result: $k_{1}=43.51\;\left(
m^{-1}\right) $ and $k_{2}=56.81\;\left( m^{-1}\right) $. Solving the series
of differential equations described in detail in Section 3 we obtain the
following constants in the coupled Nonlinear Schr\"{o}dinger Equations: $%
\alpha _{1}=0.1756\times 10^{-2}$ , $\alpha _{2}=0.2278\times 10^{-2}$ and
respectively $\sigma _{1}=0.18272$ , $\sigma _{2}=0.5110$. The frequency of
the uniform oscillations, Eq.(\ref{eqp}) is taken $p=0.0675$ giving the
maximum poloidal number for the large scale instability, Eq.(\ref{mbar}), $%
\overline{m}=\left[ 5.89\right] =5.$ The growth rate of the modulational
instability is $\gamma =4175\;\left( s^{-1}\right) $. The expression of the
field becomes 
\begin{eqnarray*}
\phi \left( r,\theta \right) &=&\phi _{0}\left[ (1+A_{1}\left( \theta
\right) \sec h(y))\cos \left( k_{1}y-\omega _{1}t+\delta _{1}\right) \varphi
_{1}\left( r\right) \right. \\
&&\left. +(1+A_{2}\left( \theta \right) \sec h(y))\cos \left( k_{2}y-\omega
_{2}t+\delta _{2}\right) \varphi _{2}\left( r\right) \right]
\end{eqnarray*}
where $\phi _{0}$ is explained below. The initial phases $\delta _{1,2}$ are
arbitrary and in the numerical calculations are taken as random quantities
with uniform distribution on the interval $\left( 0,1\right) $.

We notice that the equations (\ref{phin}) fulfilled by $\varphi _{1,2}$ are
homogeneous and the amplitudes of the two functions are arbitrary. We fix
the amplitude $\phi _{0}$ with the condition that the \emph{uniform} profile
of the averaged square of the potential of the fluctuating field obtains a
typical value of the diffusion coefficient, $D_{0}\sim \left\langle \left|
\phi _{0}\right| ^{2}\right\rangle $, which is taken $D_{0}\sim 1\;\left(
m^{2}/s\right) $. The following two figures represent the fluctuating
potential, resulting from the combination of the factors described above.

\begin{figure}[tbp]
\centerline{
 \psfig{file=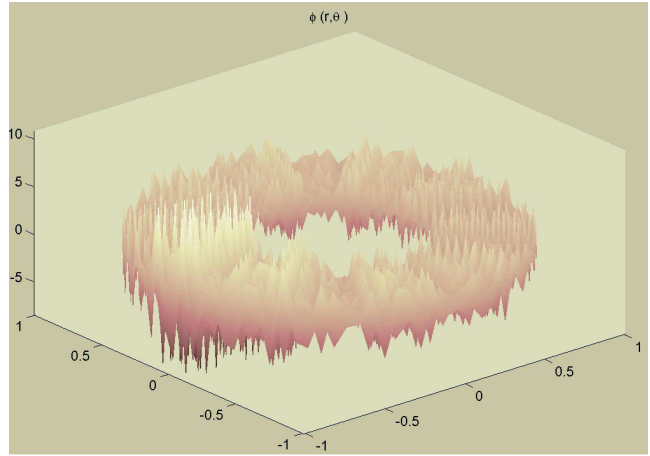,width=0.7\textwidth}}
\caption{Instantaneous amplitude of the potential fluctuation, as calculated
from the different contributions}
\end{figure}
\begin{figure}[tbp]
\centerline{
 \psfig{file=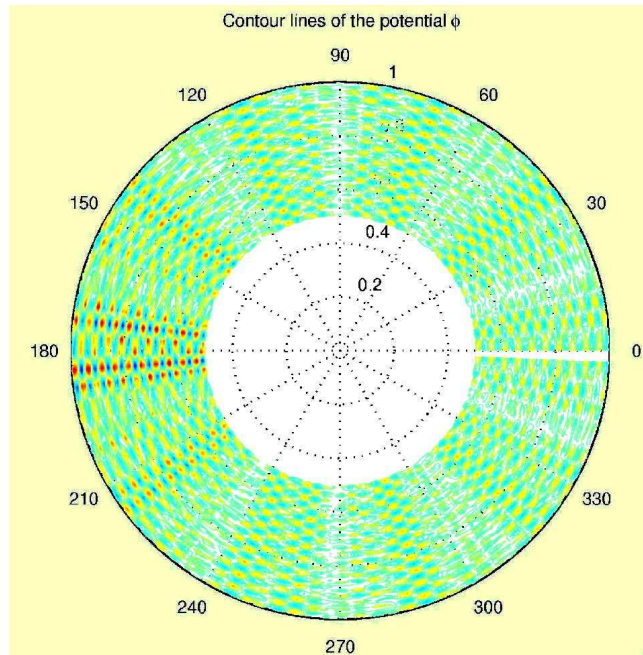,width=0.7\textwidth}}
\caption{Contour lines of the fluctuating potential; the radial extension is
artificially extended to improve the picture.}
\end{figure}

A simple way to take into account the effect of the turbulence amplitude
modulation on the diffusion coefficient is to consider that locally the
diffusion coefficient is proportional with the quantity $\left\langle \left|
\phi \right| ^{2}\right\rangle $, where the averages should be taken over a
statistical ensemble of realizations of the fluctuating field. This requires
a carefull examination when applied to numerical calculations. We have to
remember that the diffusion is a statistical process where a test particle
interacting with fluctuating background field performs random displacements
whose mean square grows linearly in time. The random displacements have to
belong to a Gaussian statistical ensemble and this requires that all kind of
elementary experiences (i.e. displacements) to be represented with the
corresponding probability. This is practically irrealisable in numerical
calculations since the statistical ensemble of the stochastic field
configurations is necessarly finite, i.e. incomplete as a Gaussian ensemble.
Any average will be made on subensembles, more or less incomplete. The
collection of field configurations are obtained from time \emph{and} from
space realizations of the field. The test particle should have the
possibility to manifest the \emph{linear} mean square displacement specific
to the Gaussian ensembles and this rises the question on what is the minimal
spatial extension on which the average should be done. We need in any case a
space region larger than the correlation length of the fluctuating field
which can be taken as the inverse of the characteristic wavenumber in the
field (minimum of $k_{1}$, $k_{2}$).

The profile of the quantity $D_{0}\left\langle \left| \phi \right|
^{2}\right\rangle $ is represented in the figure below. The exemple chosed
for these graphical representation is characterized by the imposed constrain
to obtain a rapid variation of the functions $\varphi _{1}$ and $\varphi
_{2} $ on the radial $r$-interval, with the intention to simulate short
wavelength ion turbulence, frequently observed in the numerical simulations
of the Ion Temperature Gradient driven turbulence. This is achieved by
asking the numerical procedures which solves the Schr\"{o}dinger-type
equations (\ref{phin}) to look for high order $\kappa $ of the eigenvalues.
The pictures represented in this section correspond to $\kappa =19$ for both
eigenfunctions. This generates a potential dominated by small spatial scales
and consequently, the fast variation of the local value of the diffusion
coefficient. We take, in the case $\kappa =19$ a space region containing at
least three complete oscillations of the typical wavelength. This should
allow at least several random steps in the typical displacement of the
particle. When we do this (taking arbitrarly a factor of few units larger
than the correlation length) we note that the diffusion coefficient exhibits
space variations. These are natural and are inherited from the space
dependence of the radial eigenfunctions $\varphi _{1,2}\left( r\right) $.
Since in real situations the radial eigenfunctions are superposed with
random initial phases we need to enlarge the region of averaging over most
of the radial interval considered, up to a level of saturation of these
spatial variation which can be considered convenient for the estimation of
the poloidal torque. 
\begin{figure}[tbp]
\centerline{
 \psfig{file=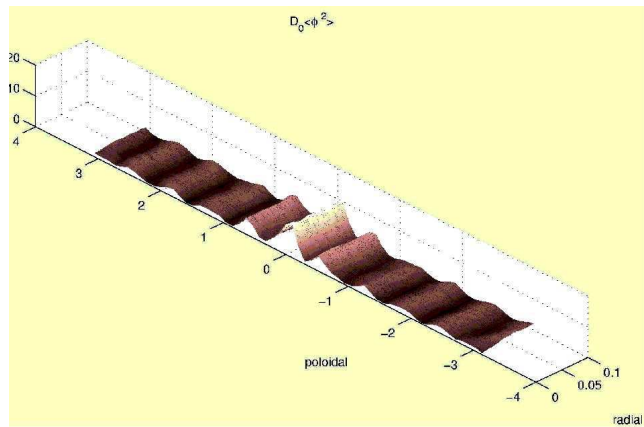,width=0.7\textwidth}}
\caption{The radial and poloidal dependence of the average diffusion
coefficient.}
\end{figure}
\begin{figure}[tbp]
\centerline{
 \psfig{file=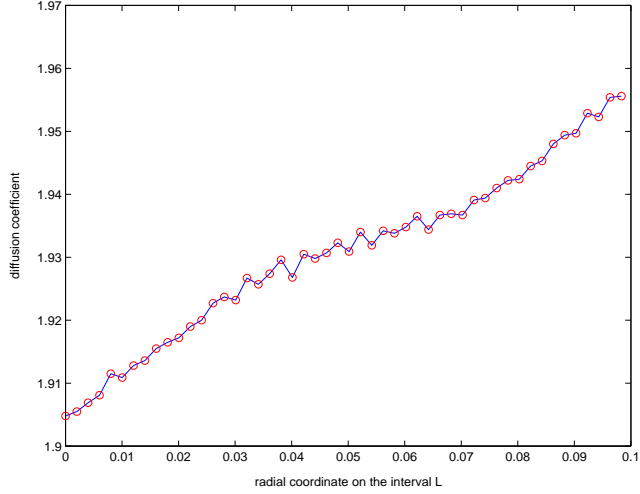,width=0.7\textwidth}}
\caption{Radial variation of the diffusion coefficient after poloidal
averaging.}
\end{figure}

\subsection{The poloidal torque applied on plasma}

We use the equations describing the plasma poloidal and toroidal rotation
derived in \cite{Hassam1}, \cite{Hassam2}: 
\begin{equation*}
\frac{\partial n}{\partial t}+\frac{1}{r}\frac{\partial }{\partial r}\left(
rn\overline{v}_{r}\right) =0
\end{equation*}
\begin{equation*}
\frac{\partial }{\partial t}\left( nV_{\varphi }\right) +\frac{1}{r}\frac{%
\partial }{\partial r}\left[ rn\left( V_{\varphi }\overline{v}%
_{r}-qV_{\theta }\widetilde{v}_{r}\right) \right] =0
\end{equation*}
\begin{equation*}
\frac{\partial }{\partial t}\left[ V_{\varphi }+\Theta \left(
1+2q^{2}\right) V_{\theta }\right] +\overline{v}_{r}\frac{\partial
V_{\varphi }}{\partial r}-\widetilde{v}_{r}\frac{\partial }{\partial r}%
\left( qV_{\varphi }\right) +\text{magnetic pumping}=0
\end{equation*}
The definition of the variables is $n=\left\langle n\left( r\right)
\right\rangle $, $nV_{\varphi }\left( r\right) \equiv \left\langle
nv_{\varphi }\frac{R}{R_{0}}\right\rangle $ and $V_{\theta }\left( r\right)
=\left\langle v_{\theta }\frac{R}{R_{0}}\right\rangle $, where $v_{\varphi }$
and $v_{\theta }$ are local toroidal and poloidal plasma velocities in the
magnetic surface. The magnetic field is 
\begin{equation*}
\mathbf{B}=\left( 0,\Theta \left( r\right) ,1\right) \frac{B_{0}\left(
r\right) }{R/R_{0}}
\end{equation*}
with $R=R_{0}+r\cos \theta $ and $q=\frac{r}{R_{0}}\frac{1}{\Theta }$. The
surface average is made according to the formula 
\begin{equation*}
\left\langle f\right\rangle =\oint \frac{d\theta }{2\pi }\frac{R}{R_{0}}\;f
\end{equation*}
In the collisional regime the \emph{magnetic pumping} term determined by
parallel viscosity is less effective and the spin up can be expected. The
decay rate of the poloidal rotation by this mechanism is \cite{Hassam3}: 
\begin{equation}
\gamma _{MP}\simeq \frac{3}{4}\left( 1+\frac{1}{2q^{2}}\right) ^{-1}\left( 
\frac{l}{qR}\right) ^{2}\nu _{ii}  \label{magnpump}
\end{equation}
where $l$ is the mean-free path. The velocities $\widetilde{v}_{r}$ and $%
\overline{v}_{r}$ arise from the diffusive flux in the radial direction, to
which is associated the local radial velocity $V_{r}$. We define the two
velocities 
\begin{equation*}
\overline{v}_{r}=\left\langle V_{r}\right\rangle
\end{equation*}
\begin{equation}
\widetilde{v}_{r}=\left\langle 2\cos \theta \;V_{r}\right\rangle
\label{avercos}
\end{equation}
If there is no poloidal variation of the local radial velocity $V_{r}$ the
average in (\ref{avercos}) is that of the trigonometric function $\cos
\theta $ and the result is zero. The average will not be zero if there is a
poloidal variation of the diffusion-generated particle radial velocity, $%
V_{r}\left( \theta \right) $. We can introduce the modified diffusion
coefficient $\widetilde{D}$ related to the nonsymmetry, 
\begin{equation}
n\widetilde{v}_{r}=\widetilde{\Gamma }_{r}=-\widetilde{D}\frac{\partial n}{%
\partial r}  \label{vrtilda}
\end{equation}
with 
\begin{equation}
\widetilde{D}=\left\langle D_{0}\left\langle \left| \phi \right|
^{2}\right\rangle _{st}2\cos \theta \right\rangle _{pol}  \label{dtilda}
\end{equation}
where the first averaging is statistical and the second is on the poloidal
surface. The equation for the plasma poloidal velocity is 
\begin{equation}
\Theta \left( 1+2q^{2}\right) \left( \frac{\partial V_{\theta }}{\partial t}%
+\gamma _{MP}V_{\theta }\right) +qV_{\theta }\frac{1}{nr}\frac{\partial }{%
\partial r}\left( nr\widetilde{v}_{r}\right) =0  \label{eqrot}
\end{equation}
The average velocity Eq.(\ref{vrtilda}) still dependes on the radial
coordinate via $D_{0}$, the density $n\left( r\right) $ and its derivatives
and, more important, via the space dependence of the potential $\varphi
\left( x\right) $ left after the statistical averaging needed for the
calculation of the diffusion coefficient (\ref{dtilda}). The following
quantity must be numerically calculated 
\begin{equation*}
\gamma \equiv \frac{-q}{\Theta \left( 1+2q^{2}\right) }\frac{1}{nr}\frac{%
\partial }{\partial r}\left( nr\widetilde{v}_{r}\right) =\frac{q}{\Theta
\left( 1+2q^{2}\right) }\frac{1}{nr}\frac{\partial }{\partial r}\left(
r\left\langle D_{0}\left\langle \left| \phi \right| ^{2}\right\rangle
_{st}2\cos \theta \right\rangle _{pol}\frac{\partial n}{\partial r}\right)
\end{equation*}
and this directly provides the rate of change of the poloidal velocity due
to the asymmetry. This must be compared to the magnetic pumping term (\ref
{magnpump}) and we expect rotation instability when 
\begin{equation*}
\gamma -\gamma _{MP}>0
\end{equation*}
In this formula the soliton profile of the poloidal asymmetry is considered
saturated and the motion of the solitons (which would give periodic
variation of the torque) is neglected. As is shown in Ref.\cite{Hassam1} the
high positive second derivative of the density profile, combined with a
dependence of the diffusion coefficient $D_{0}$ on the shear of the poloidal
velocity are the favorable circumstances for the asymmetry-induced torque to
overcome the magnetic pumping term and a spontaneous spin up to occur. We do
not take into account the variation of $D_{0}$ with the velocity shear, but
notice that the space variation of (\ref{dtilda}) also contribute to obtain
a positive value for $\gamma $ sufficient for being comparable to $\gamma
_{MP}$ and even to be larger. The most important contribution comes however
from the positive-valued second density gradient, which can be seen as an
indication that the spin-up can be expected in such region on the minor
radius. 
\begin{figure}[tbp]
\centerline{
 \psfig{file=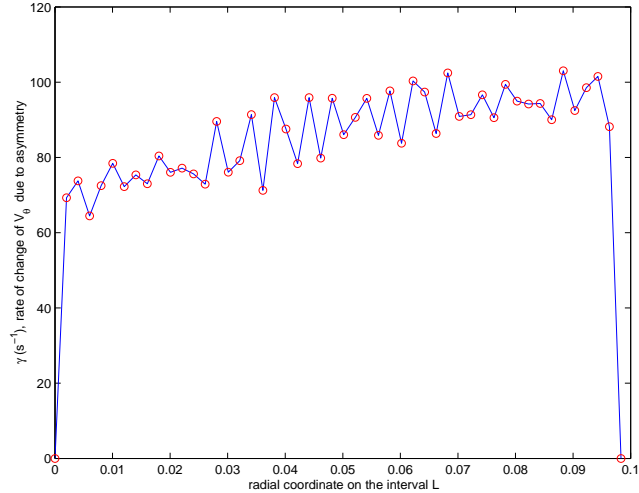,width=0.7\textwidth}}
\caption{Radial variation of the rate of time variation $\protect\gamma $
applied on plasma in the poloidal direction, due to the asymmetry.}
\end{figure}

The values represented in the figure have been calculated using a gaussian
profile for the density. The decay due to magnetic pumping is of the order
of $\gamma _{MP}\sim 12\;\left( s^{-1}\right) $. We note that the torque
applied to plasma shows a radial variation essentially inherited from the
space variation of the statistically averaged potential fluctuations, which
in general are not monotone functions of the radius on time scales of
several eddy turn-over times. These spatial variations of the diffusion
coefficient generate the profile of the torque which change the shear of the
seed velocity and provide a characteristic which is expected for zonal
flows. Even outside\ the domain of the parameters where the spin up is
efficient compared to the magnetic pumping, the soliton asymmetry and the
Stringer torque can contribute to the action of other mechanisms to generate
poloidal rotation.

\section{Discussion}

We have examined the problem of stability of the envelope of the ion wave
turbulence. Starting from a simple model for the ion dynamics and retaining
the full nonlinear contribution of the ion polarization current we have
derived an equation of the same form as the \emph{barotropic} model of the
atmosphere. The multiple space-time scale analysis shows that the envelope
of the turbulence results from a set of coupled cubic nonlinear Schrodinger
equations. The simplest solution is the space-constant envelope
corresponding to a symmetric average amplitude of the turbulence on the
poloidal circumference. In order to examine the stability of this solution
we first review the method of inverse scattering transform in the
geometric-algebraic framework, able to obtain the exact solution developing
from a certain class of initial conditions (finite-band potentials). In this
framework appears clear the origin of the sensitivity to the perturbations,
connected to the topology of a hyperelliptic Riemann surface. The uniform
solution is unstable and evolves to one of the soliton-type solution. The
nonuniformity of the turbulence average profile generates a torque on the
plasma, which may be sufficient to overcome the poloidal damping.

The simple physical model we have employed may be partly justified by the
observation that the analysis is essentially independent of the other
possible sources of poloidal asymmetry: the neoclassical variation of
parameters on the magnetic surface, the toroidal effect on the nonlinearity
represented as mode coupling, the effect of the particle drifts on the
linear and renormalized propagators, etc. The cubic nonlinear Schrodinger
equation has many solutions on the periodic domain and the relevance for the
self-modulation of the envelope of the turbulence should be examined. Here
we only considered the uniform envelope since its instability can be
sufficient to induce plasma poloidal rotation. This rotation has an origin
different of the usual sources of rotation: neoclassical, ion-loss, Reynolds
stress or momentum injection. Our approach is essentially mathematical and
so it operates at a more formal level. The physical origin of the
instability seems to be the tendency of the system (averaged amplitude of
the turbulence) to evolve to profiles characterised by the balance of the
dispersion and the nonlinearity. What remains to be examined is the
quantitative constraint on the minimal initial perturbation needed to
produce a particular evolution: for a weak perturbation the solution may be
simple ``radiation'' while for initially larger amplitude a soliton emerges
ensuring higher robustness and a localised form. It is in this case that the
asymmetry can induce plasma rotation. This question can be answered by
examining the exact solution, with specific numerical tools for the
determination of the main spectrum. It might be possible that the Riemann
curve is not a hyperelliptic surface but a manifold close to a finite-genus
curve.

\textbf{Aknowledgements}. The authors gretefully aknowledge the useful
discussions and encouragements form J. H. Misguich and R. Balescu. The
hospitality of the D\'{e}partement de Recherche sur la Fusion Control\'{e}e
(Cadarche, France) is aknowledged. This work has been partly supported by
the NATO Linkage Grant PST.CLG.977397.

\end{document}